\documentclass{pinchcr}   

\usepackage{graphicx}
\usepackage{epsfig}
\usepackage{epstopdf}
\usepackage{amsmath,amssymb}

\newcommand{\be}{\begin{equation}}
\newcommand{\ee}{\end{equation}}
\newcommand{\ba}{\begin{eqnarray}}
\newcommand{\ea}{\end{eqnarray}}
\newcommand{\nn}{\nonumber}

\newcommand{\ex}{{\rm e}}
\newcommand{\Tr}{{\rm Tr}}
\newcommand{\hTr}{\hat{\rm T}{\rm r}}
\newcommand{\Qb}{\bar{Q}}
\newcommand{\re}{{\rm Re}}
\newcommand{\im}{{\rm Im}}
\newcommand{\ho}{\hat{\omega}}
\newcommand{\psib}{\bar{\psi}}

\def\lsi{\raise0.3ex\hbox{$<$\kern-0.75em\raise-1.1ex\hbox{$\sim$}}}
\def\gsi{\raise0.3ex\hbox{$>$\kern-0.75em\raise-1.1ex\hbox{$\sim$}}}
\newcommand{\lsim}{\mathop{\lsi}}
\newcommand{\gsim}{\mathop{\gsi}}
\newcommand{\eq}{Eq.~}
\newcommand{\eqs}{Eqs.~}
\newcommand{\fig}{Fig.~}

\def\bfx{{\bf x}}
\def\bfy{{\bf y}}
\def\bfp{{\bf p}}
\def\slash#1{#1\!\!\!\!/\!\,\,} 
\def\Dslash{\slash D}

\title{Lattice QCD at non-zero temperature and baryon density}

\author{Owe Philipsen}
\affiliation{Institut f\"ur Theoretische Physik, Westf\"alische Wilhelms-Universit\"at M\"unster, \\
Wilhelm-Klemm-Str.9, 48149 M\"unster, Germany}
\authors{1}

\begin{document}

\maketitle


\preface
One of the key features of QCD is asymptotic freedom. 
While the theory is amenable to perturbation theory at large momenta, it
is non-perturbative for energy scales $\lsim 1$ GeV and lattice QCD is the only known method for
first principles calculations in this regime. 
The running of the coupling with momentum scale immediately implies the existence of
different states of nuclear matter at asymptotically high densities or temperatures: when
the coupling on the scales of temperature $T$ or baryon chemical potential $\mu_B$ is
sufficiently weak, confinement gets lost.  
At a critical temperature $T_c\sim 200$~MeV, QCD predicts a transition between the familiar confined
hadron physics and a deconfined phase of quark gluon plasma (QGP).
At the same temperature, chiral symmetry gets
restored. A thermal environment with sufficiently high temperatures 
for a QCD plasma has certainly existed during the early stages of the universe, 
which passed through the quark hadron transition on its way to its present state. 
On the other hand, for high densities and low temperatures, there is an attractive
interaction between quarks to form Cooper pairs, and exotic non-hadronic 
phases such as a colour superconductor have been predicted. 
Such physics might be realised in the cores of compact stars.

Current and future heavy ion collision experiments are
attempting to create hot and dense quark gluon plasma 
at RHIC (BNL), LHC (CERN) and FAIR (GSI).
These studies will have a bearing far beyond QCD in the context of 
early universe  and astro-particle physics.
Many other prominent features of the observable universe, such as the baryon asymmetry or the seeding
for structure formation, have been determined primordially 
in hot plasmas described by non-abelian gauge theories. The QCD plasma serves as a prototype 
also for those, since it is the only one we can hope to produce in 
laboratory experiments. There is thus ample motivation to provide controlled theoretical
predictions for the physics of hot and dense QCD. As we shall see, this turns out to be even 
more challenging than QCD in the vacuum. 

In these lectures, the focus is on basic
concepts and methods, with a few exemplary results for illustration only and a very incomplete
list of references. For summaries of the latest results and literature lists, see the annual
proceedings of the LATTTICE conferences.

\acknowledgements
I would like to thank the organizers of the school for the invitation to lecture here, and all the students
for the excellent working atmosphere, their  interest and stimulating questions. I also thank Philippe de Forcrand for continued collaboration and availability for discussion over many years, 
as well as Jens Langelage and Lars Zeidlewicz for calculational input and checking parts of the manuscript.

\tableofcontents

\maintext

\chapter{Aspects of finite temperature field theory in the continuum}

\section{Statistical mechanics reminder}

We wish to describe a system of particles in some volume $V$ which is in thermal
contact with a heat bath at temperature $T$. Associated with the particles  may be a set
of conserved charges $N_i, i=1,2,\ldots$ (such as particle number, electric charge, baryon number etc.).
In statistical mechanics there is a choice of ensembles to describe this situation.
In the canonical ensemble, $V$ and the $N_i$ are kept fixed while the system exchanges energy with the
heat bath. In the grand canonical description, also exchange of particles with the heat bath is allowed.
Of course, both ensembles provide the same description for physical observables in the thermodynamic
limit, $V\rightarrow\infty$, but for specific situations one or the other may be more appropriate.
In quantum field theory, the most direct description is in terms of the grand canonical ensemble.
Its density operator and partition function are given as
\be
\rho= \ex^{-\frac{1}{T}(H-\mu_iN_i)}\;,\qquad Z=\hTr \rho\;, \quad 
\hTr(\ldots)=\sum_n\langle n|(\ldots)|n\rangle\;,
\ee
where $\mu_i$ are chemical potentials for the conserved charges, and the quantum mechanical trace
is a sum over all energy eigenstates of the Hamiltonian. Thermodynamic averages
for an observable $O$ are then obtained as $ \langle O\rangle=Z^{-1}\hTr (\rho O)\;.$

From the partition function, all other thermodynamic equilibrium quantities follow by taking
appropriate derivatives. In particular, the (Helmholtz) free energy, pressure, entropy, 
mean values of charges and energy are obtained as

\begin{minipage}{6cm}
\ba
F&=&-T\ln Z\;,\nn\\ 
p&=&\frac{\partial (T\ln Z)}{\partial V}\;,\nn\\
S&=&\frac{\partial (T\ln Z)}{\partial T}\;,\nn
\ea
\end{minipage}
 \begin{minipage}{6cm}
 \ba \label{thermo}
\bar{N}_i&=&\frac{\partial (T\ln Z)}{\partial \mu_i}\;,\nn\\
 E&=&-pV+TS+\mu_i\bar{N}_i\;.\\ 
&&
\nn
\ea 
\end{minipage}
\vspace*{3mm}

\noindent
Since the free energy is known to be an extensive quantity, $F=fV$, and we are interested in the thermodynamic limit, it is often more convenient to consider the corresponding densities, 
\be
f=\frac{F}{V},\quad p=-f, \quad s=\frac{S}{V},\quad n_i=\frac{\bar{N}_i}{V},\quad \epsilon=\frac{E}{V}\;.
\ee 

\section{QCD at finite temperature and quark density}

Let us now consider the grand canonical partition function of QCD. 
The derivation of its path integral representation is discussed in detail 
in the textbooks \cite{Kapusta:2006pm} and 
requires discretised Euclidean time followed by a continuum limit. 
We shall thus derive it more conveniently in its lattice version in Sec.~\ref{latqcd} 
and just quote the result here,
\be  
Z(V,{\mu_f},T;g,{m_f})=\hTr \left(\ex^{-(H-\mu_f {Q_f})/T}\right)=
\int DA \,D\bar{\psi}\,D\psi \; \ex^{-S_g[A_\mu]} \,\ex^{-S_f[\bar{\psi},\psi,A_\mu]},
\label{part}
\ee
with the Euclidean gauge and fermion actions
\ba
S_g[A_\mu]&=& \int\limits_0^{1/T} d\tau \int\limits_V d^3x \;
\frac{1}{2} {\rm Tr}\; F_{\mu\nu}(x) F_{\mu\nu}(x), \nn\\
S_f[\bar{\psi},\psi,A_\mu]&=& \int\limits_0^{1/T} d\tau \int\limits_V d^3x \;
 \sum_{f=1}^{N_f} \bar{\psi}_f(x) \left( \gamma_\mu
D_\mu+ m_f- \mu_f \gamma_0 \right) \psi_f(x) .
\label{lagrangian}
\ea
The index $f$ labels the different quark flavours, and 
the covariant derivative contains the gauge coupling $g$,
\ba
D_\mu&=&(\partial_\mu-igA_\mu),\quad A_\mu=T^aA_\mu^a(x), \quad a=1,\ldots N^2-1,\nn\\
F_{\mu\nu}(x)&=&\frac{i}{g}[D_\mu,D_\nu]\;.
\ea
The thermodynamic limit is obtained by sending the spatial three-volume $V\rightarrow \infty$.
The difference to the Euclidean path integral at $T=0$ is that the temporal direction is compactified
and kept finite, i.e.~the theory lives on a torus whose compactification radius defines the inverse temperature, $1/T$.
The path integral is to be evaluated with periodic and anti-periodic boundary conditions in the temporal
direction for bosons and fermions, respectively,
\be
A_\mu(\tau,\bfx)=A_\mu(\tau+\frac{1}{T},\bfx),\qquad
\psi_f(\tau,\bfx)=-\psi_f(\tau+\frac{1}{T},\bfx)\;,
\label{bc}
\ee
which ensures Bose/Einstein statistics for bosons and the Pauli principle for fermions.
Clearly, the  path integral for vacuum QCD on infinite four-volume is 
smoothly recovered from this expression for $T\rightarrow 0$.

The partition function depends on 
the external macroscopic parameters $T,V,\mu_f$, as 
well as on the microscopic parameters like quark masses and the coupling constant. 
The conserved quark 
numbers corresponding to the chemical potentials $\mu_f$ are 
\be
Q_f=\bar{\psi}_f\gamma_0\psi_f\;.
\ee
Since the QCD phase transition happens on a scale $\sim 200$ MeV, we neglect
the $c,b,t$ quarks or treat them non-relativistically when needed. 
We will thus consider mostly two and three flavours of quarks,
and always take $m_u=m_d$. The case $m_s=m_{u,d}$ is then denoted by $N_f=3$, while $N_f=2+1$
implies $m_s\neq m_{u,d}$.
Furthermore, we will couple all flavours to the same chemical potential
$\mu$ unless otherwise stated.  

The only, but fundamental, difference compared to the $T=0$ path integral is due to the compactness of the time direction.
For example, the Fourier expansion of the fields on a finite volume $V=L^3$ is
\ba
A_\mu(\tau,\bfx)&=&\frac{1}{\sqrt{VT}}\sum_{n=-\infty}^{\infty}\sum_{\bfp} 
\ex^{i(\omega_n \tau + \bfp\cdot \bfx)}\,A_{\mu,n}(p)\;,\quad \omega_n=2n\pi T\;, \nn\\
\psi(\tau,\bfx)&=&\frac{1}{\sqrt{V}}\sum_{n=-\infty}^{\infty}\sum_{\bfp} 
\ex^{i(\omega_n \tau + \bfp\cdot \bfx)}\,\psi_{n}(p)\;,\quad \omega_n=(2n+1)\pi T\;.
\ea
The allowed momenta are $p_i=(2\pi n_i)/L$, where $n_i\in \mathbb{Z}$. The normalisation factors in front
are chosen such that the Fourier modes, called the Matsubara modes, are dimensionless. In the thermodynamic limit
the momenta become continuous
\be
\frac{1}{V}\sum_{n_1,n_2,n_3}\stackrel{V\rightarrow \infty}{\longrightarrow} \int\frac{d^3p}{(2\pi)^3},
\ee
but the Matsubara frequencies $\omega_n$ 
stay discrete due to the compactness of the time direction. 
Note that bosons/fermions have even/odd Matsubara frequencies, respectively, to ensure  the bondary
conditions \eq(\ref{bc}). We thus obtain modified
Feynman rules \cite{Kapusta:2006pm} compared to the vacuum. 
The zero components of four vectors contain the Matsubara frequencies $\omega_n$, and the 
four-momentum integration associated with internal lines gets replaced by a 3d integral and a 
Matsubara sum,
\be
\sum_{n=-\infty}^{\infty}\int \frac{d^3p}{(2\pi)^3}\;.
\label{matsum}
\ee

\section{Perturbative expansion}

Similar to $T=0$, we can take the path integral as starting point for a perturbative expansion.
For some generic field $\phi$, we proceed just as in the vacuum and
split the action in a free and an interacting part, $S=(S_0+S_i)$, in order to expand in powers
of the interaction, and hence the coupling constant,
\be
Z=N\int D\phi\;\ex^{-(S_0+S_i)}=N\int D\phi\; \ex^{-S_0}\sum_{l=0}^\infty\frac{(-1)^l}{l!}S_i^l\;.
\ee
Thus we get for the log of the partition function 
\be
\ln Z=\ln Z_0 + \ln Z_i=\ln\left(N\int D\phi\;\ex^{-S_0}\right)+
\ln \left(1+\sum_{l=1}^{\infty}\frac{(-1)^l}{l!}\frac{\int D\phi\;\ex^{-S_0}S_i^l}{\int D\phi\;\ex^{-S_0}}\right)\;.
\label{series}
\ee
We shall evaluate the ideal gas part, $\ln Z_0$, in the next section. The corrections to the non-interacting 
case are the sum of all loop diagrams without external legs. When evaluating loop diagrams, UV 
divergences are encountered and the renormalisation program has to be performed. 
Whatever regularisation and renormalisation is necessary and sufficient at zero temperature is also 
necessary and sufficient at finite temperature. This is not surprising, since the ultraviolet structure of 
the theory, i.e.~the microscopic short distance regime, 
is unchanged by the introduction of macroscopic parameters like temperature and baryon chemical potential. 

On the other hand, the infrared structure of the theory does get changed.
This is easily seen by considering the inverse propagator for a bosonic degree of freedom,
\be
p^2+m^2=\omega_n^2+\bfp^2+m^2=(2n\pi T)^2+\bfp^2+m^2\;.
\ee 
The Matsubara frequencies act like effective thermal masses $\sim T$ for all modes with $n\neq 0$.
In the case of fermions, there are only non-zero modes. Hence, all fermionic and all non-zero bosonic
modes are infrared-safe even in the limit of vanishing bare mass, $m=0$. By contrast, for the bosonic 
zero mode the inverse propagator is  $\bfp^2+m^2$, which is identical to that of a 3d field theory.
Thus, 4d Yang-Mills theory at finite temperature contains in the zero mode sector the 3d Yang-Mills theory,
which is a confining theory. 
Its propagator is infrared divergent for $m=0$, and the divergence is worse than in 4d. 
This points at non-perturbative behaviour and is at the heart of the Linde problem of finite temperature perturbation theory, Sec.~\ref{sec:linde}.

Let us briefly discuss qualitatively the self-energy corrections to the gluon propagator.
When computing the self energy diagrams, one finds that for the colour electric field $A_0$ a gluon mass is generated, the electric or Debye mass. To leading order it is
\be
m_E^{LO}= \left(\frac{N}{3}+\frac{N_f}{6}\right)^{1/2}\,gT\;.
\label{mel}
\ee
Thus, colour electric fields get screened by the medium at finite temperature, whereas for colour magnetic
fields $A_i$ one finds the corresponding magnetic mass $m_M=0$ at this order. However, there are 
contributions to $m_M\sim g^2T$ starting at the two-loop level. As we shall see in Sec.~\ref{sec:linde},
all loops contribute equally to the coefficient, such that the calculation of the magnetic screening mass is an 
entirely non-perturbative problem.

\section{Ideal gases} \label{ideal}

Ideal, i.e.~non-interacting, gases of particles are important model systems to guide our intuition.
It is therefore instructive to see how their thermal properties are derived from 
the path integral. Let us consider a bosonic system of real scalar fields, for simplicity.
After Fourier transformation, the action reads
\be
S_0=\frac{1}{2T^2}\sum_{n=-\infty}^{\infty}\sum_\bfp(\omega_n^2+\omega^2)\phi_n(p)\phi^*_n(p)
\label{actf}
\ee
where we abbreviate $\omega=\sqrt{\bfp^2+m^2}$ and $\phi^*_n(p)=\phi_{-n}(-p)$ for a real scalar field.
The partition function then factorises into a product over all Matsubara modes,
\ba
Z_0&=&N\prod_{n=-\infty}^\infty \prod_\bfp\int d\phi_n\;\exp[-\frac{1}{2T^2}(\omega_n^2+\omega^2)\phi_n(p)\phi^*_n(p)]\nn\\
&=&N
\prod_{n\geq 0}\prod_{\bfp \geq 0}\int d\phi_n\,d\phi_n^*\;\exp[-\frac{1}{2T^2}(\omega_n^2+\omega^2)\phi_n(p)\phi^*_n(p)]\nn\\
&=&N'
\prod_{n>0}\prod_{\bfp \geq 0}\int d|\phi_n|\,|\phi_n| \exp[-\frac{1}{2T^2}(\omega_n^2+\omega^2)|\phi_n|^2]\nn\\
&=&N'
\prod_{n\geq0}\prod_{\bfp \geq 0}\left(\frac{\omega_n^2+\omega^2}{T^2}\right)^{-1}=N'\prod_{n=-\infty}^{\infty}
\prod_\bfp\left(\frac{\omega_n^2+\omega^2}{T^2}\right)^{-\frac{1}{2}}\;,\nn
\ea
where we have changed to polar coordinates in the third line and absorbed constant numerical factors
into the normalisation in front. 
Thus, just as for zero temperature, the partition function of the free theroy can be formally written as
\be
Z_0=N\int D\phi\;\ex^{-S(\phi)}=N'(\det \Delta^{-1})^{-1/2}\;,
\ee 
where $\Delta^{-1}=(\omega_n^2+\omega^2)/T^2$ is the inverse propagator in momentum space.

Since $N'$ is $(V,T)$-independent,  it will not contribute to 
thermodynamics, \eqs(\ref{thermo}), and may be dropped. (It will contribute e.g.~to the entropy as an 
additive constant, which we are allowed to set to zero by the third law of thermodynamics). Thus we 
obtain
\be
\ln Z _0=-\frac{1}{2}\sum_{n=-\infty}^\infty\sum_\bfp\ln\frac{\omega_n^2+\omega^2}{T^2}\;.
\ee
The Matsubara sum is performed using the formulae
\ba
\ln\left[(2\pi n)^2+\frac{\omega^2}{T^2}\right]&=&\int_1^{\omega^2/T^2}\frac{d\theta^2}{\theta^2+(2\pi n)^2}
+\ln(1+(2\pi n)^2)\;,\nn\\
\sum_{n=-\infty}^\infty\frac{1}{n^2+(\frac{\theta}{2\pi})^2}&=&\frac{2\pi^2}{\theta}
\left(1+\frac{2}{\ex^\theta-1}\right)\;,
\ea
leading to the expression
\ba
\ln Z _0&=&-\sum_\bfp\int_1^{\omega/T}d\theta\; \left(\frac{1}{2}+\frac{1}{\ex^\theta-1}\right) + \mbox{T-indep.}
\nn\\     &\stackrel{V\rightarrow \infty}{\longrightarrow}&V\int \frac{d^3p}{(2\pi)^3}\left[\frac{-\omega}{2T}
        -\ln\left(1-\ex^{-\frac{\omega}{T}}\right)\right]\;.
\ea         
One observes that the integral over the first term diverges in the UV since $\omega\sim |\bfp|$. This however
is familiar from the quantum mechanical harmonic oscillator. The term corresponds to
the zero point energy and gives a divergent vacuum contribution to the energy and pressure for 
$T\rightarrow 0$. The 
renormalisation condition is that  the vacuum has zero pressure,
\be
p_{\rm phys}(T)=p(T)-p(T=0)\;.
\ee
The final result then is the familar form of the partition function for a free gas of spinless bosons,
\be
\ln Z_0=-V\int\frac{d^3p}{(2\pi)^3}\;\ln\left(1-\ex^{-\frac{\omega}{T}}\right).
\label{freeb}
\ee
For the massless case, $m=0$, the momentum integral can be done exactly and one 
finds the famous result for the pressure of one bosonic degree of freedom at zero chemical
potential,
\be
p=\frac{\pi^2}{90}T^4\;.
\label{psb}
\ee
Doing the same calculation for non-abelian gauge fields $A_\mu^a$, 
each field component corresponds to one
bosonic mode and we have to sum over $a=1,\ldots N^2-1$ and $\mu=1\ldots 4$ in \eq(\ref{actf}).
Hence we get a factor of $4(N^2-1)$ in front of \eq(\ref{freeb}).
Going through a similar sequence of steps for a free Dirac field, one finds instead
\be
\ln Z_0=2V\int\frac{d^3p}{(2\pi)^3}\;\left[\ln\left(1+\ex^{-\frac{\omega-\mu}{T}}\right)
+\ln\left(1+\ex^{-\frac{\omega+\mu}{T}}\right)\right].
\label{freef}
\ee
The factor of two in front accounts for the two spin states of a fermion, and there are two terms
from fermion and anti-fermion in the brackets.
Recall that for gauge theories gauge fixing is necessary in order to invert the 
two-point function. Thus, for the free photon or Yang-Mills gas, two of the four bosonic Lorentz 
degrees of freedom get cancelled by
the corresponding ghost contributions, such that we obtain two polarisation states per massless
vector particle, as it should be.

All of this can be summarised by the one-particle partition function
\be
\ln Z_i^1(V,T)=\eta V\nu_i\int \frac{d^3p}{(2\pi)^3}\; \ln(1+\eta\,\ex^{-(\omega_i-\mu_i)/T})\;,
\ee
where $\eta=-1$ for bosons and $\eta=1$ for fermions and 
$\nu_i$ gives the number of degrees of freedom for the particle $i$.
The momentum integration for one massless fermionic degree of freedom gives the pressure
\be
p=\frac{7}{8}\frac{\pi^2}{90}T^4\;.
\ee
To compute the pressure of an ideal gas of gluons and massless quarks, 
we now simply have to count the degrees of freedom to obtain (two polarisation states for each of $(N^2-1)$ gluons, 
two polarisation states per quark, three colours per quark and the same for anti-quarks)
\be
\frac{p}{T^4}=\left(2(N^2-1)+4N N_f\frac{7}{8}\right)\frac{\pi^2}{90}\;.
\ee 
This is the Stefan-Boltzmann limit of the QCD pressure which is valid for vanishing coupling, 
i.e.~in the limit $T\rightarrow \infty$.

\section{The hadron resonance gas model}

There is another ideal gas system that is useful to model the low temperature behaviour
of QCD. The modelling assumption is that, at low temperatures when we still have confinement, 
the QCD partition function is close to that of a free gas of hadrons. 
While strong coupling effects are responsible for  
confinement of the quarks and gluons, the interactions between the 
hadrons are considerably weaker and may even be
neglected if the gas is sufficiently dilute. In this case, the QCD partition function factorises into
one-particle partition functions $Z_i^1(V,T)$,
\be
\ln Z(V,T)\approx \sum_i\ln Z_i^1(V,T)\;. 
\ee 
The particles to be inserted are the known hadrons
and hadron resonances (for QCD purposes electroweak decays are to be neglected).
Taking the $\nu_i$ and the energies from the particle data booklet, one can 
supply this formula with hundreds of hadron resonances, do the momentum integral and 
obtain a thermodynamic pressure that compares remarkably well with Monte Carlo simulations
of QCD \cite{Karsch:2003vd}.  We shall see in Sec.~\ref{lhrg} that this is no accident, but that 
the hadron resonance gas model can actually be derived from lattice QCD
as an effective theory for the strong coupling regime.

\section{The Linde problem} \label{sec:linde}

\begin{figure}[t]
\begin{center}
\includegraphics[width=0.4\textwidth]{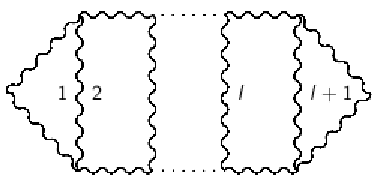}
\end{center}
\caption[]{$l+1$-loop Feynman diagram contributing to the pressure.}
\label{diagLinde}
\end{figure}
As an example to illustrate the Linde problem, consider the $l+1$-loop contribution to the QCD pressure,
\fig\ref{diagLinde}, with $2l$ three gluon vertices and $3l$ propagators. 
The Matsubara sums over the internal lines, \eq(\ref{matsum}), contain a term coming
exclusively from the zero modes. Dispensing with the index structures, its contribution is given by
the 3d loop integral
\be
I\sim g^{2l}\left(T\int d^{3}p\right)^{l+1}p^{2l}(p^{2}+{m}^{2})^{-3l}\;,
\ee 
where we have introduced a mass $m$ by hand as an infrared regulator. The momentum integrals have
to be performed up to a scale $T$, which appears as an effective UV cut-off after doing 
the Matsubara sum over the other modes. Parametrically, the integral then evaluates to
\be
I\sim\left\{
\begin{array}{ll}
g^6T^4\ln(T/m)  & \mbox{for}\quad l=3\\
g^{6}T^{4}(g^{2}T/{m})^{l-3} & \mbox{for} \quad l>3 \;.
\end{array} \right. 
\ee
It is therefore infrared divergent for $m\rightarrow 0, l>2$ and the usual bare perturbation 
theory breaks down. However, as we discussed before, mass scales are generated
dynamically by loop corrections to the propagators. Evaluating the integral with those mass
scales corresponds to an effective resummation of the perturbative series (the diagram now contains
loop insertions that would make it formally higher order in bare perturbation theory).
If our internal lines contain $A_0$ fields, we need to do so with $m_E\sim gT$ and observe that 
as a consequence our series contains odd powers of $g$ in addition to the logarithms. 
Summing up mass corrections for 
the $A_i$ amounts to an insertion of $m_M\sim g^2T$, and we see that the gauge coupling drops out
of the effective expansion parameter entirely for $l>3$.  Hence, { \it all} loop orders contribute 
to the pressure at order $g^6$, which thus is a fully non-perturbative problem.   

The same problem occurs when calculating other quantities, with the order of the breakdown depending
on the observable. For example, for the electric mass $m_E$ it appears already at NLO and for
the magnetic mass $m_M$ even at the leading non-zero order.
Thus at finite temperature, perturbation theory only works to some finite order which depends on the 
observable. Note that this is true no matter how weak the coupling $g$. Even electroweak physics
at finite temperatures is inherently non-perturbative, and perturbative answers are only useful to the 
extent that the calculable orders are sufficient for a good approximation of physical results.

\chapter{The lattice formulation for zero baryon density}

\section{Action and partition function}\label{latqcd}

\label{latte}

Let us now consider the lattice formulation of $SU(N)$ pure gauge theory on a
hypercubic lattice, $N_s^3\times N_\tau$, with lattice spacing $a$ and
the standard Wilson gauge action
\be
S_g[U]=\sum_{x}\sum_{1\leq\mu<\nu\leq 4}\beta\left(1-\frac{1}{3}{\rm Re}
\Tr U_p\right), 
\ee
where
$U_p=U_\mu(x)U_\nu(x+a\hat{\mu})U^\dag_\mu(x+a\hat{\nu})U^\dag_\nu(x)$ is
the elementary plaquette, and the bare lattice and continuum gauge couplings 
are related by $\beta=2N/g^2$.
As usual, we impose periodic boundary conditions in all directions, 
$U_\mu(\tau,\bfx)=U_\mu(\tau+N_\tau,\bfx), U_\mu(\tau,\bfx)=U_\mu(\tau,\bfx+N_s)$.
The connection between zero and finite temperature physics is most easily 
exhibited by the transfer matrix, which relates the 
path integral representation of a Euclidean lattice field theory to the 
Hamiltonian formulation. A transfer matrix element between
two times slices $\tau$ and $\tau+1$ is given by \cite{Montvay:1994cy},
\be
T[U_i(\tau+1),U_i(\tau)] = \ex^{-aH}=\int DU_0(\tau)\;\exp-L[U_i(\tau+1),U_0(\tau),U_i(\tau)],
\label{transfer}
\ee
where the action is written as a sum over time slices,
\ba
S_g&=&\sum_\tau L[U_i(\tau+1),U_0(\tau),U_i(\tau)],\nn\\
L[U_i(\tau+1),U_0(\tau),U_i(\tau)]&=&\frac{1}{2}L_1[U_i(\tau+1)]+\frac{1}{2}L_1[U_i(\tau)]\nn\\
&&+L_2[U_i(\tau+1),U_0(\tau),U_i(\tau)]\,\nn\\
L_1[U_i(\tau)]&=&-\frac{\beta}{N}\sum_{p(\tau)} \re \Tr U_p,\nn\\
L_2[U_i(\tau+1),U_0(\tau),U_i(\tau)]&=&-\frac{\beta}{N}\sum_{p(\tau,\tau+1)}\re \Tr U_p,
\label{trans}
\ea
and $p(\tau),U_i(\tau)$ denote all spatial plaquettes and links contained in the timeslice $\tau$ with
arguments $\bfx$ suppressed,
while $p(\tau+1,\tau),U_0(\tau)$  are timelike plaquettes and links connecting the timeslices
$\tau$ and $\tau+1$.
The partition function is now conveniently expressed as
\be
Z=\int \prod_\tau \left( DU_i(\tau,\bfx)\;T[U_i(\tau+1),U_i(\tau)]\right)=\hTr(T^{N_\tau})=\hTr(\ex^{-N_\tau aH})\;.
\ee
Note that the periodic boundary condition in the temporal direction is necessary for the trace
operation in order to have identical states $|n\rangle$
on the time slices $1$ and $N_\tau$.
In this discretised form we can immediately see that $Z$ is equivalent to the partition function
of a thermal system if we identify 
\be
\frac{1}{T}\equiv a N_\tau\;.
\label{temp}
\ee
The thermal expectation value of an observable is then
\be
\langle O \rangle = Z^{-1}\hTr(\ex^{-\frac{H}{T}} O)
=Z^{-1}\sum_n\langle n|T^{N_\tau}O|n\rangle
=\frac{\sum_n\langle n| O|n\rangle\, \ex^{-aN_\tau E_n}}{\sum_n \ex^{-aN_\tau E_n}}\;.
\label{boltz}
\ee 
As in the continuum, we are interested in the thermodynamic limit and hence $N_s\rightarrow \infty$, while
keeping $aN_\tau=1/T$ finite.

In this form we easily see the connection to $T=0$ physics:
projection on the vacuum expectation value is achieved by taking $N_\tau$ to infinity,
\be
\langle 0|O|0\rangle = \lim_{N_\tau\rightarrow\infty}
\frac{\sum_n\langle n| O|n\rangle\, \ex^{-aN_\tau (E_n-E_0)}}{\sum_n \ex^{-aN_\tau (E_n-E_0)}}\;.
\ee
In order to describe our gauge theory at finite temperatures, we simply need to dispense with this step.
In that case the expectation value receives contributions from all eigenstates $|n\rangle$ with
Boltzmann weights $\ex^{-\frac{E_n}{T}}$, \eq(\ref{boltz}). 
Hence, all lattices with finite $N_\tau$ (in particular those used for numerical simulations!) 
correspond to a finite temperature $T=1/(aN_\tau)$,
and for a description of vacuum physics sufficiently large $N_\tau$ is required.

On a Euclidean lattice, space and time directions are in principle indistinguishable as long
as we provide them with equal boundary conditions. 
For some applications it is useful 
to define a Hamiltonian that translates states in a spatial, say the $z$-direction,
defined through a transfer matrix between adjacent $z$-slices, and write the partition function 
in terms of that Hamiltonian (with $U(z)$ denoting $\{U_\mu(z)|\mu\neq 3\}$),
\be
T[U(z+1),U(z)]\equiv\ex^{-aH_z},\quad Z=\Tr(\ex^{-aN_zH_z})\;.
\label{hz}
\ee
For thermal physics, we want to take $N_{x,y,z}\rightarrow \infty$ and keep
$N_\tau$, which is now hidden in the definition of the Hamiltonian, finite.
It is thus equivalent to the ``zero temperature'' ($N_z\rightarrow\infty$) 
physics of the Hamiltonian $H_z$, which acts on states defined
on a space with two infinite and one finite, compactified direction. Clearly, $H_z$ has reduced symmetry
compared to $H$.
In either description, from a calculational point of view finite temperature physics on a Euclidean space-time lattice is
nothing but a finite size effect: 
the Boltzmann weighted sums, i.e.~thermal effects, become
noticeable once the temporal lattice size is small enough for the system to be sensitive to the boundary.

Adding fermions is now the same as for zero temperature and does not change this picture. Once a suitable action 
\be
S_{f}= \sum_{x,y} \bar{\psi}(x) M_{xy}(m_f)\, {\psi}(y)
\ee
has been selected,
the Gauss integral can be done and,  minding the appropriate boundary conditions in the temporal
direction, we end up with the partition function
\ba
Z(N_s,N_\tau;\beta,m_f)&=&\int DU\;\prod_f\det M(m_f)\; \ex^{-S_g[U]}\;,\nn\\
U_\mu(\tau,\bfx)&=&U_\mu(\tau+N_\tau,\bfx),\nn\\
 \psi(\tau,\bfx)&=&-\psi(\tau+N_\tau,\bfx)\;.
\ea
For definiteness, the lattice action for $N_f$ degenerate Wilson fermions is given by
\ba
S_f^W&=&\frac{1}{2a}\sum_{x,\mu,f}a^4\,\psib_f(x)[(\gamma_\mu-r)U_\mu(x)\psi_f(x+\hat{\mu})
-(\gamma_\mu+r)U_\mu^\dag(x-\hat{\mu})\psi_f(x-\hat{\mu})]\nn\\
&&+\,(m+4\frac{r}{a})\sum_{x,f}a^4\,\psib_f(x)\psi_f(x).
\ea

\section{Tuning temperature and the continuum limit}\label{tunet}

According to \eq(\ref{temp}), 
one way of tuning temperature on the lattice is by choosing $N_\tau$. But this is not
satisfactory as this is only possible in discrete steps, and for realistic lattice spacings these
are much too coarse. Hence, common practice in numerical simulations is to keep $N_\tau$ fixed and instead vary the lattice spacing $a$ via the lattice coupling, $\beta=2N/g^2(a)$, thus affecting 
temperature. This is a marked difference to simulations at zero temperature. In particular, simulation 
points at different temperature correspond to different lattice spacings and thus have different cut-off effects.
  
The relation between $a$ and $\beta$ is given by the renormalisation group. 
E.g., for the Lambda-parameter in lattice regularisation we have to leading order in perturbation theory
for $SU(3)$
\ba
a\Lambda_L&=&\left(\frac{6b_0}{\beta}\right)^{-b_1/2b_0^2}\;\ex^{-\frac{\beta}{12b_0}},\nn\\
b_0&=&\frac{1}{16\pi^2}\left(11-\frac{2}{3}N_f\right), \quad b_1=\left(\frac{1}{16\pi^2}\right)^2
\left[102-\left(10+\frac{2}{3}\right)N_f\right]\;.
\ea
However, perturbation theory is 
not convergent for accessible lattice spacings. 
The way out is to express the calculated 
observables in terms of known physical quantities of the same mass dimension.
For example, if we want to compute the critical temperature of the QCD phase transition, $T_c$,
by keeping $N_\tau$ fixed and tuning $\beta$, the location of 
a phase transition will be given as a critical coupling $\beta_c$.
The critical temperature in units of a hadron mass is then
\be
\frac{T_c}{m_H}=\frac{1}{a_cm_H N_\tau}=\frac{1}{a(\beta_c)m_H N_\tau}\;.
\ee
In order to set the physical scale for $T_c$, we then have to calculate the zero temperature 
hadron mass in lattice units at the value of the critical coupling, $(am_H)(\beta_c)$.

This procedure is good as long as we are able to simulate physical quark masses. For most
practitioners, this is not yet the case, and furthermore there are many interesting theoretical questions
concerning regimes with unphysical quark masses, such as the quenched and the chiral limits.
In order to set a scale in those cases, one uses quantities that display only little sensitivity to the quark mass values. Examples are the string tension or the Sommer scale,
\be
\frac{T}{\sqrt{\sigma}}=\frac{1}{a\sqrt{\sigma}N_\tau}, \quad \sigma\approx 425 \;\mbox{MeV};\quad 
Tr_0=\frac{r_0}{aN_\tau}, \quad r^2\frac{dV(r)}{dr}=1.65\;.
\ee

In order to take the continuum limit, we now have to compute expectation values of 
our observables of interest, $\langle O\rangle(\beta,m_f)$, in the thermodynamic limit for various
lattice spacings $a$. Then we can extrapolate to $a\rightarrow 0$. Since the continuum limit has to be 
taken along lines of constant physics, i.e. keeping temperature and mass ratios fixed, this is equivalent
to taking $N_\tau\rightarrow \infty$.  A continuum extrapolation therefore requires simulations
on a sequence of lattices with different $N_\tau$. For example, at a temperature $T=200$ MeV $\sim 1$ fm$^{-1}$,  \eq(\ref{temp}) with $N_\tau=4,8,12$ implies
$a\approx 0.25,0.125,0.083$ fm.

\section{Constraints on lattice simulations and systematic errors}
\label{sec:syst}

It is important to realise from the outset that current lattice simulations
at finite temperature and density are still hampered by sizeable systematic 
errors and uncertainties. Let us discuss the origins of those. The Compton wave length of a hadron is proportional to its inverse mass $m_H^{-1}$, and the largest of those constitutes the correlation length of the statistical system. 
To keep finite size as well as discretisation errors small, we need to require
\be
a\ll m_H^{-1}\ll aN_s.
\label{latconst}
\ee
For the low T phase we thus need a lattice size of several inverse pion masses. 
The push to do physical quark masses is only just beginning to be a possibility on the most powerful machines and with the cheapest actions
(i.e.~staggered, with Wilson rapidly catching up). On the other hand, at high $T$
screening masses scale as $m_H\sim T$, thus
\be
N_\tau^{-1}\ll 1\ll N_sN_\tau^{-1}.
\ee 
Hence, while we desire large $N_\tau$, the spatial
lattice size should be significantly larger than the temporal one. 
This limits the directly accesible temperatures to several $T_c$. 

Another important subject are cut-off effects. Similar to zero temperature, one can formally expand
lattice observables in powers of the lattice spacing about their continuum limit.
According to the discussion above, this is tantamount to expanding in the dimensionless
$N_\tau^{-1}$ about zero. In order to reduce cut-off effects, one can then apply improved
actions and operators, again in complete analogy to QCD in the vacuum. 

In order to study thermodynamic behaviour, one is often interested in the temperature dependence
of certain quantities. This already implies sequences of simulations at many $\beta$-values 
for just one $N_\tau$. If one is interested in a phase transition, one is moreover confronted with 
fluctuations between different phases, increasing correlation lengths and critical slowing down, i.e.~increased auto-correlation times. For these reasons most thermal simulations require orders 
of magnitude more
Monte Carlo trajectories than typical zero temperature problems.  Simulations are thus necessarily 
run on comparatively coarse lattices, implying larger systematic errors which are so far less
controlled than for vacuum physics. 

\section{The ideal gas on the lattice}

Similar to Sec.~\ref{ideal}, we can evaluate the ideal gas for a bosonic field on the lattice.
Starting point is the equation employing the free propagator, 
\ba
\ln Z_0&=&-\frac{1}{2} \ln\det \Delta=\frac{1}{2}\Tr\ln \Delta^{-1}\nn\\
&=&V\sum_{n=-N_\tau/2}^{N_\tau/2-1}
\int_{\frac{\pi}{a}}^{\frac{\pi}{a}}\frac{d^3p}{(2\pi)^3}\ln(\hat{p}^2+(am)^2)
\nn\\
&=&V\sum_{n=-N_\tau/2}^{N_\tau/2-1}
\int_{\frac{\pi}{a}}^{\frac{\pi}{a}}\frac{d^3p}{(2\pi)^3}\ln\left(4\sin^2(\frac{a\omega_n}{2})+4\ho^2\right),
\ea 
where now we have to deal with the lattice momenta,
\ba
\hat{p}^2&=&4\sin^2(\frac{a\omega_n}{2})+4\sum_{j=1}^3\sin^2(\frac{ap_j}{2}),\nn\\
4\ho^2&=&4\sum_{j=1}^3\sin^2(\frac{ap_j}{2})+(am)^2\;.
\ea
Due to the finite lattice spacing
the Matsubara sum is only from $-N_\tau/2,....N_\tau/2-1$.
In order to perform the sum, we employ the formula \cite{Kaste:1997ks}
\be
\frac{1}{N_\tau}\sum_{n=-N_\tau/2}^{N_\tau/2-1}g(\ex^{i\omega_n})=
-\sum_{z_i}\frac{{\rm Res}(\frac{g(z_i)}{z_i})}{z_i^{N_\tau}-1}
\ee
to the derivative of the sum of logs, 
\ba
L(\omega^2)&\equiv&\frac{1}{N_\tau}\sum_n\ln\left(4\sin^2(\frac{a\omega_n}{2})+4\ho^2\right),\nn\\
\frac{dL}{d\hat{\omega}^2}&=&\frac{1}{N_\tau}\sum_n\frac{4}{4\hat{\omega}^2+2-\ex^{i\omega_n}
-\ex^{-i\omega_n}},\nn\\
\frac{g(z)}{z}&\equiv&\frac{4}{4\hat{\omega}^2+2-z-z^{-1}}\;.
\ea
This function has simple poles at $z=2\hat{\omega}^2+1\pm2\ho\sqrt{\ho^2+1}$.
It is now convenient to change variables by $\ho=\sinh(aE/2)$ to obtain
\ba
\frac{dL}{d(aE)}&=&\frac{dL}{d\ho^2}\frac{d\ho^2}{d(aE)}=2\left(\frac{1}{\ex^{N_\tau aE}-1}+1\right), \nn\\
L&=&\frac{2}{N_\tau}\ln\left(1-\ex^{-N_\tau aE}\right)+2aE\;.
\ea
We recognise again the vacuum energy contribution, which needs to be subtracted to arrive at the final
result for the pressure of a bosonic gas on the lattice,
\be
\ln Z_0=-V\int_{-\frac{\pi}{a}}^{\frac{\pi}{a}} \frac{d^3p}{(2\pi)^3}\ln(1-\ex^{-N_\tau aE})\;.
\ee

One can now study the approach to the continuum by  extending the integration range to infinity
and expanding the log in small lattice spacing. The 
leading mistake we make by the first step in the positive integration range
of the integral is
\be
I(a)=\int_{\frac{\pi}{a}}^{\infty}\frac{d^3p}{(2\pi)^3}\ln(1-\ex^{-N_\tau aE})
=I(0)+\frac{dI}{da}\,a+\ldots ,
\ee
and similarly for the negative range.
With $I(0)=0$ and $I'(a)\propto \exp{-N_\tau \pi}$, with $N_\tau=1/(aT)$, this is exponentially small and can
be neglected compared to the power corrections of the integrand. For the expansion of the integrand,
we need the corrections to the dispersion relation. Writing 
\be
E(\bfp)=E^{(0)}(\bfp)+aE^{(1)}(\bfp)+a^2E^{(2)}(\bfp)+\ldots, \qquad
E^{(0)}(\bfp)=\sqrt{\bfp^2+m^2},
\ee 
and expanding both sides of the equation
\ba
\sinh^2(\frac{aE}{2})&=&\sum_{j=1}^3\sin^2(\frac{ap_j}{2})+\frac{(am)^2}{4},\nn\\
\frac{(aE)^2}{4}+\frac{(aE)^4}{48}&=&\sum_{j=1}^3\left(\frac{(ap_j)^2}{4}-\frac{(ap_j)^4}{48}\right)
+\frac{(am)^2}{4}+O(a^6),
\ea
one finds for the $a^2$-correction to the dispersion relation
\be
E^{(2)}(\bfp)=-\frac{1}{24E^{(0)}(\bfp)}\left(\sum_{j=1}^3p_j^4+E^{(0)4}(\bfp)\right).
\ee
Changing to dimensionless variables, $x=p/T, \varepsilon=E/T$, we then have for the expansion
of the pressure
\be
 \frac{p}{T^4}=\left(\frac{p}{T^4}\right)_{\rm cont}-a^2\int\frac{d^3x}{(2\pi)^3}
 \frac{\varepsilon^{(2)}(x)}{\ex^{\varepsilon^{(0)}(x)}-1}+\ldots
\ee
This means that the pressure of
a free gas of bosons has leading lattice corrections of $O(a^2)$. 
For the massless case the integral can again be done and one arrives at
\be
\frac{p}{p_{\rm cont}}=1+\frac{8\pi^2}{21}\frac{1}{N_\tau^2}+O\left(\frac{1}{N_\tau^4}\right)\;,
\ee
where $p_{\rm cont}$ is the continuum result \eq(\ref{psb}). 

One can repeat a similar calculation for various fermion actions and discuss their cut-off effects.
For the free gas in the chiral limit this can be found in \cite{Hegde:2008nx}, leading order
interactions and mass
effects have been evaluated in \cite{Philipsen:2008gq}.

\section{The quenched limit of QCD and $Z(N)$-symmetry}\label{zn}

In the quenched limit, i.e.~when quarks are infinitely heavy, $m_f\rightarrow \infty$, the QCD partition
function reduces to that of a pure gauge theory plus static quark fields. 
Let us examine the consequences of the compact temporal direction for gauge symmetry. 
The action is of course invariant under standard gauge transformations,
\be
S_g[U^g]=S_g[U]\quad \mbox{with}\quad
U^g_\mu(x)=g(x)U_\mu(x)g^{-1}(x+\hat{\mu}), \quad g(x)\in SU(N),
\ee
with our earlier periodic boundary conditions 
\be
U_\mu(\tau,\bfx)=U_\mu(\tau+N_\tau,\bfx), \quad g(\tau,\bfx)=g(\tau+N_\tau,\bfx)\;.
\ee
However, with the temporal boundary in place, we can also consider 
gauge transformations with topologically non-trivial matrices $g'(x)$, i.e.~matrices
that cannot be taken to unity by a smooth change of the parameters characterising the group elements.
Consider a boundary condition on the transformation matrices, which is only periodic up to a constant matrix $h$,  
\be
g'(\tau+N_\tau,\bfx)=hg'(\tau,\bfx)\;,\quad h\in SU(N)\;.
\ee
Such a $g'(x)$ does not go into itself when winding around the torus once, but picks up a ``twist'' factor 
$h$.
After a gauge transformation with $g'$, the gauge links behave across the boundary as
\be
U^{g'}_\mu(\tau+N_\tau,\bfx)=h\,U^{g'}_\mu(N_\tau,\bfx)\,h^{-1}\;.
\ee
They are only consistent with the periodicity requirement if $[h,U_i^{g'}]=0$.
This is satisfied if $h$ is proportional to the unit matrix, i.e.~it is in the centre of the group,
\be
h=z{\bf 1}\in Z(N),\quad z=\exp i\frac{2\pi n}{N}, \quad n\in \{0,1,2,\ldots N-1\}\;.
\ee
Thus, pure gauge theory at finite temperature
is invariant under gauge transformations with non-trivial winding through the temporal boundary, for any
global twist factor in $Z(N)$.
Note that this is {\it not} a symmetry of the QCD Hamiltonian. The latter is acting on a particular time slice
and does not know about temporal boundary conditions. The Hamiltonian generates translations in 
Euclidean time and its symmetries are defined on the space orthogonal to that.
Rather, the centre symmetry discussed here is a symmetry of the space-wise Hamiltonian $H_z$,
\eq(\ref{hz}), which acts on a $z$-slice and thus {\it is} sensitive to the boundary in the temporal direction.
 
Since the centre symmetry is related to non-trivial gauge transformations winding through
the temporal boundary, gauge invariant observables can only be sensitive to it if 
they wind too. Such an observable is
a Wilson line in the temporal direction closing onto itself, a Polyakov loop, 
\be
L({\bfx})=\prod_{x_0}^{N_\tau}U_{0}(x)\;.
\ee 
Physically, it corresponds to the propagator of a static quark. Under gauge transformations,
\ba
L^g(\bfx)&=&g(x)L(\bfx)g^{-1}(x),\nn\\
 L^{g'}(\bfx)&=&g'(1,\bfx)L(\bfx){g'}^{-1}(1+N_\tau,\bfx)=g'(1,\bfx)L(\bfx){g'}^{-1}(1,\bfx)h^{-1}\;.
\ea
Thus we see that the traced Polyakov loop is gauge invariant under topologically trivial gauge
transformations, while it picks up a centre element when transformed with a winding transformation,
\be
\Tr L^g=\Tr L,\quad \Tr L^{g'}=z^*\Tr L\;.
\label{polyz}
\ee
The Polyakov loop emerges naturally in the QCD path integral to leading order in the hopping 
expansion. For 
example, the partition function for a pure gauge theory with a static quark sitting at $\bfx$ is
\be
Z_Q=\int DU\;\Tr L(\bfx)\;\ex^{-S_g[U]}\;.
\label{zq}
\ee
Hence,
\be
\langle \Tr L\rangle = \frac{1}{Z}\int DU \;\Tr L \; \ex^{-S_g}=\frac{Z_Q}{Z}=\ex^{-(F_Q-F_0)/T}\;,
\ee
and the expectation value of the Polyakov loop gives the free energy difference between a 
Yang-Mills plasma
with and without the static quark \cite{McLerran:1981pb}. From this we can immediately infer two limiting cases:
For $T\rightarrow 0$ Yang-Mills theory is confining and it would cost infinite energy to remove the quark
to infinity, i.e.~$F_Q=\infty$ and therefore $\langle \Tr L \rangle =0$.
On the other hand, $T\rightarrow \infty$ corresponds to $\beta\rightarrow \infty$, for which 
$U_0\rightarrow {\bf 1}$ and $\langle \Tr L\rangle \rightarrow \Tr {\bf 1}=N$. 
Clearly, a non-zero expectation value is no longer invariant under centre transformations and  signals the
spontaneous breaking of centre symmetry.
Therefore, QCD in the quenched limit has a true (non-analytic) deconfinement phase transition corresponding to the breaking of the global centre symmetry, and the average
of the Polyakov loop is the corresponding order parameter. 

If we add dynamical quark fields to the theory, they will behave as
\be
\psi^{g}(x)=g(x)\psi(x),\quad \psi(\tau+N_\tau,\bfx)=-\psi(\tau,\bfx),\quad
\psi^{g'}(\tau+N_\tau,\bfx)=-h\psi(\tau,\bfx)\;.
\ee
Since statstical mechanics requires anti-periodic boundary conditions for fermions, the trivial $h=1$ is the only permissible choice, i.e.~there is {\it no} centre symmetry in the presence of dynamical quarks. 
Physically, if there are dynamical quarks, their pair production screens the confining force (it leads
to string breaking) and $F_Q$ is finite.
Correspondingly, $\langle \Tr L\rangle\neq 0$ for all temperatures
and the Polyakov loop is no longer a true order parameter. In this case, a non-analytic
phase transition as a function of temperature is not necessary, 
confined and deconfined regions may also be analytically 
connected by a smooth crossover.

\section{The chiral limit} \label{chiral}

Chiral symmetry and its breaking on the lattice has been the subject of another set of lectures at this 
school.
Let us merely summarise what we need for the finite temperature discussion.
In the limit of zero quark masses the classical QCD Lagrangian in the continuum 
is invariant under global chiral symmetry
transformations, the total symmetry being $U_A(1)\times SU_L(N_f)\times SU_R(N_f)$. The
axial $U_A(1)$ is anomalous, quantum corrections break it down to $Z(N_f)$. The remainder
gets spontaneously broken to the diagonal subgroup, $SU_L(N_f)\times SU_R(N_f)\rightarrow 
SU_V(N_f)$, giving rise to $N_f^2-1$ massless Goldstone bosons, the pions. The order parameter
signalling chiral symmetry is the chiral condensate,
\be
\langle\bar{\psi}\psi\rangle =\frac{1}{N_s^3N_\tau}\frac{\partial}{\partial m_f} \ln Z.
\ee
It is nonzero for $T<T_c$, when chiral symmetry is spontaneously broken, and zero for $T>T_c$.
Hence there is
a non-analytic finite temperature phase transition corresponding to chiral symmetry restoration.
For non-zero quark masses, chiral symmetry is broken explicitly and the chiral condensate 
$\langle\bar{\psi}\psi\rangle\neq 0$ for all temperatures. Again, in this case there is no need
for a non-analytic phase transition.
Of course, for most lattice fermions chiral symmetry is either reduced (staggered) or broken 
completely (Wilson), rendering the behaviour in the chiral limit a most difficult subject of study.

\section{Physical QCD}

QCD with physical quark masses obviously does not correspond to either the chiral or quenched
limit. The $Z(3)$ 
symmetry as well as the chiral symmetry are explicitly broken. 
Nevertheless, physical QCD displays confinement as well as three very light pions as ``remnants''
of those symmetries. In the presence of mass terms there is no true order parameter,
i.e.~the expectation values of the Polyakov loop as well as the chiral condensate are non-zero
everywhere. Hence the deconfined or chirally symmetric phase is analytically 
connected with 
the confined or chirally broken phase, and there is no need for a non-analytic phase transition. The following questions then arise, which should be answered by 
numerical simulations: for which parameter values of QCD is there
a true phase transition, and what is its order? Are confinement and chirality changing across the same single transition or are there different transitions? If there is just one transition, which is the driving mechanism?
If there is only a smooth crossover, how do the properties of matter change in the different regions?

\section{Strong coupling expansions at finite T}

Strong coupling expansions are well known from spin models and QCD at zero temperature. 
In contrast to the asymptotic series obtained by weak coupling expansions, they yield convergent 
series in the lattice gauge coupling $\beta=2N/g^2$
within a finite radius of convergence. The series approximate the true answer the better
the lower $\beta$, i.e.~for a fixed $N_\tau$ the lower the temperature. 
In pure gauge theory, the convergence radius is bounded from above by the critical coupling of the
deconfinement transition, $\beta_c$. Hence, strong coupling series are analytical low temperature
results, complementary to weak coupling perturbation theory which is valid at high temperatures.
 
A detailed introduction to strong coupling methods in the vacuum can be found 
in \cite{Montvay:1994cy}.
Here we merely summarise the main formulae for Yang-Mills theory 
and discuss the modifications for finite temperature applications \cite{Langelage:2008dj}. 

The Wilson gauge action can be written as a sum over all single plaquette actions,
\be
S_g[U]=\sum_{x}\sum_{1\leq\mu<\nu\leq 4}\beta\left(1-\frac{1}{3}{\rm Re}
\Tr U_p\right)\equiv \sum_p S_p\;.
\ee
Note that analytically we are able to consider an infinite spatial volume, $N_s\rightarrow \infty$.
The plaquette action has an expansion in terms of characters $\chi_r(U)=\Tr D_r(U)$ 
of the representation matrices $D_r(U)$ of the group elements $U$,
\be
\exp -S_p=c_0(\beta)[1+\sum_{r\neq0} d_r c_r(\beta)\chi_r(U_p)]\;,
\ee
and $d_r$ is the dimension of the representation $r$.
With these ingredients the free energy density can be written as 
\begin{equation}
\tilde{f}\equiv-\frac{1}{\Omega}\ln Z=-6\ln\,c_0(\beta)-\frac{1}{\Omega}
\sum_{C=(X_i^{n_i})}\,a(C)\prod_i\Phi(X_i)^{n_i}.
\label{free}
\end{equation}
where $\Omega=V\cdot N_\tau$ is the lattice volume and $c_0$ is the expansion coefficient of the trivial representation, 
which has been factored out. The combinatorial factor $a(C)$ is introduced via a moment-cumulant-formalism, 
and equals $1$ for clusters $C$ which consist of only one graph or so-called polymer $X_i$. 
The contribution of a graph $X_i$ is
\be
\Phi(X_i)=\int DU \prod_{p\in X_i}d_r c_{r_p}\chi_{r_p}(U_p)\;.
\ee
The quantity in equation (\ref{free}) is customarily called a free energy, even at zero physical temperature, because
the path integral corresponds to a partition function if one 
formally identifies the lattice coupling $\beta$ with
$1/T$. One may use the fundamental representations of the gauge groups, $c_f\equiv u$, 
as the effective expansion parameter, which together with some higher ones can be expressed
as series in the lattice coupling
\be
\mbox{SU(2):}\qquad u=\frac{\beta}{4}+{O}(\beta^{2})\qquad 
\mbox{SU(3):}\qquad u=\frac{\beta}{18}+{O}(\beta^{2}).
\qquad 
\ee

Here we are interested in a physical temperature $T=1/(aN_\tau)$, realised by compactifying the
temporal extension of the lattice. 
The physical free energy is then obtained by subtracting the formal ($N_\tau=\infty$) free energy, 
which removes the divergent vacuum energy as in the continuum. 
Thus the physical free energy density reads
\begin{equation}
f(N_\tau,u)=\tilde{f}(N_\tau,u)-\tilde{f}(\infty,u)\;.
\label{freephys}
\end{equation}
The contributing polymers $X_i$ have to be objects with a closed surface, since 
\begin{eqnarray}
\int dU \;\chi_r(U)=\delta_{r,0}\;.
\end{eqnarray}
This means the group integration projects out the trivial representation at each link. To calculate the group integrals one uses the integration formula
\begin{equation}
\int dU \;\chi_r(UV)\chi_r(WU^{-1})=\frac{1}{d_r}\chi_r(VW)\;.
\end{equation}
\begin{figure}[t]
\vspace*{1cm}
\hspace*{-3cm}
\includegraphics[viewport= 60 700 660 690]{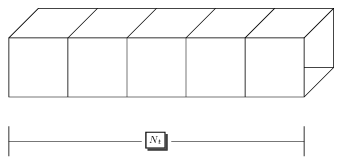}
\vspace*{2cm}
\caption[]{Graph $X_1$ contributing to the lowest order $f(N_\tau,u)$ of the expansion of the physical free energy density at finite temperature.}
\label{fig_tube1}
\end{figure}
Because of the difference in \eq(\ref{freephys}), those graphs contributing in the same way to
$\tilde{f}(N_\tau)$ and $\tilde{f}(\infty)$ drop out of the physical free energy. This is true for
all polymers with time extent less than $N_\tau$.
The calculation thus reduces to graphs with a temporal size of $N_\tau$ on the
finite $N_\tau$ lattice, and graphs spanning or extending $N_\tau$ on the infinite lattice.
Such graphs contribute either to $\tilde{f}(N_\tau)$ or to $\tilde{f}(\infty)$
(and in some cases to both).
It is therefore clear that
the strong coupling series for the physical free energy starts at a higher
order than the formal zero temperature free energy. Moreover, the order of the leading
contribution depends on $N_\tau$.

The lowest order graph existing due to the boundary condition on the finite
$N_\tau$ lattice, but not on the infinite lattice,
is a tube of length $N_\tau$ with a cross-section of one single plaquette,
as shown in \fig\ref{fig_tube1}.
It forms a closed torus through the periodic boundary and thus gives a non-vanishing
contribution, which is easily calculated to be $\Phi(X_1)=u^{4N_\tau}$.
We need to sum up all such graphs on the lattice. There are three spatial directions
for the cross section of the tube, giving a factor of 3. Translations in time take the graph into
itself and do not give a new contribution, while we get $V\Phi(X_1)$ from all spatial
translations. Together with the $1/\Omega$ in \eq(\ref{free}) this gives a factor of $1/N_\tau$.
The contribution of all tubes with all plaquettes in the fundamental representation for $SU(2)$ is thus
\be
\Phi(X_1)=\frac{3}{N_\tau}u^{4N_\tau},
\ee
which is - up to a sign - also the leading order result for the physical free energy.
For $SU(N)$ with $N\geq3$ we have an additional factor of 2 because there are also complex
conjugate fundamental representations. Thus, in the strong coupling limit $u\rightarrow 0$ the
free energy density and pressure are zero.

The leading correction comes from tubes with inner plaquettes, 
higher orders have local decorations of additional plaquettes either in the fundamental or in higher representations. For the interesting case of $SU(3)$, these contributions up to the calculated orders are \cite{Langelage:2007pi}
\begin{eqnarray}
f(N_\tau,u)=&-&\frac{3}{N_\tau}u^{4N_\tau}c^{N_\tau}\Big[1+12N_\tau u^4
+42N_\tau u^5-\frac{115343}{2048}N_\tau u^6-\frac{597663}{2048}N_\tau u^7\Big]\nonumber\\
\nonumber\\
&-&\frac{3}{N_\tau}u^{4N_\tau}b^{N_\tau}\Big[1+12N_\tau u^4+30N_\tau u^5-\frac{17191}{256}N_\tau u^6-180N_\tau u^7\Big],
\end{eqnarray}
with $b=1-3u-6v+8w, c=1+3u+6v+8w-18u^2$ and 
$v=\beta^2/432+{O}(\beta^{4}), w=\beta^2/288+{O}(\beta^{4})$.
This series is valid only for $N_\tau\geq5$, for smaller $N_\tau$ there are modifications 
coming from polymers with cross-sections larger than one plaquette. 

\section{The strong coupling regime as an ideal hadron gas}\label{lhrg}

It is now interesting to ask how the QCD pressure can be interpreted in the strong
coupling regime.
From the Wilson action it is clear that the strong coupling limit is also non-interacting.
However, as we have  noted already, in this limit the pressure is zero.
Considering strong but finite couplings, let us
recall the first orders of the $T=0$ glueball mass calculations \cite{Munster:1981es,Seo:1982jh},
\begin{eqnarray}
m(A_1^{++})&=&-4\ln\,u-3u+9u^2-\frac{27}{2}u^3-7u^4-\frac{297}{2}u^5+\frac{858827}{10240}u^6+\frac{47641149}{71680}u^7,\nn\\
m(E^{++})&=&-4\ln\,u-3u+9u^2-\frac{27}{2}u^3+17u^4-\frac{153}{2}u^5+\frac{1104587}{10240}u^6+\frac{29577789}{71680}u^7,\nn\\
m(T_1^{+-})&=&-4\ln\,u+3u+\frac{9}{2}u^3-\frac{98}{4}u^4+\frac{33}{4}u^5-\frac{36771}{1280}u^6+\frac{117897}{448}u^7,
\end{eqnarray}
where the arguments on the left side denote the representations of the point group, which map
into the different spin states \cite{Montvay:1994cy}.
We observe that the expansion of the free energy can be written 
\be
 f(N_\tau,u)=-\frac{1}{N_\tau}\left[e^{-m(A_1^{++})N_\tau} + 2e^{-m(E^{++})N_\tau} + 3e^{-m(T_1^{+-})N_\tau}\right]\left(1+ O(u^4)\right).
\ee
The prefactors before the exponentials correspond to the number of
polarisations of the respective glueball states. Note that higher spin states
start with $\sim 6\ln u$ \cite{Schor:1983py}, thus
contributing to the order $\sim u^{6N_t}$ or higher in the free energy.
Hence, through two non-trivial orders our result
is that of a free glueball gas, modified by higher order corrections.
By employing a leading order hopping expansion, the same conclusion can be 
reached when heavy quarks are added \cite{Langelage:2010yn}.
This is a rather remarkable result.
It allows to see from a first principle calculation that the pressure is
exponentially small in the confined phase, and that it is well approximated by an
ideal gas of quasi-particles which correspond to
the $T=0$ hadron excitations. 

\chapter{Some applications}

\section{The equation of state}

Energy density $\epsilon (T)$ and pressure $p(T)$ as a function
of temperature are certainly among the most
fundamental thermodynamic quantities of QCD
governing the expansion
of the plasma in the early universe as well as in heavy ion collisions. Let us use the ideal gas
results to develop some intuition about what will happen in QCD.
In the high temperature limit $T\rightarrow \infty$, we have a gas of non-interacting gluons and quarks
with
\be
\frac{p}{T^4}=\left(16+\frac{7}{8}12 N_f\right)\frac{\pi^2}{90}\;.
\ee
On the other hand, as $T\rightarrow 0$ we consider a hadron resonance gas model. 
For temperatures significantly below
$\sim 200$ MeV, only the pions are relativistic, which come in three charge states with spin zero, 
$\nu=3$. A gas of non-interacting pions has pressure
\be
\frac{p}{T^4}=3\frac{\pi^2}{90}\;.
\ee
Thus, for QCD we expect the pressure to change as a function of temperature 
from a small to a large value as a signal of deconfinement.

For the fully interacting case, we need to compute the free energy density and the pressure from the
QCD partition function.
A technical obstacle here is that, in a Monte Carlo simulation, 
one cannot compute the partition function directly, 
since all expectation values are normalised to $Z$,
\be
\langle O\rangle=Z^{-1}\Tr(\rho O)\;.
\ee
So we need to specify some observable $O$ to access the partition function.
The most frequently used detour 
is called the integral method \cite{Engels:1990vr}, in which a  
derivative of the free energy is calculated and then integrated,
\be
\frac{f}{T^4} {\biggl |}_{T_o}^{T} \; = \; - {1\over V} \int_{T_o}^{T} 
{\rm d}x \; {\partial \;\;x^{-3} \ln Z(V,x) \over \partial x }\;.
\ee
On the lattice, it is convenient to take the derivative with respect to the gauge
coupling instead of temperature,
\ba
\frac{f}{T^4} {\biggl |}_{\beta_o}^{\beta} \; &=& \;  -\frac{N_\tau^3}{N_s^3} 
\int_{\beta_o}^{\beta}
{\rm d}\beta ' \left(\left\langle \frac{\partial \ln Z}{\partial \beta'} \right\rangle - 
\left\langle \frac{\partial \ln Z}{\partial \beta'}\right \rangle_{T=0} \right)\;,\nn\\
&=&-N_\tau^4\int_{\beta_0}^{\beta}d\beta'\;\left(3\langle\Tr U_p^t+ \Tr U_p^s\rangle-6\langle \Tr U_p\rangle_{T=0}\right)\;.
\label{freediv}
\ea
Now we simply need to measure expectation values of temporal $(U_P^t)$ and spatial 
$(U_p^s)$ plaquettes with sufficient 
accuracy, so they can be integrated numerically.
Note that this introduces a lower integration constant, which needs to be fixed for the result to be
meaningful. While we do not know $f(\beta_0)$ from first principles, we can choose
$\beta_0$ corresponding to a temperature below the phase transition, where the free energy
should be well modelled by a weakly interacting hadron gas. 
For $T\lsim 130$ MeV, all hadrons become non-relativistic and we can approximate the pressure
by zero. Note however, that this procedure is not good for the chiral limit, when pions become massless
and stay relativistic down to very low temperatures.

Another difficulty is that strong discretisation effects are to be expected.
At high temperature the relevant partonic degrees of freedom
have momenta of order $\pi T\sim \pi/(aN_t)$ on the scale
of the lattice spacing, by which they are strongly affected.
For the equation of state it is therefore particularly important to gain control 
over these effects
and carry out the continuum limit $a \rightarrow 0$. 
This motivates the use of improved actions,
designed to minimise cut-off effects in the approach to the continuum.

\begin{figure}[t]
\vspace*{-0.3cm}
\centerline{
\includegraphics[height=4cm]{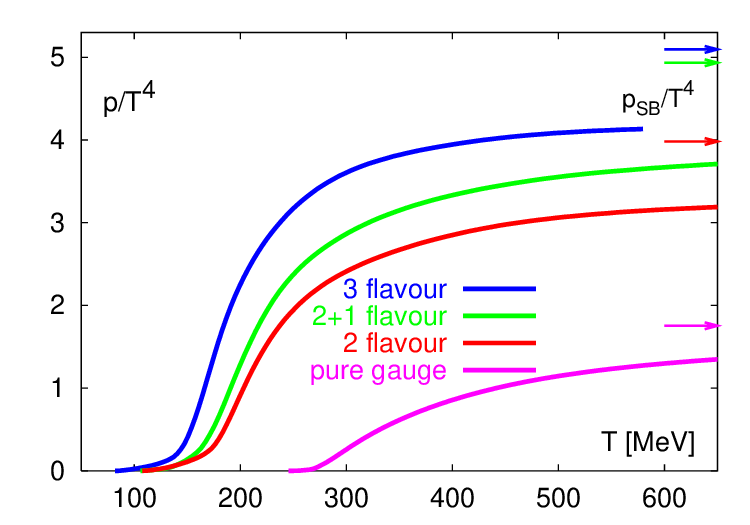}
}
\caption[]{Flavour dependence of the pressure
for $N_\tau=4$ lattices compared to a continuum extrapolated pure gauge result. From 
\cite{Karsch:2000ps}.}
\label{fig:pressure}       
\end{figure}

The results of a computation of the pressure with an improved action \cite{Karsch:2000ps}
are shown in \fig\ref{fig:pressure}. The data have been obtained for
$N_f=2,3$ with (bare) mass $m_q/T = 0.4$ as well as
for $N_f=2+1$ with a heavier mass $m^s_q/T = 1$ on a very coarse lattice, $N_\tau=4$.
Nevertheless, interesting qualitative features can be observed.
For comparison, 
continuum extrapolated pure gauge results are also included.
The figure shows a rapid rise of the pressure in a narrow transition region.
The critical temperature as well as the
magnitude of $p/T^4$ reflect the number of degrees of
freedom liberated at the transition. 
This last conclusion is firm, since the pressure also rises for fixed temperature
when light quarks are added to the theory, consistent with the behaviour 
in the Stefan-Boltzmann
limit. Another interesting feature is that the curves fall short of the ideal gas values, 
i.e.~interactions are still strong just above $T_c$. 
An important question then is whether these features survive in the continuum limit.
In pure gauge theory this can be firmly established by numerical 
extrapolation while for light dynamical quarks this is the subject of ongoing simulation programs.

\section{Screening masses}

Essentially all static equilibrium properties of a thermal
quantum field theory are encoded in its equal time correlation functions. These are quantities
that are well defined and calculable to good precision by lattice methods.
Unfortunately, these quantities are not 
directly accessible in heavy ion collision experiments. Nevertheless, their theoretical knowledge
provides us with the relevant length scales in the plasma, from which
conclusions about the active degrees of freedom and their dynamics may be drawn.

The concept of screening is most easily introduced in a QED plasma. Suppose we insert a static
external charge in a plasma of freely moving charges. These will be attracted or repelled, arranging
themselves such that the polarisation cancels out the field of the static charge, which therefore is screened. Beyond a certain distance called a screening length, a particle in the plasma will not feel
the presence of the external charge.  The screening length is defined by the spatial exponential decay
of the equal time correlator of the electric field,
\be
\lim_{\bfx \rightarrow \infty} \langle E_i(\bfx,t)E_j({\bf 0},t)\rangle = const.\; \ex^{-m_D|\bfx|},
\ee
and the inverse screening length $m_D$ is the Debye mass.
In perturbation theory, $m_D$ corresponds to the pole of the 
$A_0$ propagator at Matsubara frequency $\omega_n=0$, i.e.~it corresponds to the electric gluon mass,
$m_D=m_E$, \eq(\ref{mel}). Its Fourier transformation leads to the 
Debye-screened potential of a static charge,
\be
V(r)=Q\int \frac{d^3p}{(2\pi)^3}\frac{\ex^{i\bfp\cdot\bfx}}{\bfp^2+\Pi_{00}(0,\bfp)}=\frac{\ex^{-m_Dr}}{4\pi r}\;.
\label{debpot}
\ee

The same concept has been carried over to QCD where it is speculated to be responsible
for the loss of confinement for heavy quarkonium states by screening of colour charges in the plasma \cite{Matsui:1986dk}. 
However, there are some
conceptual difficulties with the definition of the Debye mass in QCD. First, colour-electric fields 
are no physical observables since they are gauge-dependent, and so is their correlator.
One might argue that the pole mass of this correlator is still gauge-invariant order by order in 
perturbation theory and consider the perturbative series \cite{Rebhan:1993az},
\be
m_D=m_D^{LO}+\frac{3}{4\pi}g^2 T \ln \frac{m_D}{g^2T}+c g^2T+O(g^3)\;.
\ee 
In this case one encounters the Linde problem already in next-to-leading order: the coefficiet $c$
receives contributions from all loop orders and cannot be evaluated in perturbation theory.

For a non-perturbative evaluation of screening, one therefore generalises the concept to gauge-invariant
sources, averaged over Euclidean time,
\be
\bar{O}(\bfx)=\frac{1}{N_\tau}\sum_\tau O(\bfx,\tau),\quad 
\langle \bar{O}(\bfx)\bar{O}({\bf 0})\rangle_c=\sum_n c_n\;\ex^{-M_n|\bfx|}\;.
\ee
These are correlations in space, and the connected parts therefore fall off with the eigenvalues of the space-wise Hamiltonian $H_z$, \eq(\ref{hz}).
In analogy to QED, these masses correspond to the inverse length scale over
which the equilibrated medium is sensitive
to the insertion of a static source carrying the quantum numbers of $O$.
Because of the compact Euclidean time direction at $T>0$, the continuum 
rotation symmetry of the hypertorus orthogonal to the correlation direction is broken down
from $O(3)$ to $O(2)\times Z(2)$, corresponding to rotations in the $(x,y)$-plane and reflections
of $\tau$. Correspondingly, screening masses are classified by quantum numbers $J^{PC}_R$,
where $J,P,C$ are standard spin, parity and charge conjugation, while $R$ is associated with
the Euclidean time reflection.
The appropriate subgroup for the
lattice theory is $D^4\times Z(2)=D_h^4$. The irreducible representations and the classification
of operators have been worked out for pure gauge theory \cite{Grossman:1993wm,Datta:1998eb} 
as well as for staggered quarks \cite{Gupta:1999hp}.

Just as in $T=0$ spectrum calculations, we can consider either 
glueball or meson and baryon like operators. Once again it is useful to develop some intuition
by considering the limits of low and asymptotically high temperature.
In the latter case, standard perturbation theory should apply and we can evaluate the correlators
using quark and gluon propagators. For a mesonic operator the exponential decay at large 
distances is then dominated by two quark lines, with an effective mass given by their lowest
Matsubara frequencies, $\sim \pi T$. Neglecting bare quark masses, a meson correlator in the
non-interacting limit then decays with $M_{\bar{q}q}\sim 2\pi T$. Perturbative corrections 
lift the degeneracy and appear to generally shift this towards larger values \cite{Laine:2003bd}. 
For purely gluonic operators, the lowest Matsubara
frequency is zero and the exponential decay is determined by dynamically generated mass scales. 
For operators like $\Tr A_0^2$ the propagators feature the perturbative Debye mass, and hence 
$M_{A_0^2}\sim 2m_E$. However, for operators involving $A_i$, 
we are faced with the Linde problem and a perturbative evaluation is not 
possible. 

\begin{figure}[t]
\vspace*{1cm}
\hspace*{-2.2cm}
\includegraphics[viewport= 60 700 660 690]{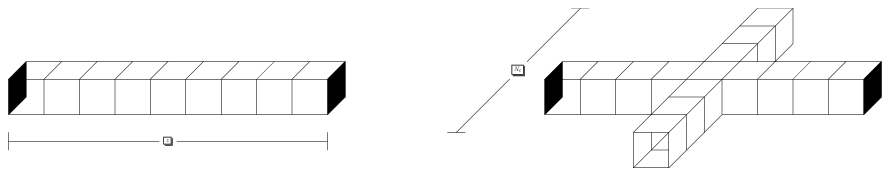}
\vspace*{2cm}
\caption[]{Left: Leading graph contributing to the plaquette correlation function. Right: Graph contributing
to the finite temperature effect of the correlator.}
\label{fig_tube}
\end{figure}
On the other hand, at low temperatures we have small $\beta$ and a calculation by strong coupling 
methods is feasible. As an example, let us consider the  colour-electric field correlator
$\langle \,\Tr F^a_{0i}(\bfx) \,\Tr F^a_{0i}(\bfy)\,\rangle$, which is in the $J^{PC}_T=0^{++}_+$
channel ($T$ denotes reflection in Euclidean time) containing the ground state and the
mass gap. On the lattice, this corresponds to a correlation of temporal plaquettes,
and the quantum numbers under the point group $D^4_h$ are $A_1^{++}$.
 Temporarily assigning separate gauge couplings to
all plaquettes, the correlator can be defined as 
\begin{equation}
C(z)=\langle\mathrm{Tr}\,U_{p_1}(0)\,\,\mathrm{Tr}\,U_{p_2}(z)
\rangle=N^2\frac{\partial^2}{\partial\beta_1\partial\beta_2}\ln\,
Z(\beta,\beta_1\beta_2)\bigg\vert_{\beta_{1,2}=\beta}.
\end{equation}
At zero temperature the exponential decay is the same as for correlations in the
time direction, and thus
determined by the glueball masses, the lowest of which may be extracted as
\begin{equation}
m=-\lim_{z\rightarrow\infty}\frac{1}{z}\ln\,C(z).\\
\end{equation}
The leading order graphs for the strong coupling series at zero temperature
are shown in \fig\ref{fig_tube} (left).
This leads to the lowest order contribution:
\begin{equation}
C(z)=A\,u^{4z}=A\mathrm{e}^{-m_sz}.
\end{equation}
Thus, to leading order the glueball mass is
$m_s=-4\ln\,u(\beta)$.

Now we switch on a physical temperature, i.e.~keep the lattice volume compact in the
time direction. As in the case of the free energy, we are here only interested in
the temperature effects, i.e.~in the mass difference
\ba
\Delta m(T)=m(T)-m(0)&=&-\lim_{z\rightarrow \infty}\frac{1}{z}[\ln C(T;z)-\ln C(0;z)]\\
&=&-\lim_{z\rightarrow \infty}\left[\ln\left(1+\frac{\Delta C(T;z)}{C(0;z)}\right)\right],
\ea
with $\Delta C(T;z)=C(T;z)-C(0;z)$. A typical graph contributing in lowest order to this
difference is shown in \fig\ref{fig_tube} (right). Summing up all leading 
order graphs gives \cite{Langelage:2008dj}
\be
\Delta m(T)=-\frac{2}{3}N_\tau\,u^{4N_t-6}\,,
\ee
i.e.~the screening masses decrease compared to their $T=0$ values.
Again, the leading order result is generic for all $SU(N)$ and quantum number channels.
As in the case of the free energy, the difference only receives $N_\tau$-dependent higher
orders of $u$, leading to weak temperature effects.
We conclude that in the confinement phase the lowest screening masses in each
quantum number channel should be close to the corresponding zero temperature particle masses,
with a significant temperature dependence showing up only near $T_c$.
This explains the findings of numerical investigations of
the lowest screening mass in $SU(3)$ gauge theory, which for temperatures
as high as $T=0.97T_c$ see very little temperature dependence,
$\Delta m(T)/m(0)\gsim 0.83$ \cite{Datta:1998eb}.

Away from the high and the low temperature limits, numerical simulations have to be performed.  
One particularly interesting aspect of screening masses is that they permit to study chiral symmetry
restoration across the quark hadron transition by looking at degeneracy patterns.  
For example, consider the lowest lying scalar, pseudo-scalar, vector and axial-vector mesons.
Chiral $SU(2)\times SU(2)$ implies degeneracy between the scalar and vector,
while an intact $U_A(1)$  implies degeneracy between the parity flipped states. 
A calculation of these masses for temperatures around $T_c$ will thus allow insight into 
the details of chiral symmetry restoration.

\vspace*{0.5cm}
\begin{minipage}{6cm}
\begin{tabular}{|c|c|c|c|c|}
\hline
 meson           & $\sigma $ & $\vec{\pi}$ & $\vec{\delta}$ & $\eta'$ \\\hline
operator         & $\bar{\psi}\psi$ & $\bar{\psi}\gamma^5\vec{\tau}\psi $ & $\bar{\psi}\vec{\tau}\psi$ &
                            $\bar{\psi}\gamma^5\psi $\\
\hline
\end{tabular}
\end{minipage}\hspace*{1cm}
\begin{minipage}{6cm}
\[
\begin{array}{ccccc}
\leftarrow & SU(2) & \times & SU(2) & \rightarrow \\
\uparrow & \sigma & & \vec{\pi}& \\
U_A(1) & & & & \\
\downarrow & \eta' & & \vec{\delta} \\
\end{array}
\]
\end{minipage}
\vspace*{0.5cm}

\section{The free energy of a static quark anti-quark pair}

Similar to a single static quark, \eq(\ref{zq}) and the discussion following it, we can consider a static quark anti-quark pair
in a Yang-Mills plasma,
\be
\langle \Tr L^\dag(\bfx)\Tr L({\bf 0})\rangle = \frac{Z_{\Qb Q}}{Z}=\exp -\frac{F_{\Qb Q}(\bfx,T)-F_0(T)}{T}\;.
\label{av}
\ee
(In the following we drop the subtraction of $F_0$ from the notation, which is always implied).
Such a system is particularly interesting because it represents the non-relativistic limit of heavy
quarkonia in the plasma, which are routinely produced in heavy ion collisions. In order to clarify the 
meaning of this quantity, let us consider its spectral decomposition. We use again the transfer matrix formalism, but with a slight modification accounting for the static sources. Let us fix to temporal gauge,
$U_0(\tau,\bfx)=1, \tau=1,...N_\tau-1$, which we are allowed to do on all temporal links but those in one time slice. On the gauge fixed time slices, we have the Kogut-Susskind Hamiltonan $H_0$, 
which acts on the Hilbert space of states with static sources, from \eq(\ref{transfer}),
\be
(T_0)_{\tau+1,\tau} \equiv \ex^{-aH_0}=\exp-L[U_i(\tau+1),1,U_i(\tau)]\;.
\ee
Now we can be rewrite the Polyakov loop correlator exactly as
\ba
\langle \Tr L^\dag(\bfx)\Tr L({\bf 0})\rangle&=&\\
&&\hspace*{-2.0cm}\frac{1}{Z}\hTr\left(
T_0^{N_\tau-1}\int DU_0(N_\tau)\;\;U_{0\alpha\alpha}^\dag(N_\tau,\bfx)U_{0\beta\beta}({N_\tau,\bf 0})\; 
\ex^{-L[U_i(1),U_0(N_\tau),U_i(N_\tau)]} \right)\nn
\ea
Next, we employ gauge invariance of the kernel of the transfer matrix, \eq(\ref{trans}), under gauge 
transformation in the upper timeslice,
\be
R(g)L=L[U^g_i(\tau+1),U_0(\tau)g^{-1},U_i(\tau)]=L[U_i(\tau+1),U_0(\tau),U_i(\tau)], \quad
\ee
where $R(g)|\psi\rangle =|\psi^g\rangle$ imposes a gauge tranformation on the states it acts on. 
Choosing $g(\bfx)=U_0(N_\tau,\bfx)$, we then obtain
\be
\langle \Tr L^\dag(\bfx)\Tr L({\bf 0})\rangle=\frac{1}{Z}\hTr(
T_0^{N_\tau}P_{\alpha\alpha\beta\beta})\;,
\ee
where we have introduced the projection operator 
\be
P_{\alpha\beta\mu\nu}=\int Dg\;\;g^\dag_{\alpha\beta}(\bfx)g_{\mu\nu}({\bf 0})R(g)
\;.
\ee
This operator annihilates all wave functions not transforming as 
\be
\psi_{\beta\mu}[U^g]=g(\bfx)_{\beta\gamma}\,\psi_{\gamma\delta}[U]\,g^\dag_{\delta\mu}({\bf 0}).
\ee
Specifically,
\be
P_{\alpha_\beta\mu\nu}|\psi_{\gamma\delta}\rangle=\frac{1}{N^2}\delta_{\beta\gamma}\delta_{\mu\delta}
|\psi_{\alpha\nu}\rangle\;.
\ee
Thus $P$ projects onto the 
subspace of states with a colour triplet sitting at $\bfx$ and an anti-triplet at $\bf 0$.
Inserting a complete set of eigenstates of the Hamiltonian $H_0$, we then find
\be
\langle \Tr L^\dag(\bfx)\Tr L({\bf 0})\rangle =\frac{1}{N^2Z}\sum_{n,\alpha,\beta}
\langle n_{\alpha\beta}|n_{\beta\alpha}\rangle\;\ex^{-\frac{E_n^{\Qb Q}(|\bfx|)}{T}}=
\frac{1}{Z}\sum_n\ex^{-\frac{E_n^{\Qb Q}(|\bfx|)}{T}}\;.
\label{lcorr}
\ee
We easily recognise this as a ratio of partition functions, the numerator being just 
the Boltzmann sum over all energy levels $E_n^{\bar{Q}Q}$, which are eigenvalues
of $H_0$ in the presence of the $\Qb Q$ pair. 
Let us now consider the zero temperature limit of this expression, 
\be
\lim_{T\rightarrow 0} \frac{\sum_n\ex^{-\frac{E_n^{\Qb Q}}{T}}}{\sum_n\ex^{-\frac{E_n}{T}}}
=\lim_{T\rightarrow 0}
\ex^{-(E^{\Qb Q}_0-E_0)/T}\frac{1+\ex^{-(E^{\Qb Q}_1(|\bfx|)-E^{\Qb Q}_0)/T}+\ldots}{1+\ex^{-(E_1-E_0)/T}+\ldots}\rightarrow \ex^{-\frac{V(|\bfx|)}{T}}\;.
\ee
Here we have identified the static potential with the ground state energy of a static quark anti-quark pair 
above the vacuum, $V(|\bfx|)=E^{\Qb Q}_0(|\bfx|)-E_0$. This is precisely the same
quantity which at zero temperature
determines the ground state of the exponential fall-off of the Wilson loop. 
Thus, we can extract the same quantity from a Polyakov loop correlator, \eq(\ref{lcorr}), in the limit 
$N_\tau\rightarrow \infty$. At finite temperatures instead, the Polyakov
loop correlator corresponds to the Boltzmann-weighted sum over all excited states of the 
static potential, and hence to a free energy.
This free energy is often called
a $T$-dependent potential, $V(r,T)\equiv F_{\bar{Q}Q}(r,T)$, even though this is not quite justified,
as we will discuss below. 

\begin{figure}
\includegraphics[width=0.5\textwidth]{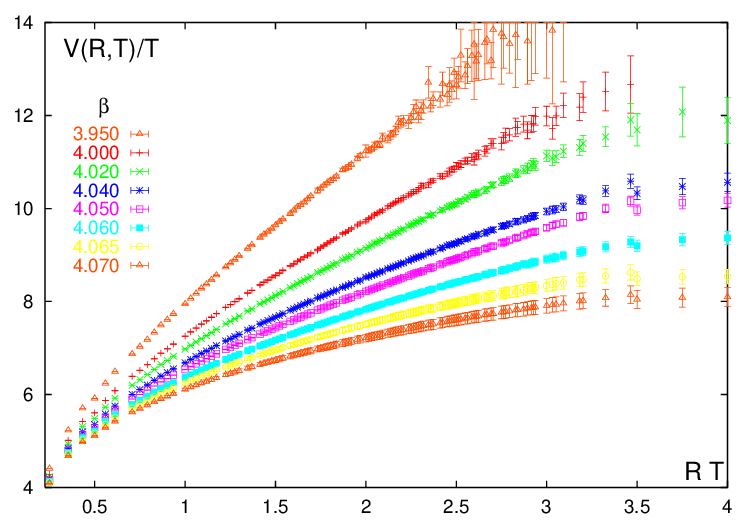}
\includegraphics[width=0.5\textwidth]{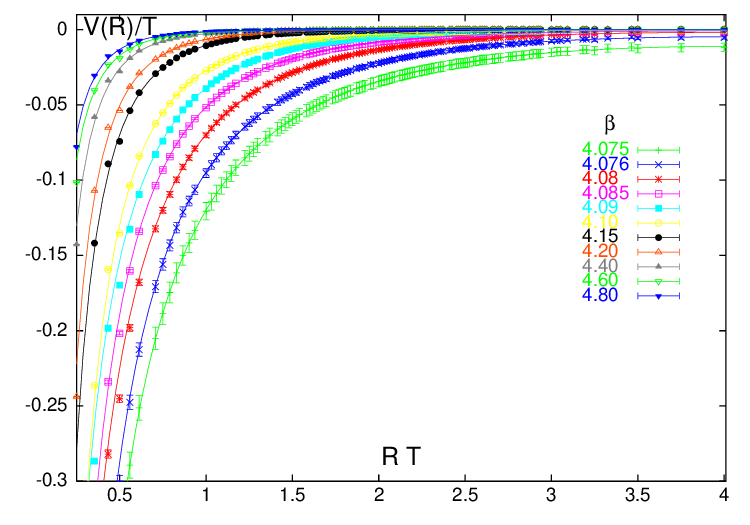}
\caption[]{Static quark anti-quark free energy/potential, \eq(\ref{av}), for
$T<T_c$ (left) and $T>T_c$ (right), on $32\times 4$ with Symanzik improved action 
\cite{Kaczmarek:1999mm}. }
\label{bie}
\end{figure}
The Polyakov loop correlator is readily simulated, with results as in \fig\ref{bie}.
It gives a linearly rising free energy in the confined phase, whose effective string tension
reduces with temperature, while in the deconfined phase the potential is 
screened, thus exhibiting clearly the phase transition from a confined to deconfined phase.
However, the screened free energy does not correspond to the Debye-screened potential 
from \eq(\ref{debpot}), as becomes apparent when considering its spatial decay at high $T$.
Fitting the screened free energy to
\be
\frac{F_{\bar{Q}Q}}{T}=-\frac{c(T)}{(rT)^d}\,\ex^{-m(T)r},
\ee
gives $d\approx 1.5$ and $m=M_{0^{++}_+}$, i.e.~the screening mass corresponds
to the lightest glueball channel \cite{Hart:2000ha}. 
This can already be seen in perturbation theory, 
where the leading term is by two-gluon exchange and thus $m=2m_D$.
Non-perturbatively it follows from the fact that the traced Polyakov loop is a gauge-invariant
operator coupling to all powers of gauge fields, hence its correlator at large distance is
dominated by the lightest gauge-invariant gluonic screening mass.

For the following, let us consider explicitly $N=3$ colours.
In order to find a thermal potential in analogy to QED Debye screening,
attempts have been made to decompose the traced and gauge invariant
Polyakov loop into colour singlet and octet configurations of the static sources,
${\bf 3}\times {\bf \bar{3}}={\bf 1}+{\bf 8}$, 
\cite{McLerran:1981pb,Nadkarni:1986as}
\ba
e^{-F_{\Qb Q}(r,T)/T }& =&
\frac{1}{9} \;  e^{- F_{1}(r,T)/T } +
\frac{8}{9}   \; e^{- F_{8}(r,T)/T },\\
\ex^{-F_1(r,T)/T}&=&
\frac{1}{3}\langle \Tr L^\dag(\bfx)L(\bfy)\rangle,\nn\\
\ex^{-F_8(r,T)/T}
&=&\frac{1}{8}\langle \Tr L^\dag(\bfx) \Tr L(\bfy)\rangle
-\frac{1}{24}\langle \Tr L^\dag(\bfx) L(\bfy)\rangle.
\label{potdef}
\ea
Note that the correlators in the singlet and octet channels are gauge
dependent, and the colour decomposition only holds perturbatively in a fixed
gauge, which has motivated many gauge fixed lattice simulations.
However, both options are unphysical at a non-perturbative level.
To understand this, let us start
from something physical and consider a meson
operator in an octet state,
$O^a=\bar{\psi}(\bfx)U(\bfx,\bfx_0)T^aU(\bfx_0,\bfy)\psi(\bfy)$, with $x_0$
the meson's center of mass.
In the plasma the colour charge can always be neutralised by a gluon.
In the correlators for the singlet and octet operators, we
integrate out the heavy quarks, replacing them by Wilson lines,
\ba
\langle O(\bfx,\bfy;0)O^{\dag}(\bfx,\bfy;N_\tau)\rangle&\propto&
\langle \Tr L^\dag(\bfx)U(\bfx,\bfy;0)L(\bfy)U^\dag(\bfx,\bfy;N_\tau)\rangle,\nn\\
\langle O^a(\bfx,\bfy;0)O^{a\dag}(\bfx,\bfy;N_\tau)\rangle&\propto&
\left[\frac{1}{8}\langle \Tr L^\dag(\bfx) \Tr L(\bfy)\rangle\right.\\
& & \left.
-\frac{1}{24}\langle \Tr L^\dag(\bfx)U(\bfx,\bfy;0)
L(\bfy)U^\dag(\bfx,\bfy;N_\tau)\rangle\right]\, . \nn
\ea
We have now arrived at gauge invariant expressions, because we used
a gauge string between the sources. The singlet correlator corresponds
to a periodic Wilson loop which wraps around the boundary. The connection
to the gauge fixed correlators is readily established, replacing
the gauge string by gauge fixing functions, $U(\bfx,\bfy)=g^{-1}(\bfx)g(\bfy)$.
Thus, in axial gauge, $U(\bfx,\bfy)=1$ (and only there),
the gauge fixed correlators are identical
to the gauge invariant ones.

Repeating the spectral analysis 
the full correlators take the form \cite{Jahn:2004qr}
\ba
\ex^{-F_1(r,T)/T}&=&\frac{1}{Z}\frac{1}{9}\sum_n\langle n_{\delta\gamma}|
U_{\gamma\delta}(\bfx,\bfy)U^\dag_{\alpha\beta}(\bfx,\bfy)|n_{\beta\alpha}\rangle
\,\ex^{-E_n^{\bar{Q}Q}(r)/T}\;,\nn\\
\ex^{-F_8(r,T)/T}&=&\frac{1}{Z}\frac{1}{9}\sum_n\langle n_{\delta\gamma}|
U^a_{\gamma\delta}(\bfx,\bfy)U^{\dag a}_{\alpha\beta}(\bfx,\bfy)|n_{\beta\alpha}\rangle
\,\ex^{-E_n^{\bar{Q}Q}(r)/T}\;.
\label{chan}
\ea
The energy levels in the exponents are identically the same in \eqs(\ref{lcorr},\ref{chan})
and correspond to the familiar gauge invariant static
potential at zero temperature and its excitations. However, while \eq(\ref{lcorr})
is purely a sum of exponentials and thus a true free energy, the
singlet and octet correlators contain matrix elements which do depend
on the operators used, thus giving a path/gauge dependent weight
to the exponentials contributing to $F_1,F_8$.
Since the spectral information contained in the average and gauge fixed
singlet and octet
channels is the same, we must conclude that any difference between those
correlators is entirely gauge dependent and thus unphysical.

The definition of a temperature-dependent static potential for bound states with the appropriate
perturbative Debye screened limit has recently been achieved. However, 
it requires a Wilson loop correlator in real time evaluated in a thermal 
ensemble \cite{Laine:2006ns,Brambilla:2008cx}. Since this cannot be done by straightforward
lattice simulations, we shall not further discuss it here.

\section{Phase transitions and phase diagrams}

\begin{figure}[t]
\centerline{
\includegraphics*[width=0.5\textwidth]{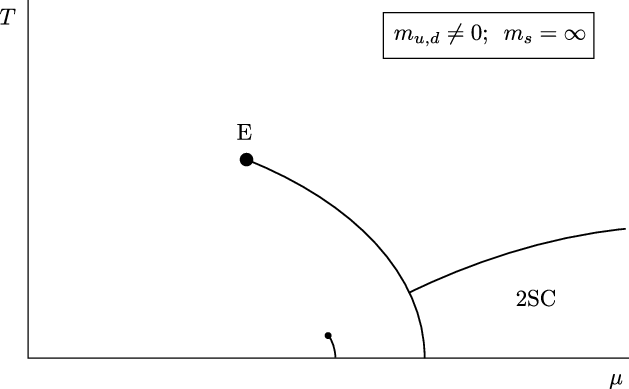}\hspace*{1cm}
}
\caption[]{
Conjectured phase diagram for $N_f=2$ QCD with finite light quark masses. The physical
case $N_f=2+1$ is believed to qualitatively look the same,
based on universality and continuity arguments as well as input from QCD-like 
models \cite{Rajagopal:2000wf}.}
\label{fig:1schem}      
\end{figure}
A question of utmost importance for experimental programs ranging from heavy ion collisions  to 
astro-particle physics are the different forms of nuclear matter for various conditions specified by
temperature and baryon chemical potential, and whether these are separated by phase transitions.
In statistical mechanics, phase transitions are defined 
as singularities, or non-analyticities, in the free energy as a function of its thermodynamic parameters.
However, on finite volumes, free energies are always analytic functions and the theorem of Lee and Yang
\cite{Yang:1952be,Lee:1952ig}
states that singularities only develop in the thermodynamic limit of infinitely many particles, or $V\rightarrow \infty$. 
This is particularly obvious in the case of lattice QCD, whose partition 
function is a functional integral
over a compact group with a bounded exponential as an integrand and without zero mass excitations. 
It is thus a perfectly analytic function of
$T,\mu,V$ for any finite $V$. 
Hence, a theoretical establishment of a true phase transition requires 
finite size scaling (FSS) studies on a series of increasing and 
sufficiently large volumes to extrapolate to the thermodynamic limit. 

Three different situations can emerge: 
a first order phase transition 
is characterised by coexistence of two phases, and hence a discontinuous jump 
of the order parameter (and other quantities), while 
a second order transition  
shows a continuous transition of the order parameter accompanied by a divergence of the correlation length and some other quantities, like the heat capacity. 
Finally, a marked change in the physical properties of a system may also occur 
without any non-analyticity of the free energy, in which case it is called 
an analytic crossover. 
A familiar system featuring all these possibilities is water, 
with a weakening first oder liquid-gas phase transition terminating in a 
critical endpoint with Z(2) universality, as well as a triple point 
where the first order liquid-gas and 
solid-liquid transitions meet. Similar structures are also conjectured to
be present in the QCD phase diagram \cite{Halasz:1998qr,Rajagopal:2000wf}, \fig\ref{fig:1schem},
where asymptotic freedom suggests the existence of at least three different regimes: the usual 
hadronic matter, a quark gluon plasma at high $T$ and low $\mu$, as well as a colour super-conducting state at low $T$ and high $\mu$.

\section{The (pseudo-) critical coupling and temperature}

The first task when investigating a potential phase transition is to locate the phase boundary.
Thus, for QCD
with a fixed quark content, we are interested in the critical (or pseudo-critical) temperature where the
transition from the confining regime to the plasma regime takes place.
The method to locate a transition in statistical mechanics usually is to look for
rapid changes of suitable observables $O(\bfx)$ and peaks in their susceptibilities. Typical examples are
the Polyakov loop, the chiral condensate or the plaquette, $O\in\{\Tr L,\bar{\psi}\psi,\Tr U_p,\ldots\}$.  
Generalised susceptibilities for $O(\bfx)$ in statistical mechanics are defined as the 
volume integral over the connected correlation function,
\be
\chi_O=\int d^3x \left(\langle O(\bfx) O(0)\rangle -\langle O(\bfx) \rangle \langle O(0)\rangle \right).
\ee
(Note that on the lattice this can be generalised to 4d. However, the thermal equilibrium system likewise lives in three dimensions,
which are the ones to be used for finite size scaling analyses in the following section).
If we instead consider the corresponding volume average,
\be
\bar{O}=\frac{1}{V}\int d^3x \;O(\bfx),
\ee
then because of its translation invariance the integration
gives trivial volume factors and we obtain on the lattice
\be
\chi_{\bar{O}}
              =N_s^3 (\langle\bar{ O}^2\rangle - \langle \bar{O}\rangle ^2)
              =  N_s^3\langle ( \delta \bar{O})^2 \rangle ,
\label{susc}              
\ee
with the fluctuations $\delta \bar{O}=\bar{O}-\langle \bar{O} \rangle$. 
At a phase transition fluctuations are maximal, hence
the locations of the peaks of susceptibilities define (pseudo-) critical
couplings, $\chi(\beta_c,m_f)=\chi_{\rm max} \Rightarrow \beta_c(m_f)$, which can be turned into temperatures as discussed in Sec.~\ref{tunet}. 
In practice, often the two-loop beta function is used as a short cut,
although this becomes valid only when the lattice spacing is fine enough to be in the perturbative regime.

\begin{figure}[t]
\vspace*{-2cm}
\centerline{
\epsfig{bbllx=105,bblly=220,bburx=460,bbury=595,file=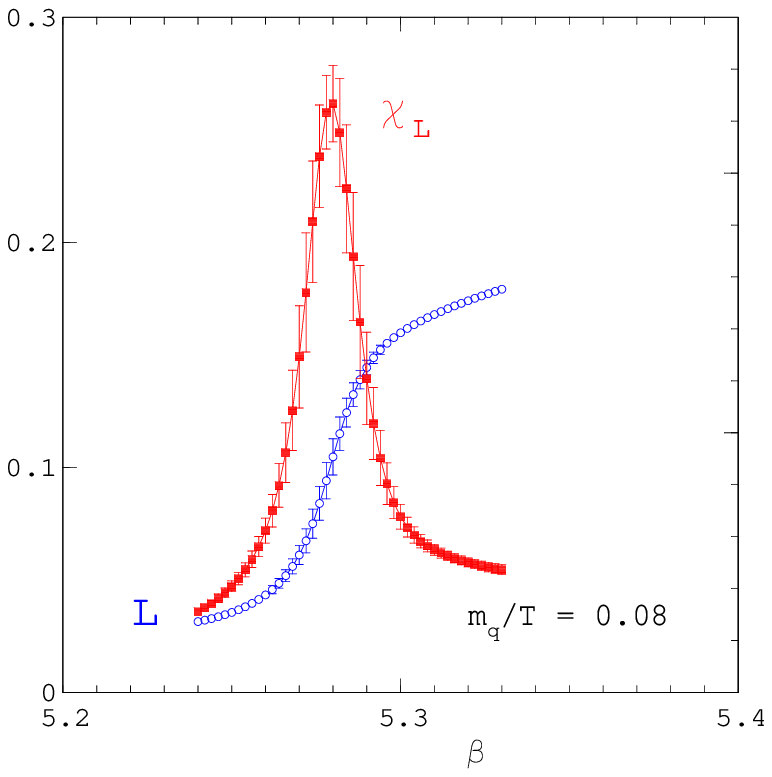,height=85mm} 
\epsfig{bbllx=105,bblly=220,bburx=460,bbury=595,file=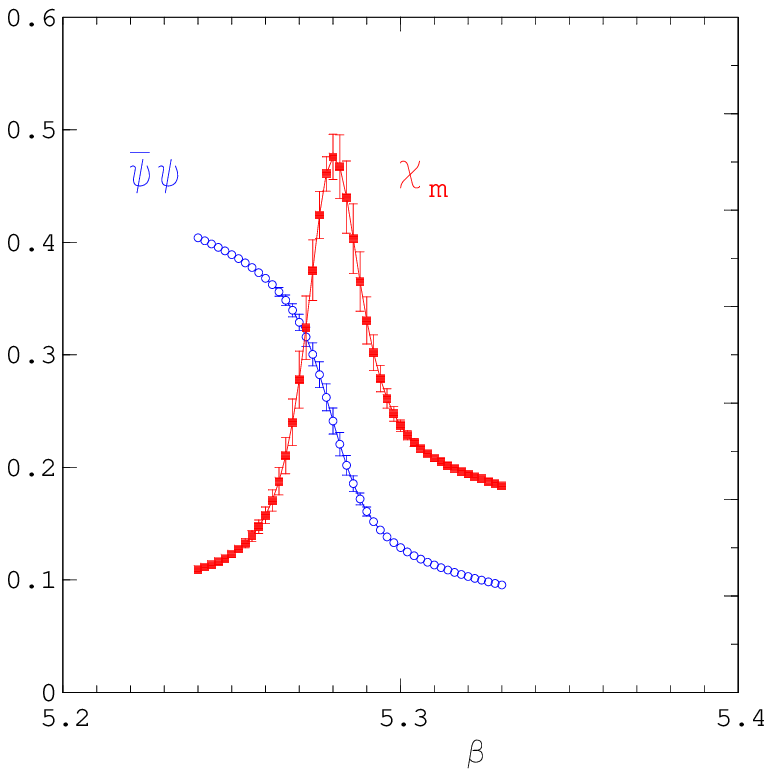,height=85mm} 
}
\vspace*{-2.5cm}
\caption[]{Polyakov loop $\langle \Tr L\rangle$ and chiral condensate $\langle \bar{\psi}\psi\rangle$ together with their susceptibilities signal a transition in two flavour QCD. From \cite{Karsch:2001cy}.}
\label{fig:tc}       
\end{figure}

\begin{figure}[t]
\hspace*{1.7cm}
\leavevmode
\epsfxsize=3cm
\epsffile{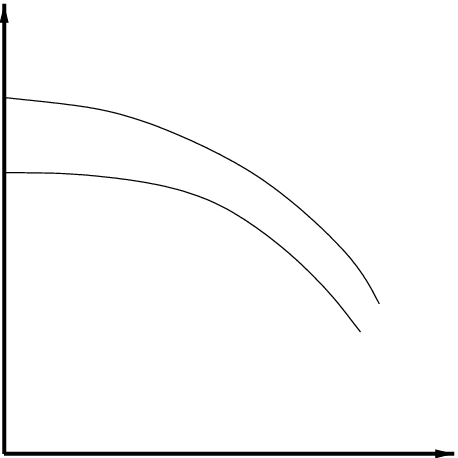}
\put(-15,-15){$\mu$}
\put(-98,80){$T$}
\put(-60,90){\large finite V}
\hspace*{3cm}
\epsfxsize=3cm
\epsffile{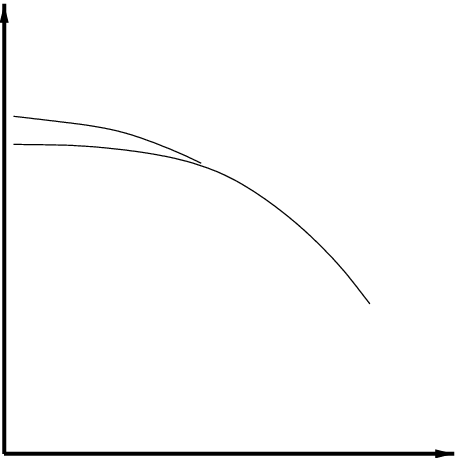}
\put(-15,-15){$\mu$}
\put(-98,80){$T$}
\put(-60,90){\large infinite V}
\caption[]{
Location of pseudo-critical parameters defined by different observables.
In the infinite volume limit, the lines merge for a true phase transition, but stay
separate for a crossover.}
\label{fig:tv}
\end{figure}
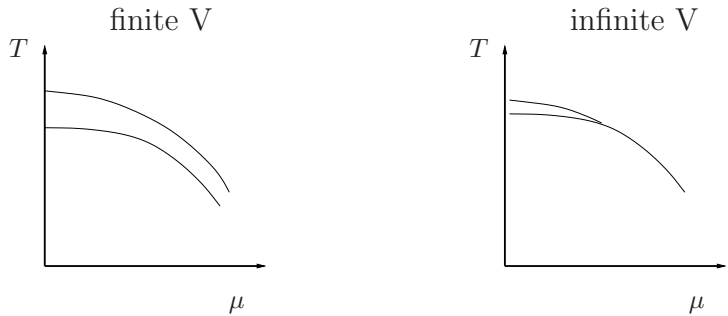

Note that for an analytic crossover the pseudo-critical couplings defined
from different observables do not need to coincide. The partition function is analytic everywhere
and there is no uniquely specified ``transition''. This holds in particular for pseudo-critical couplings extracted from finite lattices. As the thermodynamic limit is approached, the 
couplings defined in different ways
will merge where there is a non-analytic phase transition, 
and stay separate in the case of a crossover,
as illustrated in \fig\ref{fig:tv} for a putative QCD phase diagram.

\section{Universality, finite size scaling and signals for criticality}

A fascinating phenomenon in physics is the ``universality'' exhibited by physical systems
near critical points of second order phase transitions. It is due to the divergence of the
correlation length, which implies that the entire system acts as a coherent collective.
Hence, microscopic physics becomes unimportant, the collective behaviour is determined by the global symmetries and the number of dimensions of the system.
With the divergence of the correlation length any characteristic length scale 
disappears from the problem, and thermodynamic observables in the critical 
region obey scale invariant power laws. For example, at a second order 
ferromagnetic phase transition, the magnetization (the order parameter) vanishes as 
$M\sim |t|^{\beta}$ with $t=(T-T_c)/T_c$, while the specific heat, the magnetic susceptibility  
and the correlation length diverge,
$C\sim |t|^{-\alpha}$, $\chi\sim |t|^{-\gamma}$ and 
$\xi \sim |t|^{-\nu}$, respectively.
The critical exponents $\alpha,\beta,\gamma,\nu$ and similar ones for other quantities 
are the same for all systems within a universality class. 
The latter are usually labeled by spin models, since those can be readily solved numerically. 
Unfortunately for most systems, and particularly QCD, it is not obvious which 
global symmetries the system will exhibit at a critical point, 
since those are a dynamically
determined subset of the total symmetry of the theory.
Moreover, since there is no true order parameter in the case
of finite quark masses, fields and parameters of QCD map 
into those of the effective model in a non-trivial way.

Nevertheless, general scaling properties can be derived, we follow \cite{Karsch:2001nf}.
In the vicinity of a critical point, the dynamics of QCD will be governed by
an effective Hamiltonian in analogy to a spin model,
with energy-like and magnetisation-like (extensive) operators $E, M$ 
which couple to two relevant couplings,
\be
\frac{H_{eff}}{T}=\tau  E+h M.
\ee
In the case of the Ising model  $\tau$ and $h$ are proportional to
temperature and magnetic field, respectively.
The power law behaviour for thermodynamic functions near a critical point
follows from the scaling form of the singular part of the free energy, 
\be
f_s(\tau, h) = b^{-d} f_s(b^{D_\tau}\tau, b^{D_h}h),
\ee
with a dimensionless scale factor $b=LT=N_s/N_\tau$, the spatial dimension $d$ and the dimensions of the scaling fields $D_\tau,D_h$.
Using general scaling relations between those and the critical exponents,
\be
D_\tau=\frac{1}{\nu}, \quad \gamma=\frac{2D_h-d}{D_\tau},\quad \alpha=2-\frac{d}{D_\tau},
\ee 
one derives the scaling of the susceptibilities as
\ba
\chi_E&=&V^{-1}\langle(\delta E)^2\rangle=-\frac{1}{T}\frac{\partial^2f}{\partial\tau^2}\sim b^{\alpha/\nu},\nn
\\ \chi_M&=&V^{-1}\langle(\delta M)^2\rangle=-\frac{1}{T}\frac{\partial^2f}{\partial h^2}\sim b^{\gamma/\nu}\;.
\label{scale}
\ea
(Note the different volume factors between \eqs(\ref{susc},\ref{scale}), which is due
to the fact that \eq(\ref{scale}) uses extensive variables while in \eq(\ref{susc}) we use averages.)
In QCD, the operators $E,M$ will be mixtures of the plaquette action and the chiral condensate 
(and possibly higher dimension operators), while $\tau$ and $h$ are functions of $\beta,m_f,\mu$.
Thus, when measuring susceptibilities constructed from operators as in \eq(\ref{susc}), these will
in turn be mixtures of $E$ and $M$. Therefore in the thermodynamic limit all of them will show identical
FSS behaviour which is dominated by the larger of $\alpha/\nu$ and $\gamma/\nu$.
For the universality classes relevant for QCD, $Z(2)$ and $O(4),O(2)$, this is $\gamma/\nu$.

By contrast, at a first order phase transition the fluctuations diverge proportional to the spatial volume,
\be
\chi_{\bar{O}}\sim V\;,
\ee
while for an analytic crossover the peaks of susceptibilities saturate at some finite
maximal value in the thermodynamic limit. 

\begin{figure}[t]
\centerline{
\includegraphics[width=0.44\textwidth]{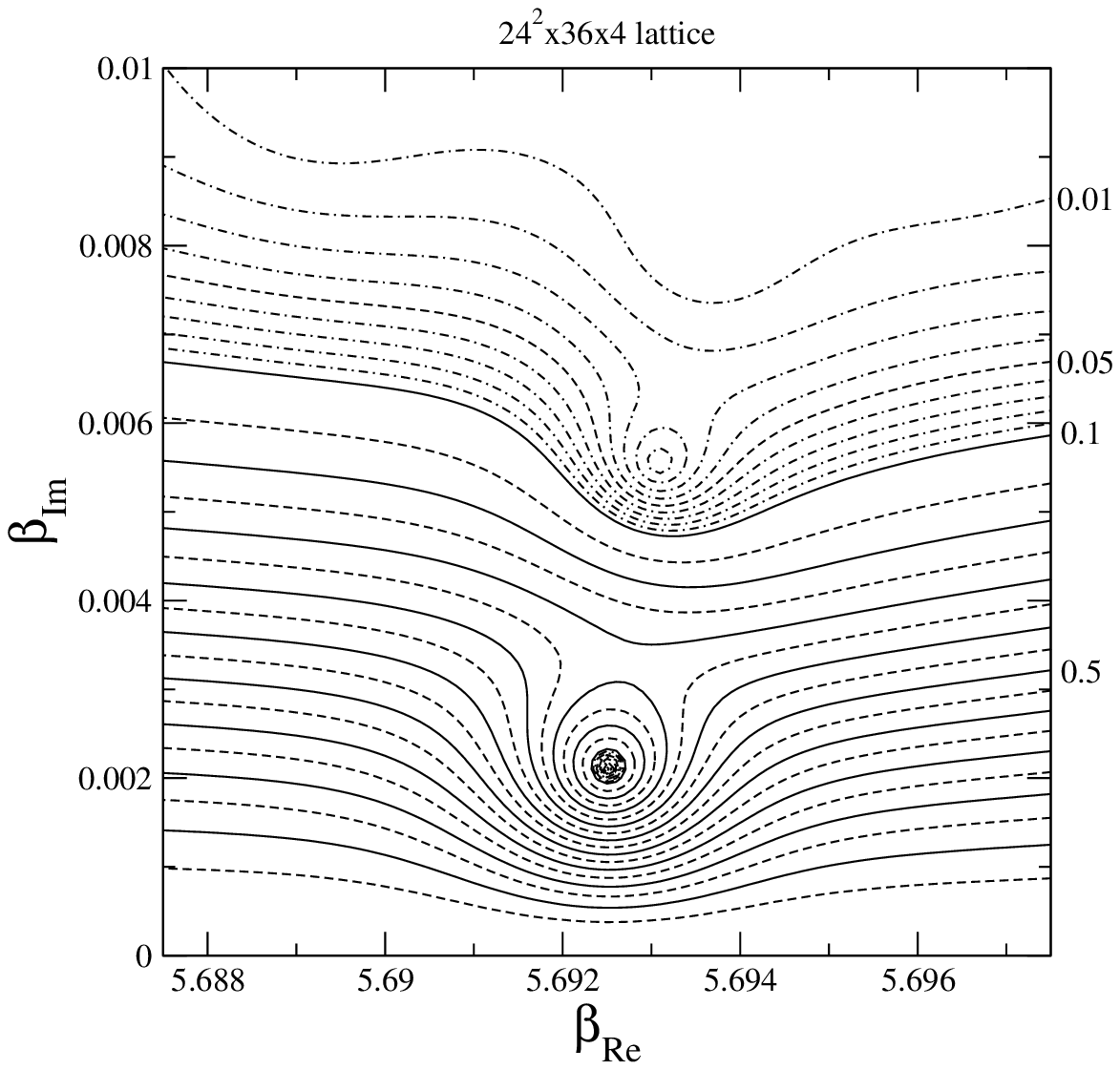}\hspace*{1cm}
\includegraphics[width=0.5\textwidth]{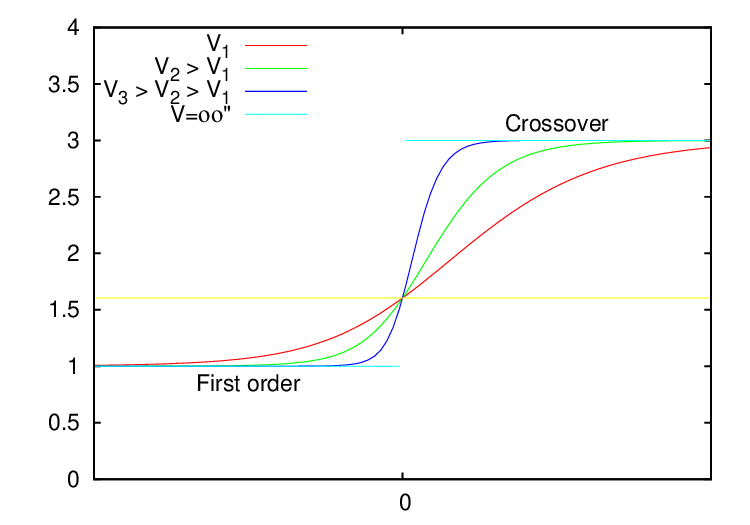}
\put(-190,120){$B_4$}
\put(-98,-6){$x-x_c$}
}
\caption[]{Left: Lee-Yang zeroes in the complex $\beta$-plane for SU(3) 
pure gauge theory \cite{Ejiri:2005ts}.
Right: The Binder cumulant as a function of some parameter $x$ which takes the phase transition from a first order to a crossover regime, passing through a critical point of 3d Ising universality.
(E.g.~$x=m_f$ in the case of $N_f=3, \mu=0$, cf.~\fig\ref{fig:2schem}) .}
\label{b4}       
\end{figure}

Another possibility to detect phase transitions is to evaluate the partition function for 
complex values of the coupling $\beta$
and look for Lee-Yang zeroes of the partition function, $Z(\beta)=0$.
On a finite volume, these will be at complex values of the coupling, since the partition function
is analytic for real parameter values. On larger volumes 
a Lee-Yang zero approaches the real axis if there is a true phase transition. 
However, such evaluations require multi-histogram
and reweighting techniques (cf.~Sec.\ref{sec:rew}), where
the values of the observable at different parameter values are calculated from one simulation
point, a technique which has to be handled with care \cite{Ejiri:2005ts}.
\fig\ref{b4} (left) shows an example of Lee-Yang zeroes for $SU(3)$ pure gauge theory
at complex values of the coupling.

A more straightforward but numerically very expensive observable to determine the order of a phase transition is the Binder cumulant
constructed from moments of fluctuations of the order parameter,
\be
B_4(m,\mu)=\frac{\langle(\delta \bar{O})^4\rangle}
{\langle(\delta \bar{O})^2\rangle^2}\;.
\ee
It is to be evaluated on the phase boundary $\beta=\beta_c(m_f,\mu)$, 
where the third moment of the fluctuation vanishes, $\langle (\delta \bar{O})^3\rangle(\beta_c)=0$.
This observable is particularly well suited to locate the change from a first order transition to 
a crossover regime as a function of some parameter, like quark mass or chemical potential. 
In the infinite volume limit, $B_4\rightarrow 1$ or $3$ for a first order transition or crossover, respectively, whereas
it approaches a value characteristic of the universality class at a critical point. For $Z(2)$ (or 3d Ising)
universality one has $B_4\rightarrow 1.604$.
Hence, when examining the change of a phase transition from a weakening first order transition to a 
critical end point and a crossover as a function of $x=am_f, a\mu$,  
$B_4$ is a non-analytic step function, which gets smoothed out to an 
analytic curve on finite volumes,
with a slope increasing with volume to gradually approach the step function, \fig\ref{b4} (right).
Near a critical point the correlation length diverges 
as $\xi\sim r^{-\nu}$, where $r$ is the distance to the critical point in the plane of temperature
and magnetic field like variables. Since the gauge coupling is tuned to $\beta=\beta_c(x)$,  
$r\sim |x-x_c|$ and $B_4$ is a function 
of the dimensionless ratio $L/\xi$, or equivalently, $(L/\xi)^{1/\nu}$. 
It can therefore be expanded as
\be
B_4(x,N_s)=b_0+b \, N_s^{1/\nu}(x-x^c)+\ldots.
\ee
Close to the thermodynamic limit all curves intersect at a critical $B_4$-value, and moreover $\nu$  
takes its universal value. Thus calculation of $B_4$ around a critical point gives access to two 
independent pieces of information governed by universality.

\section{The nature of the QCD phase transition for $N_f=2+1$ at $\mu=0$}

\begin{figure}[t]
\centerline{
\includegraphics*[width=0.5\textwidth]{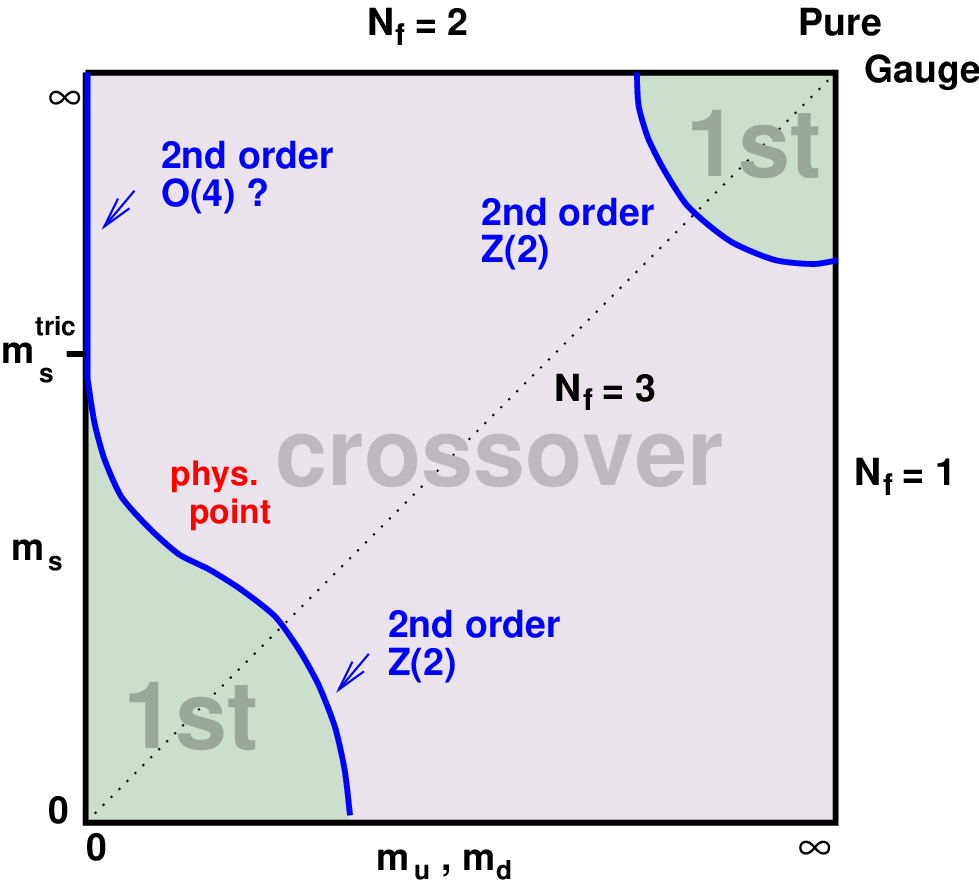}
}
\caption[]{
Order of the QCD phase transition as function of quark masses $(m_{u,d},m_s)$ at $\mu=0$.}
\label{fig:2schem}      
\end{figure}
The qualitative picture for the order of the QCD phase transition at zero baryon density
as a function of the quark masses is outlined schematically in 
\fig\ref{fig:2schem}.
As discussed in Secs.~\ref{zn},\ref{chiral}, for $N_f=3$
in the limits of zero and infinite quark masses
(lower left and upper right corners),
order parameters corresponding to the breaking of a symmetry can be defined, implying true phase 
transitions.
One finds numerically at small and large quark masses
that a first-order transition takes place at a finite
temperature $T_c$. On the other hand, one observes an analytic crossover at
intermediate quark masses. Hence, each corner must be surrounded by a region of
first-order transition, bounded by a second-order line. The line bounding the chiral transitions is
commonly called the chiral critical line, the one bounding the deconfinement transition 
is accordingly the deconfinement critical line. We know from simulations on 
$N_\tau=4$ with staggered fermions \cite{deForcrand:2006pv}  
that this line is to the lower left of the physical
point, and simulations on $N_\tau=6$ \cite{deForcrand:2007rq,Endrodi:2007gc} show that it
shrinks towards the lower left corner on finer lattices. A continuum extrapolation for the chiral critical
line is not yet available. However, there are continuum extrapolated simulations with staggered quarks
with physical masses that show that the $\mu=0$ transition in QCD is a crossover \cite{Aoki:2006we}.

The situation is not yet settled for the chiral
limit of the two flavour theory in the upper left corner.
If the transition is second order, then chiral symmetry
$SU(2)_L\times SU(2)_R \sim O(4)$ puts it in the universality
class of 3d $O(4)$ spin models.
In this case there must be a tri-critical strange quark mass $m_s^{tric}$,
where the second order chiral
transition ends and the first order region begins. The exponents at such a
tri-critical point would
correspond to 3d mean field \cite{tric}. On the other hand,
a first order scenario for the chiral limit
of $N_f=2$ so far has not been conclusively ruled out. In fact, it has been shown in 
a model with the same chiral symmetries as $N_f=2$ QCD and a tuneable $U_A(1)$ anomaly
that both scenarios are possible and the order of the transition depends on the strength of the 
anomaly at the critical temperature \cite{Chandrasekharan:2007up}. 
 
\chapter{Lattice QCD at non-zero baryon density}

\section{Implementing chemical potential}

So far we have considered lattice QCD at finite temperature with net baryon number zero, i.e.~with a 
complete balance between baryons and anti-baryons. This situation is realised during the evolution of the 
early universe because the matter anti-matter asymmetry is so tiny there.
However,  there are interesting questions 
about the nature of dense nuclear matter in the centre of compact stars,
and heavy ion collision experiments obviously also operate at finite baryon number.
Thus, we now wish to use the grand canonical ensemble with a chemical potential $\mu$ for quark
number $Q$ (recall that each quark carries 1/3 of baryon number, $Q=B/3, \mu=\mu_B/3$),
\be
Z= \hTr\;\ex^{-(H-\mu Q)},\quad Q=\int d^3x \;\bar{\psi}(x)\gamma_0\psi(x)=\int d^3x \;\psi^\dag(x)\psi(x)\;.   
\ee
For the following sections we recall that
in Euclidean space time $\gamma_\mu=\gamma_\mu^\dag, \{\gamma_5,\gamma_\mu\}=0$. 
Under charge conjugation, $C=\gamma_0\gamma_2$, fermion number changes sign,
\be
A_\mu^C=-A_\mu^*, \quad 
\psi^C=\gamma_0\gamma_2\bar{\psi}^T,\quad \bar{\psi}^C\gamma_0\psi^C=-\bar{\psi}\gamma_0\psi\;.
\ee
Thus, $\mu>0$ represents a net baryon number and $\mu<0$ a net anti-baryon number. 
From this one derives an important symmetry of the partition function. Since the measure
and the fermion action for $\mu=0$ are invariant under charge conjugation, we find that the partition 
function is an even function of chemical potential,
\ba
Z(\mu)
&=&\int DA^C \,D\bar{\psi}^CD\psi^C \,
\exp-\left[S^C_g+S_f^C(\mu=0)-\mu\int_0^{1/T} dx_0\, Q^C\right]\nn\\
&=&\int DA \,D\bar{\psi}D\psi \,\exp-\left[S_g+S_f(\mu=0)+\mu\int_0^{1/T} dx_0 \,Q\right]\nn\\
&=&Z(-\mu)\;.
\label{even}
\ea

In a straightforward implementation of chemical potential on the lattice we would simply replace the integral by a lattice sum,
\be
S_f[M(\mu)]=S_f[M(0)]+a\mu \sum_{x} \psi(x)\gamma_0\psi(x)\;.
\label{muac}
\ee
However, when computing the energy density in lattice perturbation theory, one finds it to diverge
in the continuum limit,
\be
\epsilon=\frac{1}{V}\frac{\partial}{\partial (\frac{1}{T})}\ln Z\stackrel{a\rightarrow 0}{\longrightarrow}\infty\;.
\ee
This happens after duly subtracting the divergent vacuum contribution, and thus the 
chemical potential term in \eq(\ref{muac}) appears to spoil renormalisability. This would violate our 
earlier observation that renormalisation of the vacuum theory is also sufficient
for the theory at finite temperature and density. The reason is that the discretisation violates a
generalised symmetry of the continuum theory \cite{Hasenfratz:1983ba}. 
Note that the term $\mu Q$ in the continuum looks
like the zero component of the fermion current, $j^0=\bar{\psi}\gamma^0\psi$, coupling to an external
$U(1)$ gauge field,
\be
\mu Q=-ig\int d^3x\,A_0j_0\quad \mbox{with}  \quad A_0=i\frac{\mu}{g}\;.
\label{u1}
\ee
Chemical potential for quark number is equivalent to a classical electromagnetic field $A_0$
with a constant imaginary value. The quark number term can therefore be absorbed into the covariant derivative of the Dirac action, which then is invariant under $U(1)$ gauge transformations of the quark fields and $A_0$. This symmetry protects against new divergences and gets broken by a lattice 
implementation as in \eq(\ref{muac}). The problem is solved by implementing chemical potential 
in the same way as an external $U(1)$ gauge field, namely as an additional temporal link variable
\be
U_{0,\rm ext}=\ex^{iagA_0}=  \ex^{-a\mu}\;,
\ee
which multiplies all non-abelian temporal links. For example, the Wilson action with finite chemical
potential then takes the form
\ba
S_f^W&=&a^3\sum_x  \left( \bar{\psi}(x)\psi(x)\right. \nn\\
&&-\kappa\left[\ex^{a\mu}\bar{\psi}(x)(r-\gamma_0)
U_0(x)\psi(x-\hat{0})+\ex^{-a\mu}\bar{\psi}(x+\hat{0})(r+\gamma_0)U_0^\dag(x)\psi(x)\right] \nn\\
&&\left.-\kappa\sum_{j=1}^3 \left[\bar{\psi}(x)(r-\gamma_j)U_j(x)\psi(x+\hat{j})+\bar{\psi}(x+\hat{j})(r+\gamma_j) U_j^\dag(x)\psi(x)\right]\right)\;.
\ea
As usual on the lattice, this discretisation is not unique, only the continuum limit is. For alternative
ways of implementing chemical potential, see \cite{Gavai:1985ie}.
It is now easy to verify that
\be
\det(\Dslash(U^\dag)+m+\gamma_0\mu)=\det(\Dslash(U)+m-\gamma_0\mu)\;,
\ee
and since $S_g[U^\dag]=S_g[U]$ and $DU^\dag=DU$ we have $Z(\mu)=Z(-\mu)$ on the lattice as well.

\section{The sign problem}

Let us consider what happens to the Dirac operator in the presence of chemical potential. 
Reality of the fermion determinant follows from the $\gamma_5$-hermiticity of  the Dirac operator,
\be
(\Dslash+m)^\dag=\gamma_5(\Dslash+m)\gamma_5\;.
\ee
This relation is satisfied in the continuum as well as for all standard lattice Dirac operators.
Now consider a complex chemical potential,
\be
\gamma_5(\Dslash+m-\gamma_0\mu)\gamma_5=(-\Dslash+m+\gamma_0\mu)=(\Dslash+m+\gamma_0\mu^*)^\dag\;.
\ee
For the fermion determinant this implies,
\be
\det(\Dslash + m-\gamma_0\mu)={\det}^*(\Dslash+m+\gamma_0\mu^*)\;, 
\label{detpos}
\ee
which is complex unless ${\rm Re} \,\mu=0$.
However, a complex fermion determinant cannot be interpreted as a probability measure for importance
sampling and the evaluation of the path integral by Monte Carlo methods is spoiled. This is known
as the ``sign-problem'' of QCD.

It should be stressed that there is nothing wrong with the theory at finite $\mu$. In particular,
the partition function as well as the free energy and the other thermodynamic functions are all real
after performing the path integral over the gauge fields, i.e.~the imaginary parts of the fermion determinant
in the background of gauge configurations average to zero. Being a property of the Dirac operator, the 
sign problem is generic for fermionic systems with particles and anti-particles or particles and holes, 
and
is necessary for a correct description of the physics.
For example, for the expectation value of the Polyakov loop we have
\ba
\langle \Tr L\rangle=&\ex^{-\frac{F_Q}{T}}&=\langle \re \Tr L \;\re \det M - \im \Tr L \;\im \det M\rangle_g\;,\nn\\
\langle{(\Tr L)}^*\rangle=&\ex^{-\frac{F_{\bar{Q}}}{T}}&=\langle \re\Tr L \;\re\det M + \im \Tr L \;\im \det M\rangle_g\;,
\ea
where the angular brackets denote path integration with respect to the pure gauge action,
$\langle\ldots\rangle_g=\int DU \ldots \exp-S_g[U]$.
Thus, the free energy of a static quark or an anti-quark in a plasma differ if and only if $\im\det\neq 0$, 
i.e.~for finite chemical potential $\mu$. If there is no net quark or baryon number, it costs equal amounts
of energy to insert a quark or an anti-quark into the plasma, whereas this is different if the plasma already
has a net baryon number. The origin of the sign problem may thus be traced to the behaviour 
under charge conjugation, 
\be
\det(\Dslash+m-\gamma_0\mu)\stackrel {C}{\longrightarrow} \det (\Dslash + m+\gamma_0\mu)\;.
\ee
The ``sign problem'' thus is only a problem for a Monte Carlo evaluation of the partition function with 
importance sampling methods. In some spin models, it can be solved by cluster algorithms, which 
are able to identify conjugate configurations so as to cancel imaginary parts during the simulation
\cite{Wiese:2002ws}. Unfortunately, these work only for special classes of Hamiltonians and QCD does not appear
to be one of them. Another idea is to simply avoid importance sampling and evaluate expectation values
by means of stochastical quantisation, or Langevin algorithms \cite{Karsch:1985cb}. 
However, also in this case there are difficulties, for complex actions there is no proof that the Langevin
method converges to the correct expectation value. While the method seems to work for various
model systems, there appears to be no successful application to QCD yet.

Due to the symmetry properties of the determinant, there are a number of interesting special cases permitting a Monte Carlo treatment. As long as the fermion determinant is real, we are able to simulate
even numbers of degenerate flavours with it.
\eq(\ref{detpos}) shows that this is the case for purely imaginary $\mu=i\mu_i, \mu_i\in \mathbb{R}$, 
and we shall discuss later how this can be used to learn something about real $\mu$.

\begin{figure}[t]
\begin{center}
\includegraphics*[width=0.45\textwidth]{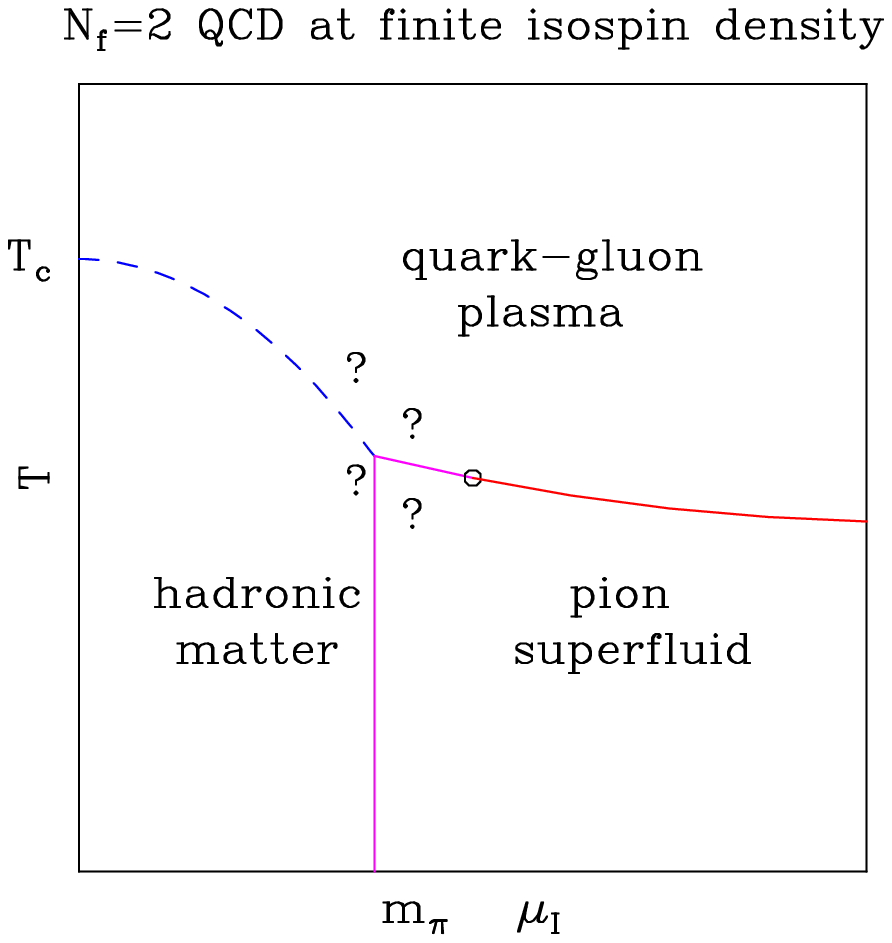}
\caption[]{Schematic phase diagram for QCD at finite isosipin density.}
\label{iso}
\end{center}
\end{figure}

Another interesting situation for $N_f=2$ is when the chemical potential for $u$ and $d$ quarks are opposite, i.e.~$\mu_u=-\mu_d\equiv \mu_I$, which corresponds to a chemical potential for 
isospin \cite{Son:2000xc}. In this case we have for the determinants
\be
\det(\Dslash + m -\gamma_0\mu_I)\det(\Dslash+m+\gamma_0\mu_I)=
|\det(\Dslash+m-\gamma_0\mu_I)|^2\geq 0\;.
\ee
Note that isospin is not conserved in the real world due to the electroweak interactions.
Nevertheless, one can investigate interesting finite density effects with this theory by actual 
simulations \cite{Kogut:2007mz}. For example,  one obtains a condensate of 
charged pions $\langle\pi^-\rangle\neq 0$ when $|\mu_I|>m_\pi$, \fig\ref{iso}.  

Finally, the determinant is real for a two-colour version of QCD with gauge group $SU(2)$.
This is because this group has real representations, i.e.~there is a matrix $S$ such that
$ST^aS^{-1}=-T^{a*}$ with $S=\sigma^2$, the Pauli matrix. 
Now consider $S=C\gamma_5 \sigma^2$, and using $C\gamma_\mu C^{-1}=-\gamma_\mu^T$ we have
\ba
S[\Dslash+m-\gamma_0\mu]S^{-1}&=&
C\gamma_5 \sigma^2[\gamma_\mu(\partial_\mu -igA_\mu)+m-\gamma_0\mu]\sigma^2\gamma_5C^{-1}
\nn\\
&=&C\gamma_5 [\gamma_\mu(\partial_\mu+igA_\mu^*)+m-\gamma_0\mu]\gamma_5C^{-1}\nn\\
&=&[\gamma_\mu^T(\partial_\mu+igA_\mu^*)+m-\gamma_0^T\mu]\nn\\
&=&[\Dslash+m-\gamma_0\mu^*]^*\;.
\ea
For real $\mu$ we have $\det M={\det}^*M$, likewise permitting simulations to study qualitative
density effects \cite{Hands:2007uc}.  

In the following sections, the main concern will be with methods for the theory of immediate 
experimental interest, i.e.~three-colour QCD with real quark number chemical potential.
As will become apparent, all those methods side-step the sign problem by introducing 
approximations which work only for sufficiently small chemical potential, empirically one finds
these methods to be good as long as $\mu\lsim T$. 

\section{Reweighting} \label{sec:rew}

\begin{figure}[t]
\begin{center}
\includegraphics[width=0.44\textwidth]{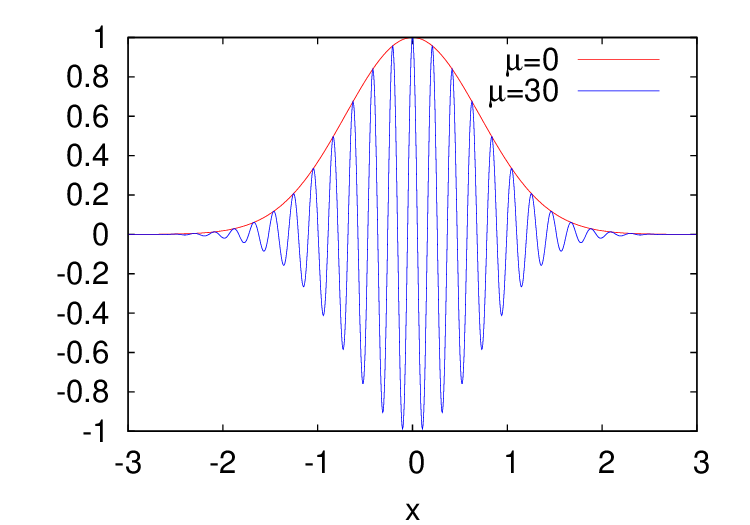}\hspace*{0.5cm}
\includegraphics[width=0.3\textwidth]{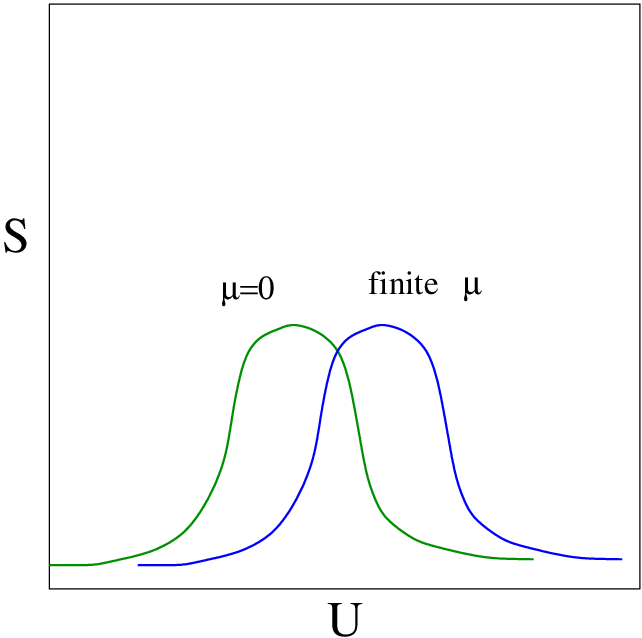}
\end{center}
\caption[]{Left: The integrand of \eq(\ref{toyz}) for different $\mu$. Right: In order for reweighting
to work, the probability distribution of the Monte Carlo ensemble must have sufficient ``overlap'' with the 
target integrand.}
\label{osc}
\end{figure}

Reweighting techniques are widely used in Monte Carlo simulations with importance sampling, in
particular in order to interpolate data between simulation points when studying phase transitions 
\cite{Ferrenberg:1988yz}. In the context of finite density physics, this technique is used to extrapolate
to finite chemical potential. The basic idea is to rewrite the path integral exactly,
\ba
Z(\mu)&=&\int DU\;\det M(\mu)\;\ex^{-S_g[U]}=\int DU\;\det M(0)\;\frac{\det M(\mu)}{\det M(0)}\;\ex^{-S_g[U]}
\nn\\ &=&Z(0) \left\langle \frac{\det M(\mu)}{\det M(0)}\right\rangle_{\mu=0}\;.
\label{rew}
\ea
The angular brackets now denote averaging over an ensemble with the 
same measure $DU \;\det M(0)$ as for zero density, i.e.~the path integral can be interpreted as
an expectation value of a reweighting factor given by the ratio of determinants. 
While this is a mathematical identity, its practical evaluation by Monte Carlo methods
turns out to be impressively difficult.
First, a numerical evaluation of this path integral requires the calculation of the reweighting factor
configuration by configuration and is expensive. This is aggravated by a signal to noise ratio
which worsens exponentially with volume because the free energy is an extensive quantity $F=V f$,
\be
\frac{Z(\mu)}{Z(0)}=\exp-\frac{F(\mu)-F(0)}{T}=\exp-\frac{V}{T}\left(f(\mu)-f(0)\right)\;.
\ee
Hence the reweighting factor gets exponentially small as we increase $V$ or $\mu/T$, 
requiring exponentially increased statistics for its determination.
The problem can be illustrated with a one-dimensional Gaussian integral with a complex ``action'',
\be
Z(\mu)=\int dx \;\ex^{-x^2+i\mu x}\;.
\label{toyz}
\ee
\fig\ref{osc} (left) shows the real part of the integrand for $\mu=0$ and $\mu=30$. 
It is obvious that the oscillatory bevahiour
of the integrand will rapidly become prohibitive for Monte Carlo integration with growing $\mu$.

The second problem is known as the overlap problem. When we evaluate the integral with importance
sampling, the integrand will be calculated the more frequently the larger its 
contribution to the integral. However, when we employ reweighting, the configurations will be generated
with the measure $DU\; \det M(0)$, whose probability distribution is shifted compared to that of 
$DU \;\det M(\mu)$, \fig\ref{osc} (right). 
If we had infinite statistics, this would not matter as long as our algorithm is ergodic 
and the entire configuration space is covered. The point of Monte Carlo methods is, however, to get
away with a few ``important'' configurations on which we evaluate our determinant and observables.
But if the difference between the integrands at $\mu=0$ and $\mu\neq 0$ gets too large, 
the most frequent configurations are the unimportant ones.
It is apparent that we only get a useful estimate for observables as long as there is sufficient
overlap between the two integrands. The difficulty with this method is to realise when 
it fails. The statistical errors are determined by the fluctuations within the generated ensemble and will 
appear small, unless there are configurations that have tunneled far into the tail where the actual
integrand is important. 

When searching for a phase transition at some fixed $\beta_c(\mu)$, one-parameter reweighting uses an  ensemble at $\beta_c(\mu),\mu=0$, which is non-critical since $T_c(\mu)<T_c(0)$, 
thus missing important dynamics.
Significant progress enabling finite density simulations was made a few years ago, by a generalisation to reweighting in two parameters \cite{Fodor:2001au}.  
The partition function is now rewritten as
\be 
Z(\mu,\beta)=Z(0,\beta_0)\left \langle \frac{\ex^{-S_g(\beta)}\det(M(\mu))}
{\ex^{-S_g(\beta_0)}\det(M(\mu=0))}\right\rangle_{\mu=0,\beta_0},
\ee
where the ensemble average is generated at $\mu=0$ and a lattice gauge coupling 
$\beta_0$, while a reweighting factor takes us to the values
$\mu,\beta$ of interest. In this way one can use an ensemble on the phase boundary at $\mu=0$, 
which actually samples both phases, and then reweight along the (pseudo-)critical line to the desired value of $\mu$.
On physical grounds, this is drastically improving the overlap when attempting to describe a phase 
transition.

Clearly, the choice of reweighting factors is not unique and one may ask for an optimal choice.
A useful criterion is to minimise fluctuations in the reweighting factor, \eq(\ref{rew})
\cite{deForcrand:2002pa}.
The optimal setup with a feasible Monte Carlo implementation is to split the determinant into 
modulus and phase, $\det M=|\det M| \ex^{i\theta}$,
and employ the modulus for the generation of the ensemble. Note that this is equivalent to an ensemble
with finite isospin chemical potential. Thus,
\be
Z(\mu)=Z(\mu_I)\langle \ex^{i\theta}\rangle_{\mu_I}\;.
\ee
Numerical experiments show that for small to moderate chemical potentials, this choice is 
advantageous \cite{deForcrand:2002pa} compared to the standard procedure. Moreover, it
serves for interesting theoretical insights. 
There are indications in QCD \cite{Splittorff:2005wc} and
firm results in random matrix models \cite{Han:2008xj} that the phase transition to the pion condensate
in the theory with finite isospin chemical potential is directly related to the strength of the sign problem,
which becomes maximal with an average sign near zero in the neighbourhood of that transition, thus
signalling a breakdown of reweighting methods there. 

\section{Finite density by Taylor expansion}

Another straightforward way to employ Monte Carlo ensembles generated at $\mu=0$ to learn 
about finite $\mu$ is by Taylor-expanding the partition function in $\mu/T$. 
The grand canonical partition function is an analytic function of its parameters
away from phase transitions. The Lee-Yang theorem tells us that on finite volumes there are no
non-analytic phase transitions, and so the partition function must be analytic everywhere.
It is then natural to consider power series in $\mu/T$ for thermodynamic observables like the pressure
\cite{Allton:2002zi}, 
\be
{p\over T^4}=
\sum_{n=0}^\infty c_{2n}(T) \left({\mu\over T}\right)^{2n}\equiv \Omega(T,\mu).
\label{press}
\ee
Note that there are only even powers of $\mu/T$ because of the reflection symmetry, \eq(\ref{even}).
The leading Taylor coefficient is just the pressure at zero density, and all higher coefficients
are derivatives evaluated at $\mu=0$,
\be
c_0(T)=\frac{p}{T^4}(T,\mu=0),\quad c_{2n}(T)=\frac{1}{(2n)!}\left.
\frac{\partial^{2n}\Omega}{\partial (\frac{\mu}{T})^{2n}}\right |_{\mu=0}\;.
\ee
These can of course be calculated by
standard Monte Carlo techniques. Once the coefficients are available, 
the series for other thermodynamic quantities follow,
like the quark number density or the quark number susceptibility, respectively,
\ba
\frac{n}{T}&=&\frac{\partial\Omega}{\partial(\frac{\mu}{T})}=2c_2\frac{\mu}{T}+
4c_4\left(\frac{\mu}{T}\right)^3+\ldots,\nn\\
\frac{\chi_q}{T^2}&=&\frac{\partial^2\Omega}{\partial(\frac{\mu}{T})^2}=2c_2+12c_4\left(\frac{\mu}{T}\right)^2
+30c_6\left(\frac{\mu}{T}\right)^4+\ldots
\label{tobs}
\ea
Clearly, this strategy can be applied to any observable of interest and is well-defined
computationally. 
As one needs to calculate coefficient by coefficient, the full functions are only 
well approximated as long as the series converges sufficiently rapidly, i.e.~for sufficiently small $\mu/T$.

In practice, what needs to be evaluated during a simulation are the derivatives of a particular
observable with respect to chemical potential,
\be
\frac{\partial \langle O \rangle}{\partial \mu}=\left\langle \frac{\partial O}{\partial \mu}\right\rangle
+ N_f\left(\left\langle O\frac{\partial \ln\det M}{\partial \mu}\right\rangle-
\langle O\rangle\left\langle\frac{\partial \ln\det M}{\partial \mu}\right\rangle\right)\;.
\ee
Using $\det M=\exp\Tr\ln M$, this is converted into expressions like
\ba
\frac{\partial\ln\det M}{\partial \mu}&=&\Tr\left(M^{-1}\frac{\partial M}{\partial \mu}\right) \nn\\
\frac{\partial^2\ln\det M}{\partial \mu^2}&=&\Tr\left(M^{-1}\frac{\partial^2M}{\partial \mu^2}\right)
-\Tr \left(M^{-1}\frac{\partial M}{\partial \mu}M^{-1}\frac{\partial M}{\partial \mu}\right)\nn\\
&\mbox{etc.}&
\ea
The advantage is that the derivatives of the fermion determinant are represented by local operators,
which can be evaluated using random noise vectors.
But it is clear that higher order derivatives quickly turn into very complex expressions involving
many cancellations, and hence are very cumbersome to 
evaluate with controlled accuracy. For a discussion of this point, see \cite{Gavai:2004sd}.

\section{QCD at imaginary chemical potential}

The hermiticity relation \eq(\ref{detpos}) tells us that the QCD fermion determinant is
real for imaginary chemical potential $\mu=i\mu_i$. For such a 
parameter choice we can simulate without sign problem with no more complications than at $\mu=0$,
and moreover we can do so with an ensemble that is actually sensitive to chemical potential.
In order to get back to real chemical potential, we can use once more the Taylor expansion,
\be
\langle O \rangle (\mu_i)=\sum_{k=1}^Nc_k\left(\frac{\mu_i}{T}\right)^{2k}\;,
\ee
with only even powers for observables without explicit $\mu$-dependence.
Note that in this case the Monte Carlo results represent the left side of the equation and contain no
approximation beyond the usual finite size and cut-off effects. Thus, if there are sufficiently many and
accurate data points, one can test whether the observable is well-represented by a truncated Taylor
expansion, and if this is the case one may analytically continue, $\mu_i\rightarrow -i\mu_i$.

In order to apply this technique, it is necessary to discuss a few general properties of the partition function.
From the form of the grand canonical density operator, $Z= \hTr\;\exp{-(H-\mu Q)}$, it is clear that
the partition function is going to be periodic for imaginary $\mu$. Is this a sensible theory? Modifying
the discussion around \eq(\ref{u1}), imaginary chemical potential is formally equivalent to a real external $U(1)$ field, so this parameter choice is well-defined. Next, let us examine the periodicity 
\cite{Roberge:1986mm}. Since 
we always have a compactified temporal lattice direction, we can eliminate the quark number term
from the action and absorb it by the modified boundary condition,
\be
Z^{(1)}(i\mu_i)=\int DU \det M(0) \ex^{-S_g},\quad \mbox{b.c.:} \;\;\;
\psi(\tau+N_\tau,\bfx)=-\ex^{i\frac{\mu_i}{T}}
\psi(\tau,\bfx)\;.
\label{z1}
\ee
Now let us consider gauge transformed fields $\psi^{g'}(x), U_\mu^{g'}(x)$, using the large 
gauge transformations discussed in Section \ref{zn}, i.e.
\be
g'(\tau+N_\tau,x)=\ex^{-i\frac{2\pi n}{N}}g'(\tau,\bfx)\;.
\label{gprime}
\ee
The measure and the QCD action for $\mu=0$ are invariant under such a transformation, so the new 
form of the partition function is
\be
Z^{(2)}(i\mu_i)=\int DU \det M(0) \ex^{-S_g},\quad \mbox{b.c.:} \;\;\;
\psi(\tau+N_\tau,\bfx)=-\ex^{-i\frac{2\pi n}{N}}\ex^{i\frac{\mu_i}{T}}
\psi(\tau,\bfx).
\label{z2}
\ee
We now observe that
\be
Z^{(2)}\left(i\frac{\mu_i}{T}+i\frac{2\pi n}{N}\right)=Z^{(1)}\left(i\frac{\mu_i}{T}\right)\;.
\ee
Since we have obtained one partition function from the other by a gauge transformation, the two
give completely equivalent descriptions of physics, so we can conclude for the QCD partition function
\be
Z\left(i\frac{\mu_i}{T}+i\frac{2\pi n}{N}\right)=Z\left(i\frac{\mu_i}{T}\right)\;.
\label{zper}
\ee
Comparing \eqs(\ref{z1},\ref{gprime}), we see that for discrete values 
$\mu_i/T=2\pi n/N$ an imaginary chemical potential is equivalent to a centre transformation, and 
\eq(\ref{zper}) tells us that the partition function is symmetric under such transformations, and therefore
periodic.
We have thus established two remarkable properties: the period of the QCD partition function in the presence of an imaginary chemcial potential is $2\pi/N$, and
the $Z(N)$ centre symmetry is a good symmetry even in the presence
of finite mass quarks! 

Let us now specialise to the physical case, $N=3$.
We recall from Sec.~\ref{zn} that the Polyakov loop closes through the temporal boundary and 
hence picks up phases when changing the centre sector, \eq(\ref{polyz}).
Therefore the phase of the Polyakov loop is an observable to identify the different $Z(3)$-sectors
as the imaginary chemical potential is increased.
There are $Z(3)$ transitions between neighbouring centre sectors for all 
$(\mu_i/T)_c=\frac{2\pi}{3} \left(n+\frac{1}{2}\right), n=0,\pm1,\pm2,...$. It has been numerically verified that these transitions are first order for high temperatures and a smooth crossover for low temperatures 
\cite{deForcrand:2002ci,D'Elia:2002gd}. 

\begin{figure}[t]
\centerline{
\includegraphics[width=0.4\textwidth]{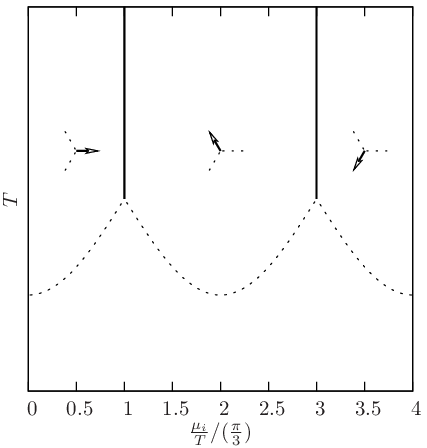}
\includegraphics[width=0.4\textwidth]{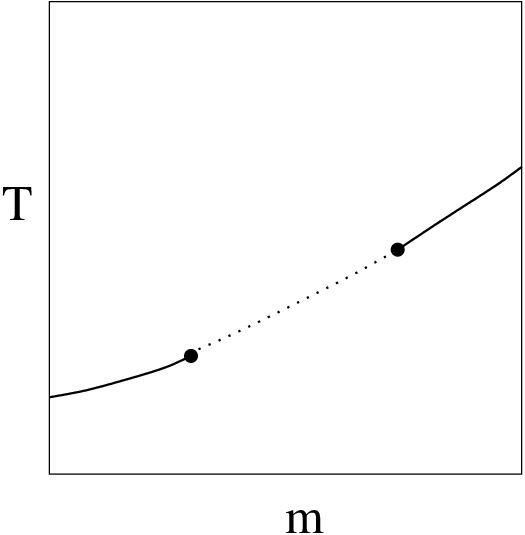}
\put(-133,120){coexistence of two phases of $L$}
\put(-105,35){averaged phases of $L$}
}
\caption[]{Left: Periodic phase diagram for imaginary chemical potential. Vertical lines correspond to 
first order transitions between the different $Z(3)$-sectors, the arrows indicate the corresponding phase
of the Polyakov loop. The $\mu=0$ chiral/deconfinement transition continues to imaginary chemical
potential, its order depends on $N_f$ and the quark masses. Right: Nature of the junction at 
fixed $\mu_i/T=\pi/3 \;\mbox{mod.}\; 2\pi/3$. Solid lines are lines of triple points
ending in tri-critical points,  connected by a $Z(2)$-line.}
\label{schem}
\end{figure}
\fig\ref{schem} (left) shows a schematic phase diagram.
The vertical lines represent the high temperature, first order $Z(3)$-transitions. On the other hand,
at $\mu=0$ we also have the chiral/deconfinement transition, which analytically continues to imaginary
chemical potential, as indicated by the dotted lines. As discussed before, its order depends on 
the quark mass combinations. It joins the $Z(3)$ transitions in their endpoint, whose 
nature has recently been investigated by explicit simulations for $N_f=2,3$ \cite{D'Elia:2009qz,deForcrand:2010he}.
For small and large quark masses, where there is a first order chiral and deconfinement transition,
respectively, three first order lines join up at this point, rendering it a triple point.
For intermediate quark masses the quark/hadron transition is only a crossover, and the
$Z(3)$ transition features an endpoint with 3d Ising universality.  Tuning the quark masses
interpolates between these situations, with tri-critical points marking the changes, \fig\ref{schem} (right).
If we want to learn about certain observables for real $\mu$ by means of analytic continuation,
the radius of convergence is given by the closest non-analyticity due to
phase transitions. At high temperatures and for the chiral/deconfinement transition line itself, 
we are thus limited to the first sector, or $\mu/T<\pi/3$.

\section{The canonical partition function}

The grand canonical partition function for imaginary chemical potential appears again in the 
formulation of finite density QCD by means of the
canonical partition function \cite{Roberge:1986mm}. Consider the Fourier transform of $Z(i\mu_i/T)$, 
\ba
\tilde{Z}(\bar{Q})&=& \frac{1}{2\pi}\int_{-\pi}^{\pi}d\left(\frac{\mu_i}{T}\right)\;\ex^{-i\frac{\mu_i}{T}\bar{Q}}\;
Z\left(i\frac{\mu_i}{T}\right)\nn\\
&=& \frac{1}{2\pi}\int_{-\pi}^{\pi}d\left(\frac{\mu_i}{T}\right)\;\ex^{-i\frac{\mu_i}{T}\bar{Q}}\;
Z\left(i\frac{\mu_i}{T}+i\frac{2\pi}{N}\right)\nn\\
&=& \frac{1}{2\pi}\int_{-\pi}^{\pi}d\left(\frac{\mu_i}{T}\right)\;\ex^{-i\left(\frac{\mu_i}{T}-\frac{2\pi}{N}\right)
\bar{Q}}\;
Z\left(i\frac{\mu_i}{T}\right)\;.
\ea
In the second line we have made use of the periodicity of the grand canonical partition function, 
\eq(\ref{zper}), and in the third we have shifted $\mu_i/T\rightarrow \mu_i/T+2\pi/N$.
The first and the third line can only be equal with  $\tilde{Z}\neq0 $ if
\be
\frac{\bar{Q}}{N}=n, \quad n=0,1,\ldots
\ee
Inserting the grand canonical partition function with 
the quark number operator $Q$, we can write
\ba
 \tilde{Z}(\bar{Q})&=& \frac{1}{2\pi}\int_{-\pi}^{\pi}d\left(\frac{\mu_i}{T}\right)\;\ex^{-i\frac{\mu_i}{T}\bar{Q}}\;
\Tr\left(\ex^{-\frac{(H-i\mu_iQ)}{T}}\right)\nn\\
&=&\Tr\left(\ex^{-\frac{H}{T}}\delta(\bar{Q}-Q)\right)\;,
\ea
where we have used the integral representation of the delta-function, 
\be
\delta(\bar{Q}-Q)=\frac{1}{2\pi}\int d\left(\frac{\mu_i}{T}\right) \exp \left(i\frac{\mu_i(\bar{Q}-Q)}{T}\right)\;.
\ee
We thus arrive at the conclusion that $\tilde{Z}(\bar{Q})$ corresponds to the QCD partition function
evaluated with a delta-constraint fixing quark number to be $\bar{Q}$. This is just the canonical
partition function for QCD,
\be
\tilde{Z}(\bar{Q})=Z_C(\bar{Q})\;.
\ee
Since quark number is constrained to be $\bar{Q}=nN$, baryon 
number $B=N\bar{Q}$ must be an integer.

The descriptions in terms of the canonical or the grand canonical partition function are identical
in the thermodynamic limit. The grand canonical ensemble can be obtained from the canonical
one by the fugacity expansion,
\ba 
Z(V,T,\mu)&=&\sum_{B=1}^{\infty}\ex^{\frac{\mu_BB}{T}}Z_C(V,T,B)\nn\\
Z(T,\mu)&\stackrel{V\rightarrow\infty}{=}&\int_{-\infty}^\infty d\rho\;\ex^{VN\rho\frac{\mu}{T}}\;Z_C(T,\rho)\nn\\
&=&\lim_{V\rightarrow\infty}\int_{-\infty}^\infty d\rho\;\ex^{-\frac{V}{T}(f(\rho) - \mu N \rho)}\;,
\ea
where $B=\rho V$ and $f$ has now been defined by the canonical ensemble.
 It is then possible to convert back from baryon number to 
chemical potential by
\be
\mu(\rho)=\frac{1}{N}\frac{\partial f(\rho)}{\partial \rho}\;.
\ee
Since the grand canonical partition function at imaginary chemical potential can be evaluated by
Monte Carlo methods, this offers another approach to deal with finite density QCD \cite{Alford:1998sd}.
Numerical data for the grand canonical partition function can be Fourier transformed numerically to 
give the canonical partition function at fixed baryon number. 
However, in this approach the sign problem re-enters at the stage of the Fourier transform. 
The latter clearly
has an oscillatory integrand and thus the same problem we encountered before. The integration can 
only be done for sufficiently small quark numbers $Q$, whereas the thermodynamic limit requires
$Q\rightarrow \infty$ with a fixed quark number density $Q/V=const.$ 
Nevertheless, this is an interesting alternative approach which does not require reweighting or
truncation of Taylor series, and thus is explored numerically \cite{Kratochvila:2005mk,Alexandru:2005ix}. 

\section{Plasma properties at finite density}

Having developed computational tools for finite density, 
one can apply them to the studies discussed in the
previous sections and see how finite baryon densities affect the screening 
masses, the equation of state or the static potential. In all those cases the influence of the 
chemical potential is found to be rather weak. These calculations appear to be
well under control, and we will not further discuss them here.
Instead we outline recent attempts to determine the QCD phase diagram at finite 
density, where the order of the phase transition is expected to change  
as $\mu$ is increased, \fig\ref{fig:1schem}.

\chapter{Towards the QCD phase diagram}

\section{The critical temperature at finite density}

As in the case of zero density, let us first discuss the phase boundary, 
$T_c(\mu)$, before dealing with the order of the phase transition. 
The (pseudo-)critical line has been calculated for a variety of flavours and 
quark masses using different methods. 
Its computation is most straightforward by reweighting, where one directly evaluates susceptibilities
of observables, \eq(\ref{susc}), and locates their maxima to identify the critical couplings. 
With the Taylor expansion method, one computes instead the coefficients of such susceptibilites,
cf.~\eq(\ref{tobs}), which then is known to some order in $\mu/T$. Similarly one can evaluate
susceptibilities at imaginary chemical potentials and locate their maxima as a function of $\mu_i$.  

\begin{figure}[t!]
\centerline{\includegraphics[angle=-90,width=7cm]{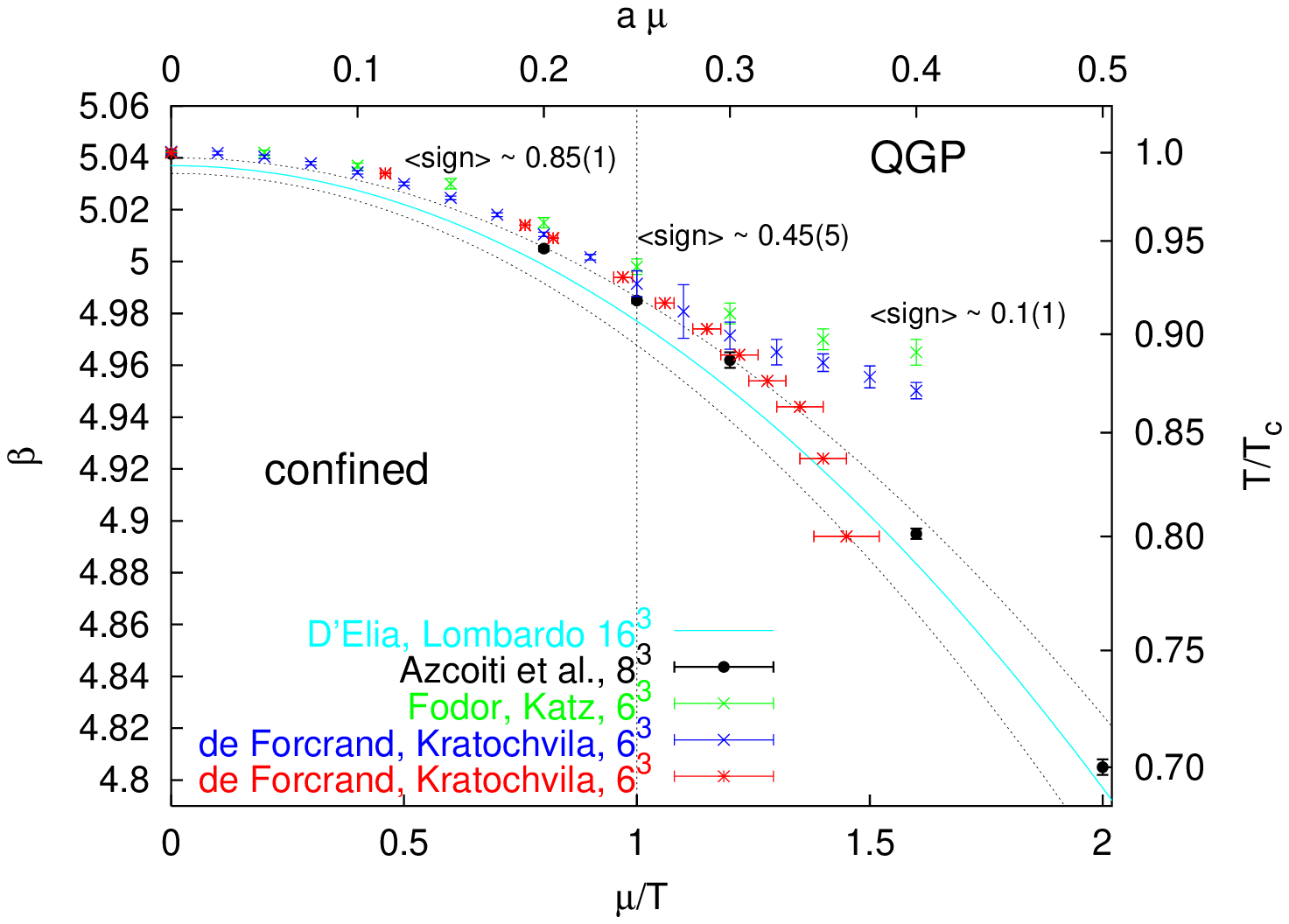}}
\caption[]{Comparison of different methods to compute the critical couplings
\cite{Kratochvila:2005mk}.}
\label{fig:tccomp}
\end{figure}

On the other hand, the (pseudo-)critical gauge coupling  can itself be expressed as a Taylor series. 
It  was defined as an implicit function to be the coupling 
for which a generalised susceptibility peaks. 
The implicit function theorem then guarantees that, if $\chi(\beta,\mu)$ is an analytic function, so is
$\beta_c(\mu)$, and hence
\be
\beta_c(m_f,\frac{\mu}{T})=\sum_{n}b_{2n}(m_f)\left(\frac{\mu}{T}\right)^{2n}\;.
\ee
Using some form of the renormalisation group beta-function, this can be converted into an expansion for the (pseudo-)critical temperature,
\be
\frac{T_c(m_f,\mu)}{T_c(m_f,0)}=1+t_2(m_f)\left(\frac{\mu}{T}\right)^2
+t_4(m_f)\left(\frac{\mu}{T}\right)^4+\ldots
\ee
Thus, the Taylor expansion method as well as simulations at imaginary chemical potential
followed by fits to polynomials allow for independent determinations of the coefficients, providing
valuable cross checks to control the systematics. 
For a quantitative comparison one 
needs data at one fixed parameter set and also eliminate the uncertainties of 
setting the scale. Such a comparison is shown
for the critical coupling in \fig\ref{fig:tccomp}, 
for $N_f=4$ staggered quarks with the same action and quark mass $m/T\approx 0.2$.
(For that quark mass the transition is first order
along the entire curve).
One observes quantitative 
agreement up to $\mu/T\approx 1.3$, after which the different results start 
to scatter. Thus we conclude that all methods discussed here appear 
to be reliable for $\mu/T\lsim 1$. 

For quark mass values close to the physical ones,
one observes that the critical temperature is decreasing only very 
slowly with $\mu$ on coarse lattices. These calculations can be repeated on finer lattices. While the cost for reweighting becomes formidable,
the other methods should allow for a reliable determination of the leading coefficients in the continuum
limit, and thus the physical phase boundary between the hadron and quark gluon plasma phase for $\mu\lsim T$, in the near future. 

\section{The QCD phase diagram for $\mu\neq 0$ and the critical point}

\begin{figure}[t!]
\vspace*{-1cm}
\centerline{
\scalebox{0.65}{\includegraphics{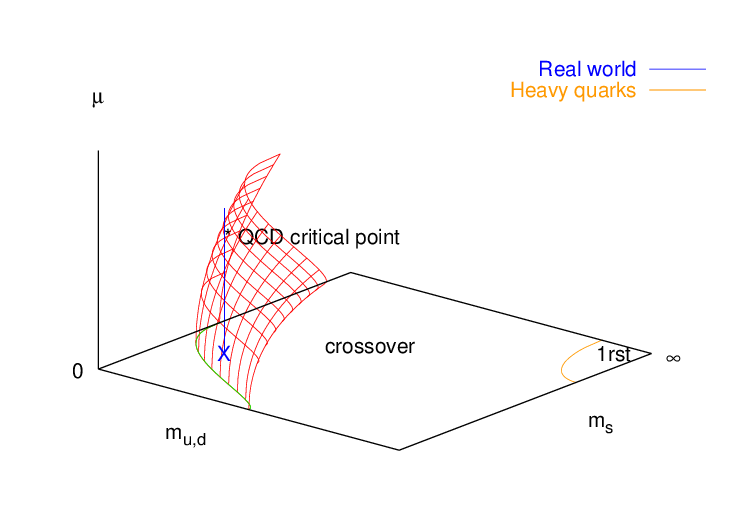}}
\scalebox{0.65}{\includegraphics{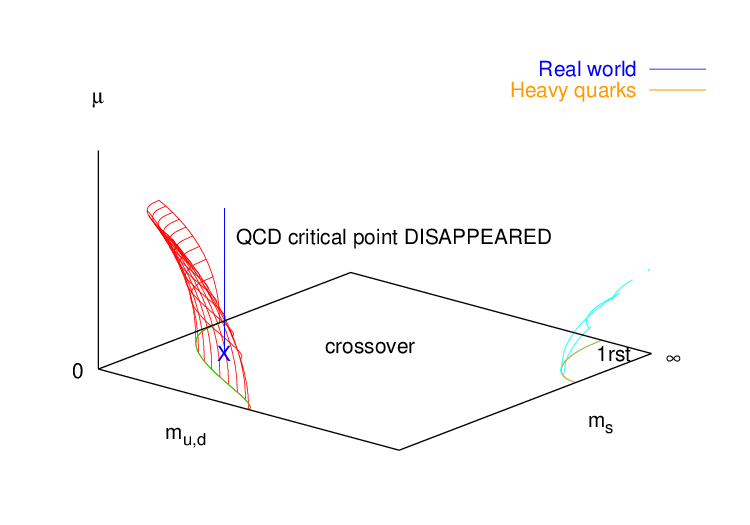}}
}
\centerline{
\scalebox{0.58}{\includegraphics{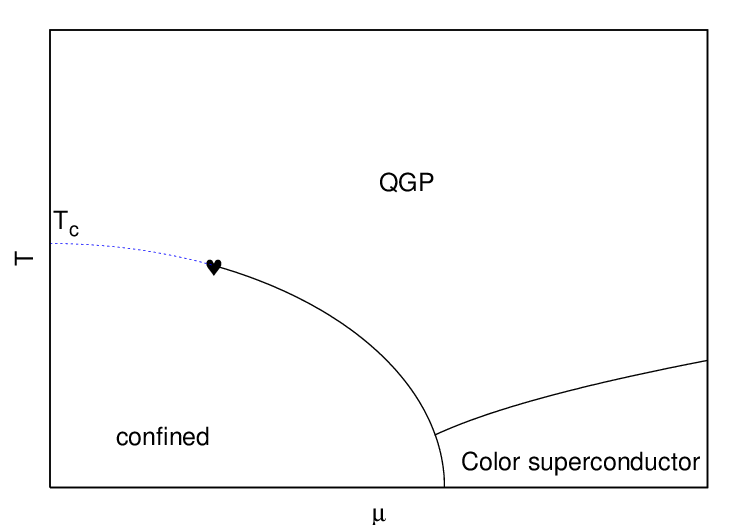}}
\scalebox{0.58}{\includegraphics{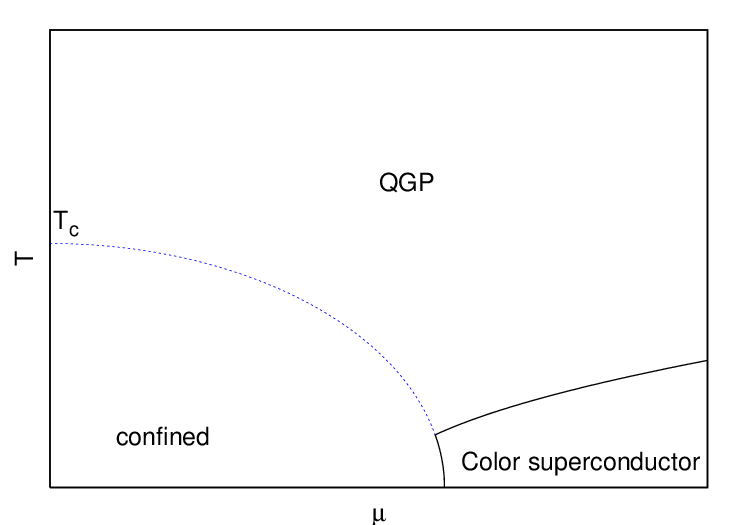}}
}
\caption{Upper panel: 
The chiral critical surface in the case of positive (left) and negative (right) curvature.
If the physical point is in the crossover region for $\mu=0$, a finite $\mu$
chiral critical point will only arise in the scenario (left) with positive curvature,
where the first-order region expands with $\mu$.
Note that for heavy quarks, the first-order region shrinks with $\mu$,
cf.~Sec.~\ref{sec:potts}.
Lower panel: phase diagrams for fixed quark mass (here $N_f=3$) corresponding to 
the two scenarios depicted above.}
\label{fig:3schem}
\end{figure}

As in the case of $\mu=0$, a determination of the order of the transition, and hence the search for the critical endpoint, is much harder, and we begin by discussing the qualitative picture.
If a chemical potential is switched on for the light quarks, there is an additional parameter requiring
an additional axis for our phase diagram characterising the order 
of the transition, \fig\ref{fig:2schem}.
This is shown in \fig\ref{fig:3schem}, where the horizontal plane is spanned by the $\mu=0$ phase diagram in $m_s, m_{u,d}$ and the vertical axis represents $\mu$. 
The critical line separating the first order section from the crossover will now extend to finite $\mu$ and
span a surface. A priori the shape of this surface is not known.
%
However, the expected QCD
phase diagram corresponds to the left scenario of \fig\ref{fig:3schem}. 
The first order region expands as $\mu$ is turned on, so that the
physical point, initially in the crossover region, eventually belongs to the chiral
critical surface. At that chemical potential $\mu_E$, the transition is second order:
that is the QCD chiral critical point. Increasing $\mu$ further makes the transition
first order. In this scenario the transition is generally strenghtened with real chemical potential.
A completely different scenario arises if instead the transition weakens and the first order
region shrinks as $\mu$ is turned on. In that case the physical
point remains in the crossover region and there is no chiral critical point 
for moderate $\mu$, \fig\ref{fig:3schem} (right). 

There are then two different strategies to learn about the QCD phase diagram. 
One can fix a particular set of quark masses and for that theory switch on and increase the chemical potential to see whether a critical surface is crossed or not,  
Sec.~\ref{sec:cp}.
Alternatively, Sec.~\ref{sec:crit} discusses how to start from the known critical line at $\mu=0$
and study its evolution with a finite $\mu$. 

\section{Critical point for fixed masses: reweighting and Taylor expansion}
\label{sec:cp}

\begin{figure}[t]    
\vspace*{-1cm}
\begin{center}
\includegraphics[width=0.5\textwidth]{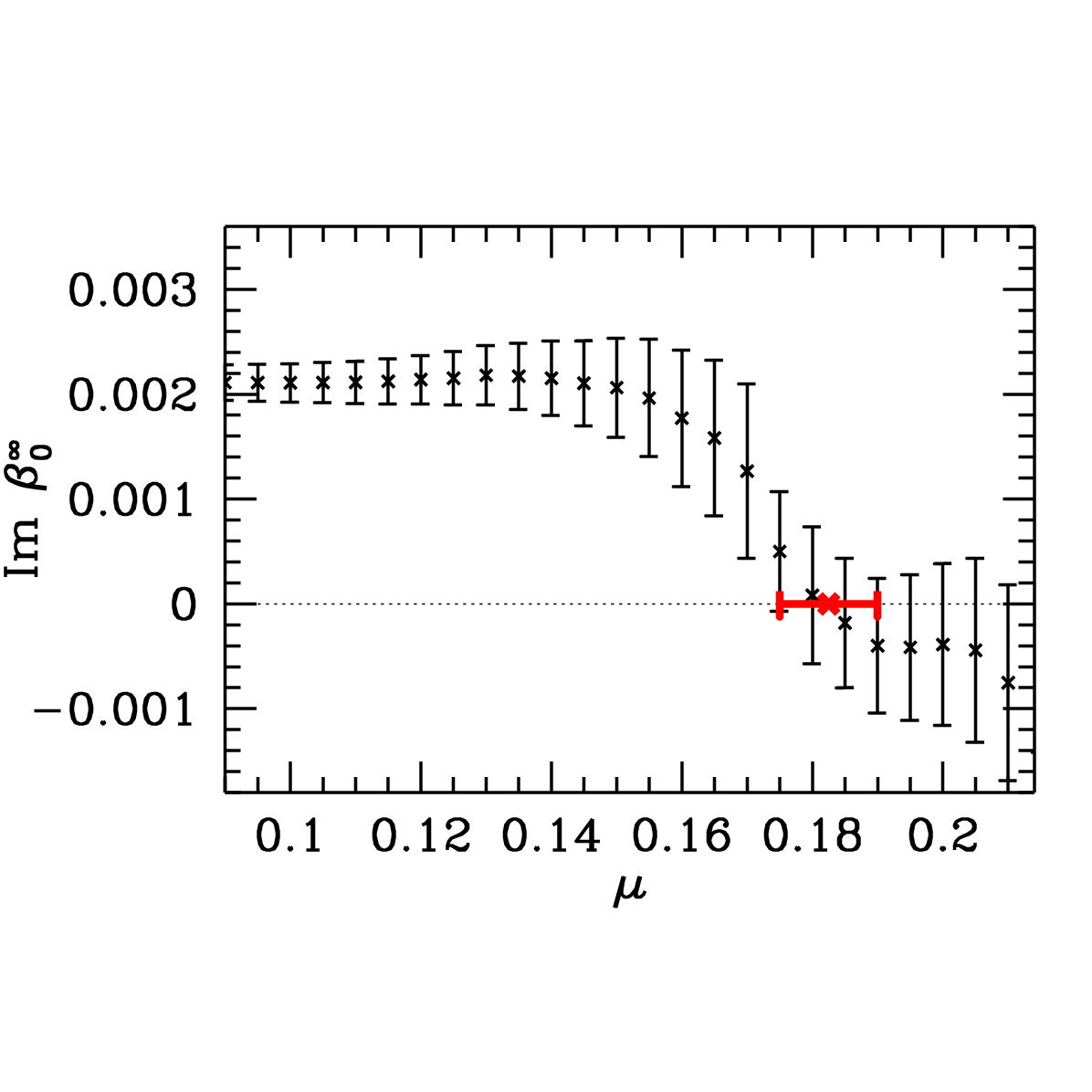}
\end{center}
\vspace*{-1.5cm}
\caption[]{ 
Imaginary part of the Lee-Yang zero closest to the real axis as a 
function of chemical potential \cite{Fodor:2004nz}. }
\label{fk}
\end{figure}

Reweighting methods at physical quark masses get a signal for a critical point 
at $\mu_B^E\sim 360$ MeV \cite{Fodor:2004nz}. 
In this work $L^3\times 4$ lattices with $L=6-12$
were used, working with the standard staggered fermion action. 
Quark masses were tuned to $m_{u,d}/T_c\approx 0.037, m_s/T_c\approx 1$, 
corresponding to the mass ratios 
$m_{\pi}/m_{\rho}\approx 0.19, m_{\pi}/m_K\approx 0.27$, 
which are close to their physical values.
A Lee-Yang zero analysis was employed in order to find the 
change from crossover behaviour at $\mu=0$ to a first order 
transition for $\mu>\mu_E$.
This is shown in \fig\ref{fk}. For a crossover the partition 
function has zeroes only off the real axis,
whereas for a phase transition the zero moves to the real axis 
when extrapolated to infinite volume.  
For a critical discussion of the use of Lee-Yang zeros in 
combination with reweighting, see \cite{Ejiri:2005ts}.
A caveat of this calculation is the observation that the critical point is found in the 
immediate neighbourhood of
the onset of pion condensation in the phase quenched theory, which is where
sign problem becomes maximally severe \cite{Splittorff:2005wc,Han:2008xj}, cf.~the discussion in 
Sec.~\ref{sec:rew}. 
Therefore, one would like to confirm this result with independent methods.

In principle the determination of a critical point is also possible 
via the Taylor expansion.
In this case true phase transitions will be signalled by an 
emerging non-analyticity, or a finite radius of convergence for the pressure series about $\mu=0$,
\eq(\ref{press}), as the volume is increased. 
The radius of convergence of a power series gives the distance between the expansion point and the nearest singularity, and may be extracted from the high order behaviour of the series. Possible 
definitions are
\be
\rho,r = \lim_{n\rightarrow\infty}\rho_n,r_n\qquad \mbox{with}\quad
    \rho_n=\left|\frac{c_0}{c_{2n}}\right|^{1/2n},
     \qquad
   r_n=\left|\frac{c_{2n}}{c_{2n+2}}\right|^{1/2}.
\label{rad}
\ee

General theorems ensure that if the limit exists and asymptotically all coefficients of the series are positive, then there is a singularity on the real axis. 
More details as well as previous applications to strong coupling expansions 
in various spin models can be found in \cite{series}.
In the series for the pressure such a singularity would correspond to the critical point in the $(\mu,T)$-plane.
The current best attempt is based on four consecutive coefficients, 
i.e.~knowledge of the pressure to eighth order, and
a critical endpoint for the $N_f=2$ theory was reported  in \cite{Gavai:2004sd}. 
There are also difficulties in this approach. Firstly, there are different definitions for the radius of convergence, which are only unique in the asymptotic limit, but differ by numerical factors at finite $n$.
Furthermore, the estimated $\rho_n,r_n$ at a given order is neither an upper nor a lower bound on an
actual radius of convergence. Finally, at finite orders the existence of a finite radius of convergence is
a necessary, but not a sufficient condition for the existence of a critical point. For example, 
one also obtains finite estimates for a radius of convergence
from the Taylor coefficients of the hadron resonance gas model, even though that model does not 
feature a non-analytic phase transition. 

\section{The change of the chiral critical line with $\mu$} \label{sec:crit}

Rather than fixing one set of masses and considering the effects of $\mu$, 
one may map out the critical surface in \fig\ref{fig:3schem} by measuring
how the $\mu=0$ critical boundary line changes under the influence of $\mu$.
For example, for a given value of the strange quark mass the corresponding critical $m_{u,d}$ has 
again a Taylor expansion,
\be
\frac{m_{u,d}^c(\mu)}{m_{u,d}^c(0)}=1+c_1\left(\frac{\mu}{T}\right)^2 +c_2 \left(\frac{\mu}{T}\right)^4+\ldots
\label{mexp}
\ee
The curvature of the critical surface in lattice units is directly related to the behaviour of the Binder cumulant via the chain rule,
\be
\frac{dam_c}{d(a\mu)^2}=-\frac{\partial B_4}{\partial (a\mu)^2}
\left(\frac{\partial B_4}{\partial am}\right)^{-1}\,,
\ee
and similar expressions for the higher derivatives.
While the second factor is sizeable and easy to evaluate, the $\mu$-dependence
of the cumulant is excessively weak and requires enormous statistics to extract. In order to guard
against systematic errors, this derivative can be evaluated in two independent ways.
One is to fit the corresponding Taylor series of $B_4$ in powers of $\mu/T$ to data generated at 
imaginary chemical potential, the other to compute the derivative directly and without
fitting via the finite difference quotient,
\be
\frac{\partial B_4}{\partial (a\mu)^2}=\lim_{(a\mu)^2\rightarrow 0}\frac{B_4(a\mu)-B_4(0)}{(a\mu)^2}.
\ee 
Because the required shift in the couplings is very small,
it is adequate and safe to use the original Monte Carlo ensemble 
for $am^c(0),\mu=0$ and reweight the results by the standard 
Ferrenberg-Swendsen method. 
Moreover, by reweighting to imaginary $\mu$
the reweighting factors remain real positive and close to 1.

On coarse $N_\tau=4$ lattices, the first two coefficients in \eq(\ref{mexp}) are found to be negative, 
\cite{deForcrand:2008vr},
hence the region featuring a first order chiral transition is shrinking when a real chemical potential is 
turned on. This implies that, for moderate $\mu\lsim T$, there is no critical point belonging to the chiral
critical surface. An open questions here is what the higher order corrections are. For example,
a sudden change of behaviour as in \fig\ref{fk} would be difficult to capture with only few Taylor 
coefficients. 
Also, there might be 
a chiral critical point at larger chemical potentials and finally  we do not yet know yet how the curvature of the critical surface behaves in the continuum limit.
In addition, these findings do not exclude a critical point that is 
not associated with the chiral critical surface. 

\section{The change of the deconfinement critical line with $\mu$}
\label{sec:potts}

In light of these results, it is interesting to compare with the situation in
the heavy quark corner of the schematic phase diagram, \fig\ref{fig:3schem}.
As the quark mass goes to infinity, quarks can be integrated out and QCD reduces to a gauge theory of Polyakov lines. 
At a second order phase transition, universality allows us to neglect the 
details of gauge degrees of freedom, so the theory reduces to the 
3d three-state Potts model, which is in the appropriate 3d Ising universality class. 
Hence, studying the three-state Potts model should teach us about the behaviour of QCD in the neighbourhood of the deconfinement critical line separating the quenched first order region from the crossover region.

\begin{figure}
\centerline{
\includegraphics*[width=0.45\textwidth]{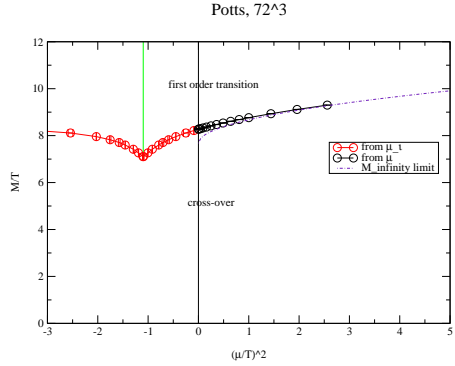}
}
\caption[]{The critical heavy quark mass separating first order from crossover as a function of $\mu^2$ from the Potts model \cite{Kim:2005ck}.}
\label{potts}
\end{figure}

Here we are interested in simulations at small chemical potential.  In this case, the sign problem is mild enough for brute force simulations at real $\mu$ to be feasible. 
In \cite{Kim:2005ck}, the change of the critical heavy quark mass is 
determined as a function of real as well as imaginary $\mu$, 
as shown in \fig\ref{potts}. 
Note that $M^c(\mu)$ rises with real chemical potential, such that the first order region 
in \fig\ref{fig:2schem} 
shrinks as finite baryon density is switched on. 
The calculation also gives some insight in the problem of analytic continuation: \fig\ref{potts} clearly endorses the approach with $M_c(\mu)$ passing analytically through $\mu=0$. 
However, high order fits might be required in practice in order to reproduce the data
on both sides of $\mu^2=0$. 

The same qualitative behaviour is observed in a combined strong coupling and hopping parameter expansion \cite{Langelage:2009jb}. While such a calculation is only accurate for coarse lattices, the universal features of the critical surface should be cut-off independent.
Thus, QCD with heavy quarks is an example of the non-standard scenario 
discussed in the previous section. 

\section{The critical surfaces at imaginary $\mu$}

The chiral and deconfinement critical surfaces, upper \fig\ref{fig:3schem} (right), continue to 
imaginary $\mu$, which has been used to determine their curvature.  Rather than studying the
neighbourhood of $\mu=0$ with an aim to analytic continuation, it is also illuminating
to follow those surfaces into the imaginary regime, until the phase boundary to the 
next $Z(3)$-sector, \fig\ref{schem} (left), is hit. This has already been done in the case of the deconfinement transition within the Potts model, cf.~the curve at negative $\mu^2$,
\fig\ref{potts}. The meeting point of the
deconfinement critical line with the critical endpoint of the $Z(3)$ transition is then tri-critical, and 
corresponds precisely to the tri-critical point in the large mass region of \fig\ref{fig:2schem} (right). 
\begin{figure}[t!!!]
\centerline{
\includegraphics[width=0.3\textwidth]{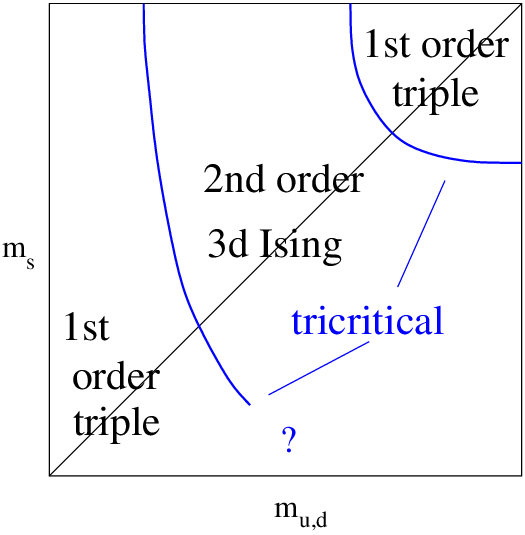} \hspace*{1cm}
\includegraphics[width=0.5\textwidth]{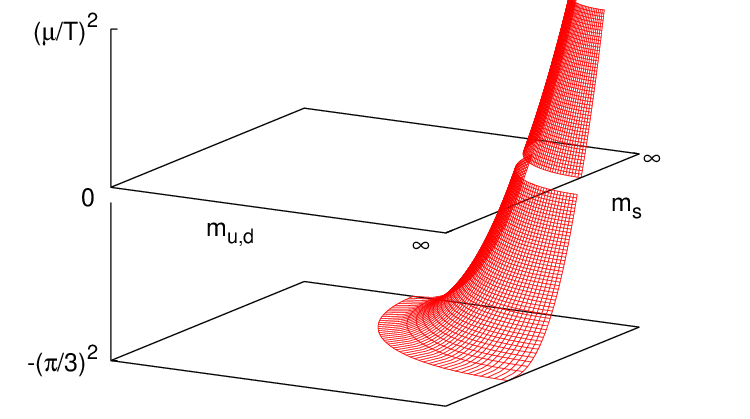}
}
\caption[]{Left: Nature of the $Z(3)$-transition endpoint at $\mu/T=i\pi/3$. 
\cite{deForcrand:2010he}. Right: Deconfinement critical surface for real and imaginary $\mu$.}
\label{col}
\end{figure}

For three flavours of quarks and varying quark masses, one can now draw a schematic diagram
analogous to  \fig\ref{fig:1schem}, showing the nature of the junction at
$\mu=i\pi T/3$. 
At the time of writing, only the two tri-critical points on the $N_f=3$ diagonal and the light mass one
for $N_f=2$ have been determined on coarse lattices. 
A natural extension would then be that there are two tri-critical lines which delimit the 
quark mass sections for which the junction
corresponds to triple points. The area in between corresponds to a second order endpoints
of the $Z(3)$-transition, with no connection to the chiral/deconfinement transition which is 
merely a crossover there. The tri-critical lines represent the boundaries in which the chiral and deconfinement
critical surfaces end at imaginary $\mu=i\pi T/3$, before they periodically repeat. 
On coarse lattices they are found ``outside'' the corresponding $\mu=0$ critical lines, which demonstrates
an increase of the first order area in the imaginary direction. In fact, for heavy quarks this change 
is monotonic.  
The functional form of the curve in \fig\ref{potts}, and hence the surface in \fig\ref{col} (right),
was found to be determined by tri-critical scaling for the whole range up to real $\mu\sim T$.
This would explain the 
curvature of the critical surfaces, i.e.~the weakening of the chiral and deconfinement transitions with
real chemical potential.

\section{Discussion}

It thus appears that first order transitions are weakened when a chemical potential for baryon number is
switched on. The same observation is made for finite isospin chemical potential \cite{Kogut:2007mz}.
Note however, that most of these computations are done on $N_\tau=4,6$ lattices, where 
cut-off effects appear to be larger than finite density effects. Hence, 
definite conclusions for continuum physics cannot yet be drawn. A general finding is the 
steepness of the critical surface, making the location of a possible critical endpoint extremely quark mass sensitive, and hence difficult to determine accurately.
Furthermore, the chiral first order region is also observed to shrink with decreasing lattice spacing,
such that in \fig\ref{fig:2schem} $m_s^{tric}<m_s^{phys}$.
In this case the chiral critical surface we have been discussing here
might not be responsible for a possible critical point, regardless of its curvature, but another surface emanating
from the putative $O(4)$-chiral limit, or one connected to finite density physics, or.....
Possibilities and uncertainties abound! 
To weed them out we need to conclusively understand
the situation in the $N_f=2$ chiral limit. Finally, even if those issues get settled, all the methods 
discussed here are only valid for $\mu\lsim T$. Thus, learning about high density QCD, 
relevant for compact stars, requires entirely new theoretical methods. 
In summary, it remains a challenging but also most exciting task to settle   
even the qualitative features of the QCD phase diagram. 

\bibliographystyle{OUPnamed_notitle}
\bibliography{LH}

\thebibliography{0}

\bibitem[\protect\citeauthoryear{Alexandru, Faber, Horvath and Liu}{Alexandru
  {\em et~al.}}{2005}]{Alexandru:2005ix}
Alexandru, A., Faber, M., Horvath, I., and Liu, K.-F. (2005).
\newblock {\em Phys. Rev.\/},~{\bf D72}, 114513.

\bibitem[\protect\citeauthoryear{Alford, Kapustin and Wilczek}{Alford {\em
  et~al.}}{1999}]{Alford:1998sd}
Alford, Mark~G., Kapustin, A., and Wilczek, F. (1999).
\newblock {\em Phys. Rev.\/},~{\bf D59}, 054502.

\bibitem[\protect\citeauthoryear{Allton {\em et~al.}}{Allton {\em
  et~al.}}{2002}]{Allton:2002zi}
Allton, C.~R. {\em {\em et~al.}} (2002).
\newblock {\em Phys. Rev.\/},~{\bf D66}, 074507.

\bibitem[\protect\citeauthoryear{Aoki, Endrodi, Fodor, Katz and Szabo}{Aoki
  {\em et~al.}}{2006}]{Aoki:2006we}
Aoki, Y., Endrodi, G., Fodor, Z., Katz, S.~D., and Szabo, K.~K. (2006).
\newblock {\em Nature\/},~{\bf 443}, 675--678.

\bibitem[\protect\citeauthoryear{Brambilla, Ghiglieri, Vairo and
  Petreczky}{Brambilla {\em et~al.}}{2008}]{Brambilla:2008cx}
Brambilla, N., Ghiglieri, J., Vairo, A., and Petreczky, Peter (2008).
\newblock {\em Phys. Rev.\/},~{\bf D78}, 014017.

\bibitem[\protect\citeauthoryear{Chandrasekharan and Mehta}{Chandrasekharan and
  Mehta}{2007}]{Chandrasekharan:2007up}
Chandrasekharan, S. and Mehta, A.~C. (2007).
\newblock {\em Phys. Rev. Lett.\/},~{\bf 99}, 142004.

\bibitem[\protect\citeauthoryear{Datta and Gupta}{Datta and
  Gupta}{1998}]{Datta:1998eb}
Datta, S. and Gupta, S. (1998).
\newblock {\em Nucl. Phys.\/},~{\bf B534}, 392--416.

\bibitem[\protect\citeauthoryear{de~Forcrand, Kim and Philipsen}{de~Forcrand
  {\em et~al.}}{2007}]{deForcrand:2007rq}
de~Forcrand, Ph., Kim, S., and Philipsen, O. (2007).
\newblock {\em PoS\/},~{\bf LAT2007}, 178.

\bibitem[\protect\citeauthoryear{de~Forcrand, Kim and Takaishi}{de~Forcrand
  {\em et~al.}}{2003}]{deForcrand:2002pa}
de~Forcrand, Ph., Kim, S., and Takaishi, T. (2003).
\newblock {\em Nucl. Phys. Proc. Suppl.\/},~{\bf 119}, 541--543.

\bibitem[\protect\citeauthoryear{de~Forcrand and Philipsen}{de~Forcrand and
  Philipsen}{2002}]{deForcrand:2002ci}
de~Forcrand, Ph. and Philipsen, O. (2002).
\newblock {\em Nucl. Phys.\/},~{\bf B642}, 290--306.

\bibitem[\protect\citeauthoryear{de~Forcrand and Philipsen}{de~Forcrand and
  Philipsen}{2007}]{deForcrand:2006pv}
de~Forcrand, Ph. and Philipsen, O. (2007).
\newblock {\em JHEP\/},~{\bf 01}, 077.

\bibitem[\protect\citeauthoryear{de~Forcrand and Philipsen}{de~Forcrand and
  Philipsen}{2008}]{deForcrand:2008vr}
de~Forcrand, Ph. and Philipsen, O. (2008).
\newblock {\em JHEP\/},~{\bf 11}, 012.

\bibitem[\protect\citeauthoryear{de~Forcrand and Philipsen}{de~Forcrand and
  Philipsen}{2010}]{deForcrand:2010he}
de~Forcrand, Ph. and Philipsen, O. (2010).
\newblock {\em arxiv:1004.3144\/}.

\bibitem[\protect\citeauthoryear{D'Elia and Lombardo}{D'Elia and
  Lombardo}{2003}]{D'Elia:2002gd}
D'Elia, M. and Lombardo, M.-P. (2003).
\newblock {\em Phys. Rev.\/},~{\bf D67}, 014505.

\bibitem[\protect\citeauthoryear{D'Elia and Sanfilippo}{D'Elia and
  Sanfilippo}{2009}]{D'Elia:2009qz}
D'Elia, M. and Sanfilippo, F. (2009).
\newblock {\em Phys. Rev.\/},~{\bf D80}, 111501.

\bibitem[\protect\citeauthoryear{Ejiri}{Ejiri}{2006}]{Ejiri:2005ts}
Ejiri, S. (2006).
\newblock {\em Phys. Rev.\/},~{\bf D73}, 054502.

\bibitem[\protect\citeauthoryear{Endrodi, Fodor, Katz and Szabo}{Endrodi {\em
  et~al.}}{2007}]{Endrodi:2007gc}
Endrodi, G., Fodor, Z., Katz, S.~D., and Szabo, K.~K. (2007).
\newblock {\em PoS\/},~{\bf LAT2007}, 182.

\bibitem[\protect\citeauthoryear{Engels, Fingberg, Karsch, Miller and
  Weber}{Engels {\em et~al.}}{1990}]{Engels:1990vr}
Engels, J., Fingberg, J., Karsch, F., Miller, D., and Weber, M. (1990).
\newblock {\em Phys. Lett.\/},~{\bf B252}, 625--630.

\bibitem[\protect\citeauthoryear{Ferrenberg and Swendsen}{Ferrenberg and
  Swendsen}{1988}]{Ferrenberg:1988yz}
Ferrenberg, A.~M. and Swendsen, R.~H. (1988).
\newblock {\em Phys. Rev. Lett.\/},~{\bf 61}, 2635--2638.

\bibitem[\protect\citeauthoryear{Fodor and Katz}{Fodor and
  Katz}{2002}]{Fodor:2001au}
Fodor, Z. and Katz, S.~D. (2002).
\newblock {\em Phys. Lett.\/},~{\bf B534}, 87--92.

\bibitem[\protect\citeauthoryear{Fodor and Katz}{Fodor and
  Katz}{2004}]{Fodor:2004nz}
Fodor, Z. and Katz, S.~D. (2004).
\newblock {\em JHEP\/},~{\bf 04}, 050.

\bibitem[\protect\citeauthoryear{Gavai}{Gavai}{1985}]{Gavai:1985ie}
Gavai, R.~V. (1985).
\newblock {\em Phys. Rev.\/},~{\bf D32}, 519.

\bibitem[\protect\citeauthoryear{Gavai and Gupta}{Gavai and
  Gupta}{2005}]{Gavai:2004sd}
Gavai, R.~V. and Gupta, S. (2005).
\newblock {\em Phys. Rev.\/},~{\bf D71}, 114014.

\bibitem[\protect\citeauthoryear{Grossman, Gupta, Heller and Karsch}{Grossman
  {\em et~al.}}{1994}]{Grossman:1993wm}
Grossman, B., Gupta, S., Heller, U.~M., and Karsch, F. (1994).
\newblock {\em Nucl. Phys.\/},~{\bf B417}, 289--306.

\bibitem[\protect\citeauthoryear{Gupta}{Gupta}{1999}]{Gupta:1999hp}
Gupta, S. (1999).
\newblock {\em Phys. Rev.\/},~{\bf D60}, 094505.

\bibitem[\protect\citeauthoryear{Guttman}{Guttman}{1989}]{series}
Guttman, A.~J. (1989).
\newblock In {\em Phase transitions and critical phenomena} (ed. C.~Domb and
  J.~L. Lebowitz), Number~13, p.~1. Academic Press, London.

\bibitem[\protect\citeauthoryear{Halasz, Jackson, Shrock, Stephanov and
  Verbaarschot}{Halasz {\em et~al.}}{1998}]{Halasz:1998qr}
Halasz, A.~M., Jackson, A.~D., Shrock, R.~E., Stephanov, M.~A., and
  Verbaarschot, J. J.~M. (1998).
\newblock {\em Phys. Rev.\/},~{\bf D58}, 096007.

\bibitem[\protect\citeauthoryear{Han and Stephanov}{Han and
  Stephanov}{2008}]{Han:2008xj}
Han, J. and Stephanov, M.~A. (2008).
\newblock {\em Phys. Rev.\/},~{\bf D78}, 054507.

\bibitem[\protect\citeauthoryear{Hands, Sitch and Skullerud}{Hands {\em
  et~al.}}{2008}]{Hands:2007uc}
Hands, S., Sitch, P., and Skullerud, J.-I. (2008).
\newblock {\em Phys. Lett.\/},~{\bf B662}, 405.

\bibitem[\protect\citeauthoryear{Hart, Laine and Philipsen}{Hart {\em
  et~al.}}{2000}]{Hart:2000ha}
Hart, A., Laine, M., and Philipsen, O. (2000).
\newblock {\em Nucl. Phys.\/},~{\bf B586}, 443--474.

\bibitem[\protect\citeauthoryear{Hasenfratz and Karsch}{Hasenfratz and
  Karsch}{1983}]{Hasenfratz:1983ba}
Hasenfratz, P. and Karsch, F. (1983).
\newblock {\em Phys. Lett.\/},~{\bf B125}, 308.

\bibitem[\protect\citeauthoryear{Hegde, Karsch, Laermann and Shcheredin}{Hegde
  {\em et~al.}}{2008}]{Hegde:2008nx}
Hegde, P., Karsch, F., Laermann, E., and Shcheredin, S. (2008).
\newblock {\em Eur. Phys. J.\/},~{\bf C55}, 423--437.

\bibitem[\protect\citeauthoryear{Jahn and Philipsen}{Jahn and
  Philipsen}{2004}]{Jahn:2004qr}
Jahn, O. and Philipsen, O. (2004).
\newblock {\em Phys. Rev.\/},~{\bf D70}, 074504.

\bibitem[\protect\citeauthoryear{Kaczmarek, Karsch, Laermann and
  L{\"u}tgemeier}{Kaczmarek {\em et~al.}}{2000}]{Kaczmarek:1999mm}
Kaczmarek, O., Karsch, F., Laermann, E., and L{\"u}tgemeier, Martin (2000).
\newblock {\em Phys. Rev.\/},~{\bf D62}, 034021.

\bibitem[\protect\citeauthoryear{Kapusta and Gale}{Kapusta and
  Gale}{2006}]{Kapusta:2006pm}
Kapusta, J.~I. and Gale, C. (2006).
\newblock {\em {Finite-temperature field theory: Principles and applications}}.
\newblock Cambridge, UK: Univ. Pr. (2006) 428 p.

\bibitem[\protect\citeauthoryear{Karsch}{Karsch}{2002}]{Karsch:2001cy}
Karsch, F. (2002).
\newblock {\em Lect. Notes Phys.\/},~{\bf 583}, 209--249.

\bibitem[\protect\citeauthoryear{Karsch, Laermann and Peikert}{Karsch {\em
  et~al.}}{2000}]{Karsch:2000ps}
Karsch, F., Laermann, E., and Peikert, A. (2000).
\newblock {\em Phys. Lett.\/},~{\bf B478}, 447--455.

\bibitem[\protect\citeauthoryear{Karsch, Laermann and Schmidt}{Karsch {\em
  et~al.}}{2001}]{Karsch:2001nf}
Karsch, F., Laermann, E., and Schmidt, C. (2001).
\newblock {\em Phys. Lett.\/},~{\bf B520}, 41--49.

\bibitem[\protect\citeauthoryear{Karsch, Redlich and Tawfik}{Karsch {\em
  et~al.}}{2003}]{Karsch:2003vd}
Karsch, F., Redlich, K., and Tawfik, A. (2003).
\newblock {\em Eur. Phys. J.\/},~{\bf C29}, 549--556.

\bibitem[\protect\citeauthoryear{Karsch and Wyld}{Karsch and
  Wyld}{1985}]{Karsch:1985cb}
Karsch, F. and Wyld, H.~W. (1985).
\newblock {\em Phys. Rev. Lett.\/},~{\bf 55}, 2242.

\bibitem[\protect\citeauthoryear{Kaste and Rothe}{Kaste and
  Rothe}{1997}]{Kaste:1997ks}
Kaste, P. and Rothe, H.~J. (1997).
\newblock {\em Phys. Rev.\/},~{\bf D56}, 6804--6815.

\bibitem[\protect\citeauthoryear{Kim, de~Forcrand, Kratochvila and
  Takaishi}{Kim {\em et~al.}}{2006}]{Kim:2005ck}
Kim, S., de~Forcrand, Ph., Kratochvila, S., and Takaishi, T. (2006).
\newblock {\em PoS\/},~{\bf LAT2005}, 166.

\bibitem[\protect\citeauthoryear{Kogut and Sinclair}{Kogut and
  Sinclair}{2008}]{Kogut:2007mz}
Kogut, J.~B. and Sinclair, D.~K. (2008).
\newblock {\em Phys. Rev.\/},~{\bf D77}, 114503.

\bibitem[\protect\citeauthoryear{Kratochvila and de~Forcrand}{Kratochvila and
  de~Forcrand}{2006}]{Kratochvila:2005mk}
Kratochvila, S. and de~Forcrand, Ph. (2006).
\newblock {\em PoS\/},~{\bf LAT2005}, 167.

\bibitem[\protect\citeauthoryear{Laine, Philipsen, Romatschke and
  Tassler}{Laine {\em et~al.}}{2007}]{Laine:2006ns}
Laine, M., Philipsen, O., Romatschke, P., and Tassler, M. (2007).
\newblock {\em JHEP\/},~{\bf 03}, 054.

\bibitem[\protect\citeauthoryear{Laine and Veps{\"a}l{\"a}inen}{Laine and
  Veps{\"a}l{\"a}inen}{2004}]{Laine:2003bd}
Laine, M. and Veps{\"a}l{\"a}inen, M. (2004).
\newblock {\em JHEP\/},~{\bf 02}, 004.

\bibitem[\protect\citeauthoryear{Langelage, M{\"u}nster and
  Philipsen}{Langelage {\em et~al.}}{2007}]{Langelage:2007pi}
Langelage, J., M{\"u}nster, G., and Philipsen, O. (2007).
\newblock {\em PoS\/},~{\bf LAT2007}, 201.

\bibitem[\protect\citeauthoryear{Langelage, M{\"u}nster and
  Philipsen}{Langelage {\em et~al.}}{2008}]{Langelage:2008dj}
Langelage, J., M{\"u}nster, G., and Philipsen, O. (2008).
\newblock {\em JHEP\/},~{\bf 07}, 036.

\bibitem[\protect\citeauthoryear{Langelage and Philipsen}{Langelage and
  Philipsen}{2010{\em a}}]{Langelage:2009jb}
Langelage, J. and Philipsen, O. (2010{\em a}).
\newblock {\em JHEP\/},~{\bf 01}, 089.

\bibitem[\protect\citeauthoryear{Langelage and Philipsen}{Langelage and
  Philipsen}{2010{\em b}}]{Langelage:2010yn}
Langelage, J. and Philipsen, O. (2010{\em b}).
\newblock {\em JHEP\/},~{\bf 04}, 055.

\bibitem[\protect\citeauthoryear{Lawrie and Sarbach}{Lawrie and
  Sarbach}{1984}]{tric}
Lawrie, I.~D. and Sarbach, S. (1984).
\newblock In {\em Phase transitions and critical phenomena} (ed. C.~Domb and
  J.~L. Lebowitz), Number~9, p.~1. Academic Press, London.

\bibitem[\protect\citeauthoryear{Lee and Yang}{Lee and Yang}{1952}]{Lee:1952ig}
Lee, T.~D. and Yang, C.~N. (1952).
\newblock {\em Phys. Rev.\/},~{\bf 87}, 410--419.

\bibitem[\protect\citeauthoryear{Matsui and Satz}{Matsui and
  Satz}{1986}]{Matsui:1986dk}
Matsui, T. and Satz, H. (1986).
\newblock {\em Phys. Lett.\/},~{\bf B178}, 416.

\bibitem[\protect\citeauthoryear{McLerran and Svetitsky}{McLerran and
  Svetitsky}{1981}]{McLerran:1981pb}
McLerran, L.~D. and Svetitsky, B. (1981).
\newblock {\em Phys. Rev.\/},~{\bf D24}, 450.

\bibitem[\protect\citeauthoryear{Montvay and M{\"u}nster}{Montvay and
  M{\"u}nster}{1994}]{Montvay:1994cy}
Montvay, I. and M{\"u}nster, G. (1994).
\newblock {\em {Quantum fields on a lattice}}.
\newblock Cambridge, UK: Univ. Pr. (1994) 491 p. (Cambridge monographs on
  mathematical physics).

\bibitem[\protect\citeauthoryear{M{\"u}nster}{M{\"u}nster}{1981}]{Munster:1981%
es}
M{\"u}nster, G. (1981).
\newblock {\em Nucl. Phys.\/},~{\bf B190}, 439.

\bibitem[\protect\citeauthoryear{Nadkarni}{Nadkarni}{1986}]{Nadkarni:1986as}
Nadkarni, S. (1986).
\newblock {\em Phys. Rev.\/},~{\bf D34}, 3904.

\bibitem[\protect\citeauthoryear{Philipsen and Zeidlewicz}{Philipsen and
  Zeidlewicz}{2010}]{Philipsen:2008gq}
Philipsen, O. and Zeidlewicz, L. (2010).
\newblock {\em Phys. Rev.\/},~{\bf D81}, 077501.

\bibitem[\protect\citeauthoryear{Rajagopal and Wilczek}{Rajagopal and
  Wilczek}{2000}]{Rajagopal:2000wf}
Rajagopal, K. and Wilczek, F. (2000).
\newblock In {\em At the frontier of particle physics} (ed. M.~Shifman),
  Number~3, p. 2061.

\bibitem[\protect\citeauthoryear{Rebhan}{Rebhan}{1993}]{Rebhan:1993az}
Rebhan, A.~K. (1993).
\newblock {\em Phys. Rev.\/},~{\bf D48}, 3967--3970.

\bibitem[\protect\citeauthoryear{Roberge and Weiss}{Roberge and
  Weiss}{1986}]{Roberge:1986mm}
Roberge, A. and Weiss, N. (1986).
\newblock {\em Nucl. Phys.\/},~{\bf B275}, 734.

\bibitem[\protect\citeauthoryear{Schor}{Schor}{1983}]{Schor:1983py}
Schor, R.~S. (1983).
\newblock {\em Phys. Lett.\/},~{\bf B132}, 161.

\bibitem[\protect\citeauthoryear{Seo}{Seo}{1982}]{Seo:1982jh}
Seo, K. (1982).
\newblock {\em Nucl. Phys.\/},~{\bf B209}, 200--216.

\bibitem[\protect\citeauthoryear{Son and Stephanov}{Son and
  Stephanov}{2001}]{Son:2000xc}
Son, D.~T. and Stephanov, M.~A. (2001).
\newblock {\em Phys. Rev. Lett.\/},~{\bf 86}, 592--595.

\bibitem[\protect\citeauthoryear{Splittorff}{Splittorff}{2005}]{Splittorff:200%
5wc}
Splittorff, K. (2005).
\newblock {\em hep-lat/0505001\/}.

\bibitem[\protect\citeauthoryear{Wiese}{Wiese}{2002}]{Wiese:2002ws}
Wiese, U.~J. (2002).
\newblock {\em Nucl. Phys.\/},~{\bf A702}, 211--216.

\bibitem[\protect\citeauthoryear{Yang and Lee}{Yang and
  Lee}{1952}]{Yang:1952be}
Yang, Chen-Ning and Lee, T.~D. (1952).
\newblock {\em Phys. Rev.\/},~{\bf 87}, 404--409.

\endthebibliography

\end{document}